\documentclass[showpacs,amsmath,amssymb,twocolumn,superscriptaddress,notitlepage,preprintnumbers,pra]{revtex4-1}
\usepackage[dvips]{graphicx}
\usepackage{amsmath,amssymb,amsthm,mathrsfs,amsfonts,dsfont}
\usepackage{subfigure, epsfig}
\usepackage{braket}
\usepackage{bm}
\usepackage{enumerate}
\usepackage{color,xcolor}
\usepackage{physics}
\usepackage{qcircuit}
\usepackage{graphicx}
\usepackage{algorithm}
\usepackage{algpseudocode}
\usepackage{hyperref}
\hypersetup{colorlinks=true, linkcolor=blue, citecolor=blue, urlcolor=black }
\usepackage{comment}
\usepackage[normalem]{ulem}
\usepackage{booktabs}
\usepackage{mathtools}

\newcommand{\cO}{\mathcal{O}}

\DeclareMathOperator*{\argmax}{arg\,max}
\DeclareMathOperator*{\argmin}{arg\,min}

\newcommand{\pphys}{p}

\newcommand{\mc}[1]{\mathcal{#1}}

\DeclareMathOperator{\e}{e}

\usepackage{varioref}
\labelformat{equation}{(#1)}

\labelformat{section}{#1}

\labelformat{figure}{#1}

\labelformat{proposition}{#1}

\labelformat{lemma}{#1}

\labelformat{theorem}{#1}

\labelformat{observation}{#1}

\labelformat{definition}{#1}

\labelformat{problem}{#1}

\labelformat{algorithm}{#1}

\labelformat{corollary}{#1}

\labelformat{statement}{#1}

\newcommand{\suppref}[1]{Supplementary Information~\ref{#1}}

\makeatother

\newcommand{\sun}[1]{\textcolor{black}{#1}}

\newcommand{\gyt}[1]{\textcolor{black}{#1}}

\begin{document}
 
\title{Quantum-classical crossover in fault-tolerant quantum dynamics simulation
}
 
\date{\today}

% --------------------  ABSTRACT  --------------------
% ------------  AUTHORS AND AFFILIATIONS ----------

\author{Jinzhao Sun     }
\thanks{These authors contributed equally to this work.}
% \email{jinzhao.sun.phys@gmail.com}
\affiliation{School of Physical and Chemical Sciences, Queen Mary University of London, London, E1 4NS, United Kingdom}
\affiliation{Clarendon Laboratory, University of Oxford, Parks Road, Oxford, OX1 3PU, United Kingdom}

\author{Bozhen Zhou }
\thanks{These authors contributed equally to this work.}
\affiliation{Institute of Theoretical Physics, Chinese Academy of Sciences, Beijing 100190, China}

\author{Jue Xu }
\thanks{These authors contributed equally to this work.}
\affiliation{QICI Quantum Information and Computation Initiative, Department of Computer Science, The University of Hong Kong, Pokfulam Road, Hong Kong}

\author{Yuan Yao }
\thanks{These authors contributed equally to this work.}
\affiliation{Center on Frontiers of Computing Studies, Peking University, Beijing 100871, China}

\author{Zhenyu Du }
\thanks{These authors contributed equally to this work.}
\affiliation{Center for Quantum Information, Institute for Interdisciplinary
Information Sciences, Tsinghua University, Beijing 100084, China}

\author{Zixu Zhang }
\thanks{These authors contributed equally to this work.}
\affiliation{School of Physics and Astronomy, Shanghai Jiao Tong University, Shanghai 200240, China}%

\author{Yuntian Gu}
\thanks{These authors contributed equally to this work.}
\affiliation{School of Intelligence Science and Technology, Peking University, Beijing 100871, China}

\author{Junxiang Huang}
\affiliation{Center on Frontiers of Computing Studies, Peking University, Beijing 100871, China}

\author{Shuo Zhou}
\affiliation{Center on Frontiers of Computing Studies, Peking University, Beijing 100871, China}

\author{Ziruo Wang}
\affiliation{Center on Frontiers of Computing Studies, Peking University, Beijing 100871, China}

\author{Alexander Yosifov}
\affiliation{Clarendon Laboratory, University of Oxford, Parks Road, Oxford, OX1 3PU, United Kingdom}

\author{Wenzheng Dong}
\affiliation{Department of Physics, Virginia Tech, Blacksburg, Virginia 24061, USA}

\author{Yiming Huang}
\affiliation{Institute of High Energy Physics, Chinese Academy of Sciences, Beijing 100049, China}
\affiliation{China Center of Advanced Science and Technology, Beijing 100190, China}
\affiliation{Center on Frontiers of Computing Studies, Peking University, Beijing 100871, China}

\author{Daniel Serrano}
\affiliation{Clarendon Laboratory, University of Oxford, Parks Road, Oxford, OX1 3PU, United Kingdom}

\author{Xinzhao Wang}
\affiliation{Center on Frontiers of Computing Studies, Peking University, Beijing 100871, China}

\author{Tianfeng Feng}
\affiliation{QICI Quantum Information and Computation Initiative, Department of Computer Science, The University of Hong Kong, Pokfulam Road, Hong Kong}

\author{Shreyas Sadugol}
\affiliation{Department of Physics and Engineering Physics, Tulane University, New Orleans, USA}

\author{Wenjun Yu}
\affiliation{QICI Quantum Information and Computation Initiative, Department of Computer Science, The University of Hong Kong, Pokfulam Road, Hong Kong}

% \author{...}
\author{Zhou You}
\affiliation{Center on Frontiers of Computing Studies, Peking University, Beijing 100871, China}
\affiliation{Key Laboratory for Information Science of Electromagnetic Waves (Ministry of Education), Fudan University, Shanghai 200433, China}

\author{Dayue Qin}
\affiliation{Key Laboratory for Information Science of Electromagnetic Waves (Ministry of Education), Fudan University, Shanghai 200433, China}

\author{Xiao-Ming Zhang}
\affiliation{School of Physics, South China Normal University, Guangzhou, 510006, China}

\author{Yantao Wu}
\affiliation{Institute of Physics, Chinese Academy of Sciences, Beijing 100190, China}

\author{Aditya Iyer}
\affiliation{Clarendon Laboratory, University of Oxford, Parks Road, Oxford, OX1 3PU, United Kingdom}

\author{You Zhou}
\affiliation{Key Laboratory for Information Science of Electromagnetic Waves (Ministry of Education), Fudan University, Shanghai 200433, China}

\author{Tongyang Li}
\affiliation{Center on Frontiers of Computing Studies, Peking University, Beijing 100871, China}

\author{Ying Li}
\affiliation{Graduate School of China Academy of Engineering Physics, Beijing 100193, China}

\author{Xiongfeng Ma}
\affiliation{Center for Quantum Information, Institute for Interdisciplinary
Information Sciences, Tsinghua University, Beijing 100084, China}

\author{Qi Zhao}
\affiliation{QICI Quantum Information and Computation Initiative, Department of Computer Science, The University of Hong Kong, Pokfulam Road, Hong Kong}
% \affiliation{Foresight Quantum, Shanghai, China}

\author{Pei Zeng}
\affiliation{Institute of Natural Sciences, Shanghai Jiao Tong University, Shanghai 200240, China}
\affiliation{School of Physics and Astronomy, Shanghai Jiao Tong University, Shanghai 200240, China}%

\author{Pan Zhang}
\affiliation{Institute of Theoretical Physics, Chinese Academy of Sciences, Beijing 100190, China}

\author{Xiao Yuan}
\affiliation{Center on Frontiers of Computing Studies, Peking University, Beijing 100871, China}

\begin{abstract}
% Seeking useful quantum advantage is a central goal of quantum science and technology. Here we identify the quantum–classical crossover in the simulation of quantum dynamics. We present a full-stack fault-tolerant resource analysis of quantum dynamics simulation, connecting algorithmic structure to fault-tolerant execution within a surface-code architecture. Focusing on the mixed-field Ising model, we compare two paradigms for implementing non-Clifford gates: conventional Clifford+$T$ circuits based on magic state distillation, and direct implementations of small-angle rotations using state injection. 
% Our results indicate that quantum computation can already outperform classical methods based on matrix product states for one-dimensional systems at sizes $L>30$. For the two-dimensional mixed-field Ising model, which is particularly challenging for classical computing, the estimated quantum runtime ranges from minutes to hours. 
% % By benchmarking against classical tensor network simulations, we identify a clear quantum–classical crossover regime.
% Our results establish concrete requirements for fault-tolerant quantum devices and provide a roadmap towards demonstrating clear quantum usefulness in dynamical simulations. This analysis avoids the neefd for prior assumptions on the initial state in eigenenergy estimation, which provides a clearer route for establishing practical quantum advantage. 

While quantum computers promise to solve classically intractable problems, identifying the point at which fault-tolerant quantum computation outperforms the best classical algorithms for practical applications remains an outstanding challenge. Here we establish a concrete quantum-classical crossover for quantum many-body dynamics under realistic hardware conditions. We introduce a scalable fault-tolerant framework that combines coherent observable estimation with a space-time-efficient implementation of non-Clifford rotations, suppressing the residual logical errors that limit existing partially fault-tolerant approaches. A benchmark against state-of-the-art tensor-network and variational Monte Carlo algorithms reveals a concrete crossover for mixed-field Ising dynamics at modest system sizes. For a physical error rate of $p=10^{-3}$, fault-tolerant simulation requires approximately 2 hours and $3.7 \times 10^5$ physical qubits for a 100-site 1D system, whereas tensor network approaches would require about 100 years. For 2D models, where rapid entanglement growth limits the classical evolution time, we project quantum runtimes within minutes. A physical error rate of $p=10^{-4}$ leads to at least an order of magnitude reduction in qubit count ($3.1 \times 10^4$ physical qubits) and runtime (minutes for 1D and seconds for 2D). The reduction in quantum runtime arises from our improved rotation-state injection and co-design of quantum error correction and observable-estimation protocols, which jointly suppress logical-error accumulation and reduce sampling overhead. Our results establish a scalable route towards practical quantum advantage and identify quantitative engineering targets for future fault-tolerant architectures.

% 
% This reduction in quantum runtime stems from the state-of-the-art rotation injection method and the co-design of the quantum error correction and observable estimation algorithms, which jointly minimise the overall runtime by reducing both the number of quantum error correction cycles and the sampling overhead.
% % This reduction of quantum runtime owes to the state-of-the-art rotation injection method and the co-design of the QEC and observable estimation algorithm, which minimises the overall runtime, which depends on both the QEC code cycles and the sampling overhead.
% % By delineating the strict boundaries of classical solvability, 
% Our results dictate definitive engineering targets for fault-tolerant architectures and their physical error rates, providing a rigorous roadmap for achieving quantum usefulness

\end{abstract}

\maketitle

% ===== Disable TOC for main part =====
\let\oldaddcontentsline\addcontentsline
\renewcommand{\addcontentsline}[3]{}

%\sun{NOTE: use British English to write up the manuscript.}

\section{Introduction}

Establishing quantum advantage is the central goal of quantum computing and a foundational pillar of quantum technology. 
Significant progress has been made towards this goal, ranging from random circuit sampling~\cite{arute2019quantum,boixo2018characterizing,wu2021strong} and Gaussian boson sampling~\cite{zhong2020quantum,madsen2022quantum} to practically motivated applications in many-body dynamics~\cite{kim2023evidence,king2025beyond,haghshenas2026digital} and optimisation~\cite{ebadi2022quantum}. 
While these applications hold broad relevance across condensed matter physics~\cite{childsFirstQuantumSimulation2018}, quantum chemistry~\cite{mcardleQuantumComputationalChemistry2018}, and high-energy physics~\cite{bauerQuantumSimulationHigh2022}, their implementation has so far relied on noisy quantum processors. 
Consequently, the claimed advantages~\cite{kim2023evidence,king2025beyond} have been repeatedly challenged by improved classical algorithms~\cite{beguvsic2024fast,tindall2024efficient,mauron2025challenging,tindall2026dynamics}. 
% xuClassicalSimulationNoiseless2026
These challenges arise because noisy quantum devices can often be simulated efficiently using classical resources~\cite{schuster2025polynomial,shao2024simulating,oh2024classical}, suggesting that they alone are insufficient to firmly establish robust quantum advantage. 
The central question has therefore shifted from whether quantum computers can outperform classical algorithms in principle to exactly when they will do so in practice~\cite{beverland2022assessing,preskill2025beyond,eisert2025mind}. 

Dynamics simulation (also known as Hamiltonian simulation or, more broadly, quantum simulation~\cite{lloyd1996universal,georgescu2014quantum}) is one of the most promising candidates for demonstrating quantum advantages in the early fault-tolerant regime. It is among the most natural applications of quantum computing~\cite{kivlichan2020improved,childsFirstQuantumSimulation2018,campbell2022early} and serves as a core subroutine in a wide range of quantum algorithms, such as optimisation and eigenvalue estimation~\cite{an2023linear,lin2021heisenberg,ding2024quantum}. Simulating quantum dynamics is typically hard on classical computers because the entanglement generated during time evolution grows rapidly. Furthermore, unlike ground-state energy estimation and optimisation, \sun{extensively researched recently}~\cite{chung2026partially,akahoshi2025compilation,lee2021even,ding2023even,yoshioka2024hunting,chung2026partially}, dynamics simulation does not require prior knowledge or assumptions about the initial state. These features make dynamics simulation a direct and compelling route for demonstrating an unambiguous quantum advantage.

Despite the theoretical promise, a key open challenge is to identify a concrete quantum-classical crossover point at which fault-tolerant quantum dynamics simulation outperforms the best available classical methods under realistic hardware error rates \sun{beyond non-error-corrected simulation~\cite{alam2025fermionic,alam2025programmable}.}
On the quantum side, this necessitates combining detailed algorithmic performance bounds with realistic fault-tolerant implementation costs. \sun{One must carefully account for the interplay between algorithm cost and the hardware overhead associated with implementing non-Clifford rotation gates. Existing fault-tolerant implementations can be classified into two categories}: (i) Clifford+\texttt{T} schemes based on magic-state distillation (MSD) or cultivation~\cite{hastings2018distillation,bravyi2005universal,campbell2010bound,sales2025experimental,Litinski2019gameofsurfacecodes,Litinski2019MSD,litinski2022active,gidney2025experimental_cultivation,Gidney2024Cultivation,dowling2025bridgingentanglementmagicresources}, and (ii) Clifford+$\varphi$ schemes based on space-time efficient analogue rotation (STAR) architecture target for Hamiltonian simulations~\cite{akahoshi2024partially,akahoshi2025compilation,toshio2026star,Toshio2025PracticalQuantumAdvantage,choi2023fault,ismail2026transversal,ismail2026fast}. 
% including the resources required for magic-state distillation (MSD) or cultivation~\cite{hastings2018distillation,bravyi2005universal,campbell2010bound,sales2025experimental,Litinski2019gameofsurfacecodes,Litinski2019MSD,litinski2022active,gidney2025experimental_cultivation,Gidney2024Cultivation,dowling2025bridgingentanglementmagicresources} or the rotation-state injection based on space-time-efficient analogue rotation (STAR) architecture target for Hamiltonian simulations~\cite{akahoshi2024partially,akahoshi2025compilation,toshio2026star,Toshio2025PracticalQuantumAdvantage,choi2023fault,ismail2026transversal,ismail2026fast}. 
While STAR schemes avoid \texttt{T}-gate compilation, a major challenge is their limited scalability to larger system sizes or longer time scales, as residual logical errors eventually dominate and the error-mitigation cost grows exponentially beyond a certain space-time threshold. On the classical side, a rigorous baseline must be established by analysing its resource scaling, including tensor network (TN) approaches, whose cost grows with entanglement, full-amplitude state-vector approaches, which require exponentially growing storage, \gyt{and time-dependent variational Monte Carlo (tVMC) sampling, with errors from the limited expressivity of the variational ansatz and Monte Carlo sampling.} Only by systematically comparing these detailed fault-tolerant quantum requirements against classical resources can the precise boundary of practical quantum advantage be identified.
\\

Here we present a full-stack resource analysis for simulating quantum many-body dynamics. We consider the mixed-field Ising model, a paradigmatic non-integrable system relevant to non-equilibrium dynamics, thermalisation, and quantum chaos. Our framework combines algorithmic error analysis and resource estimation with improved fault-tolerant implementations under realistic hardware conditions. \sun{We introduce a new algorithm for estimating observable expectations~\cite{Brassard_2002,Rall2023amplitudeestimation,GiurgicaTiron2022lowdepthalgorithms}, called Gaussian-sampled Chebyshev amplitude estimation (GCAE), built upon a recent work~\cite{huang2026low} sharing the same circuit structure. It allows us to decrease the maximum depth in a coherent run while surpassing the standard quantum limit, thus decreasing the requirement for fault-tolerant implementation.} We translate logical resource requirements to physical hardware by analysing the quantum error correction (QEC) overhead of two fault-tolerant architectures: Clifford+\texttt{T} synthesis~\cite{Litinski2019gameofsurfacecodes,Litinski2019MSD,litinski2022active} and Clifford+$\varphi$ synthesis via surface-code-protected injection of small-angle rotations~\cite{zeng2025error}. By quantifying the total space-time volume, we show that direct rotation-state injection suppresses logical errors to $\mathcal{O}(|\theta|\pphys^2)$ for physical error rate $p$ and angle $\theta$, thereby enabling fully fault-tolerant simulations for systems exceeding $n=100$ spins---a scale inaccessible to existing partially fault-tolerant designs due to the rapid accumulation of residual logical error~\cite{akahoshi2024partially,Toshio2025PracticalQuantumAdvantage,ismail2026fast,ismail2026transversal,chung2026partially}. 

Within this framework, we identify a concrete quantum-classical crossover by benchmarking fault-tolerant quantum resource requirements against matrix product state (MPS), projected entangled pair state (PEPS), \sun{and tVMC algorithms}. A striking finding is that classical simulation of one-dimensional (1D) systems by MPS rapidly becomes prohibitively expensive when evolved to $t=n/2$, exceeding 1,000 compute hours for $n>50$ and projecting to timescales of order 100 years for $n=100$ at error below $\epsilon = 0.01$. For two-dimensional (2D) systems with $n$, rapid entanglement growth further restricts classical evolution to just $t = \sqrt{n}$, beyond which the target error cannot be reduced below $0.1$. In stark contrast, for a physical error rate $p=10^{-3}$, the corresponding fault-tolerant simulation requires about 2 hours for a 100-site 1D system, and about 5 minutes for a 100-site 2D system.  This dramatic reduction in quantum runtime stems from our state-of-the-art rotation injection method and novel co-design of the QEC and observable-estimation protocols, which jointly reduce both the number of QEC cycles and the sampling overhead. Moreover, achieving a physical error rate of $p=10^{-4}$ leads to at least an order of magnitude reduction in both qubit count ($3.1 \times 10^4$ physical qubits) and runtime (minutes for 1D and seconds for 2D). These results establish a concrete milestone for useful fault-tolerant quantum simulation and provide quantitative engineering targets for future quantum architectures.

\section{Framework and overview of the key components}

% To that end, we adopt a full-stack approach in which each layer—model, algorithm, circuit synthesis, and quantum error correction—is explicitly connected, allowing the total space--time cost to be expressed in terms of physical qubits and error-correction cycles. The technical innovation is thus involved in three major components. 
% (1) At the algorithms level, we employed tighter analysis for the Trotter error bound.
% (2)
% QEC-level. This method extends to a larger system size. Rotation gates time cost and error, compared to distillation.
% Bridgeing (1) and (2), we develop a co-design scheme for algorithms and QEC. This balances the Depth trade-off by GDMAE. With error,  error budget and mitigation.
% (3) 
% Quantum advantages. We analysed the resources by TN. The bond dimension is computed by controllling the error and search for the requried bond dimsion, this clearly esbalished quanutm clssical creossover by 

% We begin from the target Hamiltonian $H$ and its real-time evolution operator $U(t)=e^{-iHt}$. We focus on mixed-field Ising Hamiltonians in 1D and 2D, which are non-integrable and quickly generate entanglement.
% The goal is to estimate a local observable's expectation value $\bra{\psi(t)}O\ket{\psi(t)}$ to within an additive error $\varepsilon$.

\begin{figure*}[t]
\centering
\includegraphics[width=\linewidth]{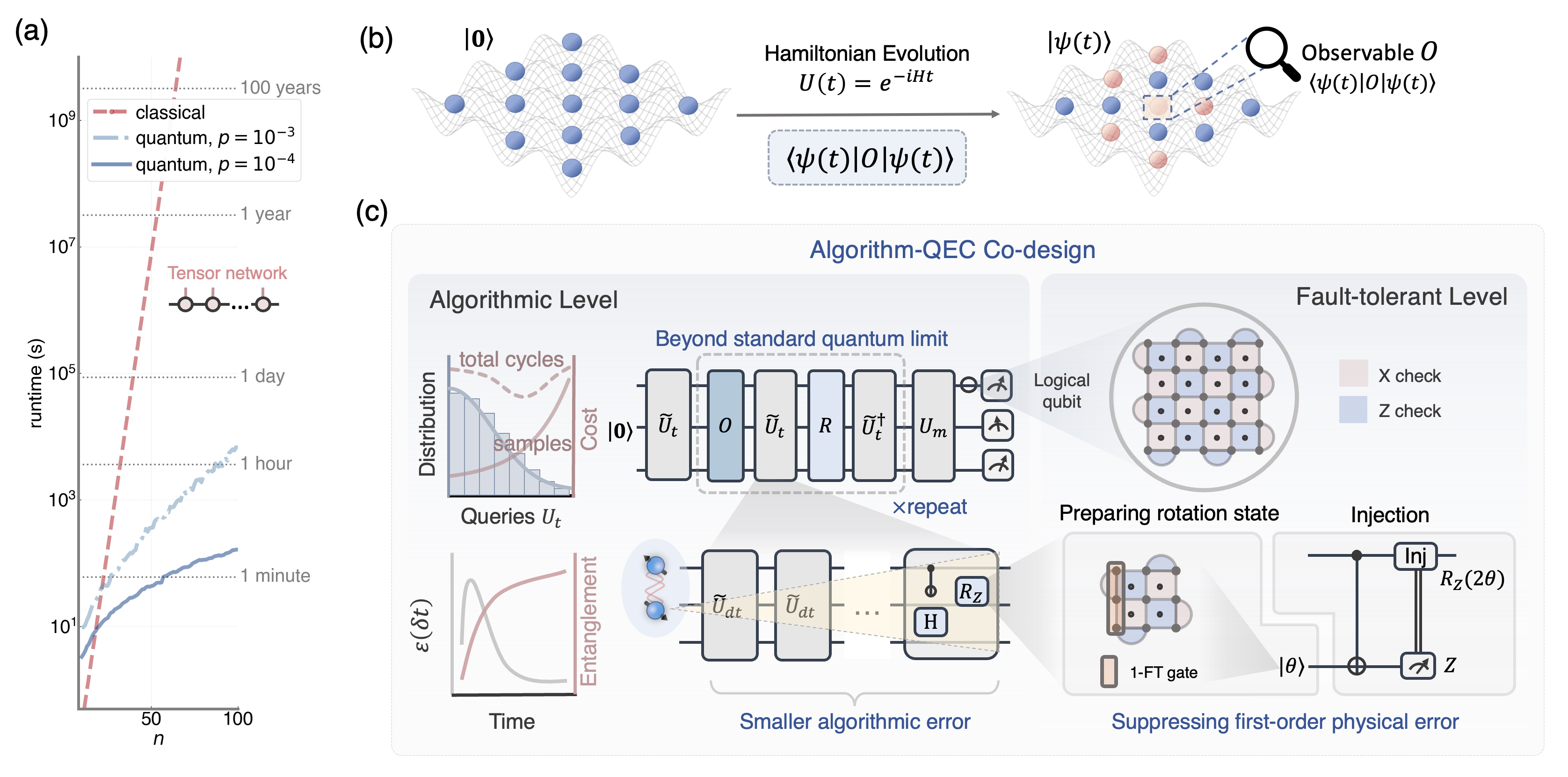}
	\caption{
\textbf{Full-stack framework for fault-tolerant quantum dynamics simulation and runtime comparison to classical approaches.} 
(a) Quantum-classical runtime crossover for 1D mixed-field Ising models. Runtimes are evaluated at physical error rates $p = 10^{-3}$ and $10^{-4}$ (detailed in Fig.~\ref{fig:1D_classical_simulation}). 
% The total runtime exhibits distinct scaling behaviours for quantum and classical methods, leading to a crossover regime beyond which fault-tolerant quantum simulation becomes advantageous. 
(b) Quantum simulation task. The goal is to estimate the expectation value $\bra{\psi(t)} O \ket{\psi(t)}$ of a time-evolved state $\ket{\psi(t)} = U(t) \ket{\mathbf{0}}$ to a specified precision. 
% Trotter method is applied.
(c) Schematic of the algorithm-QEC co-design framework.
\textbf{Left panel:} Logical-level illustration of the quantum algorithm. 
\textbf{Upper left:} Observable estimation built upon the amplitude amplification circuit~\cite{huang2026low}, surpassing the standard quantum limit for which sampling complexity scales as $\mathcal{O}(\varepsilon^{-2})$. This algorithm uses the same circuit as in~\cite{huang2026low} and can estimate observable expectation values over any range thanks to the properties of the Chebyshev expansion. This algorithm balances the maximum circuit depth against the required number of samples.  The estimation circuit relies on coherently querying the real-time evolution operator $U(t)=e^{-iHt}$ multiple times, which is approximated via a fourth-order Trotterisation $\tilde U(t)$. \sun{The circuit ($U_m$ and measurement) depends on the parity of  $m$, sampled from Gaussian distribution. Larger queries lead to higher sample costs but occur less frequently, so the total code cycles (proportional to runtime) may have an optimal value.
}
\textbf{Bottom left:} The time interval $\delta t$ is determined by a Trotter error analysis.  A tighter Trotter error bound $\varepsilon(\delta t)$ can be obtained as entanglement entropy grows. For arbitrary initial states, the error is controlled by entanglement-informed Trotter error analysis, saturating the average-case bound. For the specific initial state $\ket{\mathbf{0}}$, the required number of Trotter steps is obtained via extrapolation.  
% to with $m$ applications of the reflection operation. Lastly, we construct a target function based on the measurement result for which the peak lies on an $\epsilon$-close estimate of the target amplitude to obtain the observable dynamics. 
% This algorithm queries the real-time evolution operator $U(t)=e^{-iHt}$ 
%  This yields a circuit composed of layers of local Pauli rotations, $R_Z(\theta)$ gates acting on neighbouring qubits, with depth proportional to $\nu$. 
% (b2) A tighter Trotter bound can be obtained as entanglement entropy grows. (b3)  The distribution and the sampling overhead dependencies on the number of queries of $U(t)$. 
% \sun{Pending on new results}
\textbf{Right panel:} Fault-tolerant implementation on a surface-code architecture. 
% (c1) In the Clifford+T paradigm, arbitrary rotations are decomposed into Clifford and T gates, with non-Clifford resources supplied by magic state distillation factories and distributed to data qubits via lattice surgery. 
\textbf{Bottom right:} Realising small-angle rotation gates $R_z(\theta)$ via rotation-state injection. 
The injection procedure is repeated until a measurement outcome $0$ is obtained; upon outcome $1$, the procedure is retried with the target angle doubled to $2\theta$.
The injected rotation state is prepared using a level-1 fault-tolerant (1-FT) procedure that suppresses first-order physical errors, thereby enabling greater coherent circuit depths and reducing error-mitigation overhead.
%This method also allows parallel execution with an expected depth that scales logarithmically with system size.
% denotes recursive state injection: the measurement is repeated until outcome 0 is obtained; upon outcome 1, the entire procedure is repeated.
% These components in (c) illustrate the mapping from algorithmic circuit construction to fault-tolerant execution.
% In this work, we co-design the QEC and the residual logical error mitigation to minimise the overall runtime, which depends on both the QEC code cycles and the sampling overhead.
Ultimately, this framework co-designs the QEC implementation and the mitigation of residual logical errors to minimise the overall runtime which depends on both the QEC code cycles and the sampling overhead.
% (e) Illustration of the TN methods: MPS, PEPS and  TEBD update. Here the error source comes from the  SVD  truncation. 
}
	\label{fig:cartoon}
\end{figure*}

We consider the real-time dynamics governed by a target Hamiltonian $H$, with a focus on 1D and 2D mixed-field Ising models which are non-integrable and rapidly generate entanglement. The core computational task is to estimate the expectation value of a observable, $\bra{\psi(t)}O\ket{\psi(t)}$, to within an additive error $\varepsilon$, where $\ket{\psi(t)} \coloneqq U(t) \ket{\mathbf{0}}$ is the state evolved under the unitary operator $U(t)\coloneqq e^{-iHt}$, see Fig.~\ref{fig:cartoon}(b).

Achieving this precision $\varepsilon$ via direct sampling is constrained by the standard quantum limit, incurring a sampling overhead that scales as $\mathcal{O}(\varepsilon^{-2})$. 
Advanced observable-estimation algorithms can surpass this limit by coherently querying $U(t)$ multiple times at the cost of requiring circuits substantially deeper than those needed merely to prepare the time-evolved state $\ket{\psi(t)}$, yet it remains comparatively straightforward for classical methods. 
Independently, accurately simulating $U(t)$ itself requires extended circuit depth to suppress algorithmic errors arising from Trotter decomposition. 
This creates a tension: while deeper circuits are required to simultaneously reduce sampling complexity and algorithmic errors, the accumulation of residual logical errors ultimately limits the maximum depth \sun{in fault-tolerant dynamics simulation based on STAR architecture.  
% Consequently, estimating observables poses a significantly greater challenge for quantum computing than mere state preparation, yet it remains comparatively straightforward for classical methods. 
 Studies on Hamiltonian simulation resource estimates~\cite{ismail2026fast,ismail2026transversal,khan2026architecting} have primarily analysed the resource cost per shot rather than the total multi-shot cost, which thus may not provide a concrete assessment of computational advantage relative to classical methods.}

To address this, a key aspect of our framework is to identify algorithmic regimes where the required circuit depth remains compatible with fault-tolerant execution under realistic hardware error rates. 
For certain fault-tolerant architectures, resource estimates indicate a non-Clifford budget up to approximately $4 \times 10^4$ rotation gates~\cite{akahoshi2024partially}. 
Beyond this regime, as we shall discuss, suppressing the accumulated logical errors requires quantum error mitigation techniques~\cite{li2017efficient, PhysRevLett.119.180509,cai2023quantum}, which introduce an exponentially increasing sampling overhead—thereby negating the advantage coherent algorithms were meant to provide. 
This necessitates a co-design approach that jointly optimises the QEC cycles and the observable-estimation sampling overhead.

To that end, we adopt a full-stack framework in which each layer—from quantum algorithms for observable estimation and real-time evolution to fault-tolerant QEC—is explicitly connected [Fig.~\ref{fig:cartoon}(c)]. This allows the total space-time cost to be quantified directly in terms of physical qubits and hardware cycles. The technical advances of this work are organised into three main components:

First, at the algorithmic level, we introduce the observable-estimation framework built on a circuit in Ref.~\cite{huang2026low}, which balances the number of coherent queries to the real-time evolution operator $U(t)$ against the measurement cost for estimating physical observables.  
Crucially, this balance relies on minimising the circuit depth of Trotter decomposition per query.
Standard worst-case analytical bounds, based on operator-norm estimates~\cite{childs2021theory}, substantially overestimate the depth. Deeper circuits incur larger residual logical errors to the extent that a prohibitively large measurement overhead would be required.
To overcome this, we derive tighter Trotter-error bounds by incorporating entanglement-informed analyses~\cite{zhao2025entanglement} alongside empirical estimates that explicitly account for the evolving quantum state.
This approach significantly reduces the number of Trotter segments required, achieving a balance between circuit depth and sampling complexity compatible with current physical noise levels in quantum hardware.

Second, at the fault-tolerant level, we analyse two distinct implementations of non-Clifford operations: (i) conventional Clifford+\texttt{T} synthesis, where rotation gates are decomposed into \texttt{T} gates, prepared on the surface code, and injected into data qubits via MSD; and (ii) direct, surface-code-protected injection of small-angle rotations [Fig.~\ref{fig:cartoon}(c)]. 
By deriving the corresponding scalings for runtime, logical error accumulation, and QEC overhead, we demonstrate that the standard surface-code-based MSD method imposes prohibitively high space-time overhead, rendering it largely impractical for architectures limited to 1 million physical qubits.
To reduce cost, this work adopts a Clifford+$ {\varphi} $ hybrid fault-tolerant architecture~\cite{Litinski2019gameofsurfacecodes,akahoshi2024partially} and employs error-structure-tailored rotation-gate design~\cite{zeng2025error}, which can be regarded as an improved STAR architecture.  Here, the logical rotation gate satisfies the 1-FT gate requirement, which thus extends the simulation to larger scales than a non-FT implementation of rotation gates.
Ultimately, by bridging the algorithmic and QEC layers, our co-design framework jointly optimises the coherent depth for observable estimation and the error budget allocation, thereby significantly reducing the total circuit depth and sample complexity required for quantum simulation.

% We derive the corresponding scalings for runtime, logical error accumulation, and QEC overhead, showing that rotation-based methods significantly reduce the depth and time cost of non-Clifford operations. 
 % \sun{We compare the spatial and time resource cost for both implementations. The MSD method requires a much higher space-time overhead: it is hard to implement with 1 million qubits, thus making it less practical in the near term.}
% A direct comparison of spatial and temporal resources shows that the standard MSD method imposes prohibitively high space-time overhead, making it largely impractical for architectures limited to 1 million physical qubits.

Third, we establish the quantum-classical crossover by benchmarking against \sun{state-of-the-art classical methods. These include MPS, PEPS and tVMC based on Jastrow expansion~\cite{jastrow1955many} (1D) and PEPS (2D) ansatz.} 
The MPS cost is determined by systematically increasing the bond dimension $\chi$ to control the simulation error, allowing us to extract the runtime required to achieve the exact target precision of the quantum simulation. 
This comparison provides a direct, quantitative characterisation of the regimes in which fault-tolerant quantum computation can outperform TN, tVMC or full-amplitude state-vector methods. 
Strikingly, we find that even for 1D simulations, for which MPS is known to behave well, quantum algorithms surpass classical methods at $n > 30$ under a physical error rate of $\pphys = 10^{-3}$. This separation becomes even more pronounced for 2D simulations.

\section{Fault-tolerant observable estimation}

We first introduce the quantum algorithm development at the logical level.
The target Hamiltonian can be decomposed into a sum of local terms that admit natural Trotterisation.
We choose the fourth-order product formula as it balances well between approximation accuracy and circuit cost~\cite{childsFirstQuantumSimulation2018,childs2021theory,sun2026high}.
The real-time evolution is approximated by $\nu$ Trotter segments, where $\nu$ depends on the total evolution time $t$ and the desired accuracy. Each segment consists of layers of local non-Clifford Pauli rotations, including both single-qubit and nearest-neighbour two-qubit terms.

% The total logical depth of the circuit is therefore proportional to $\nu$, and the dominant contribution to the resource cost arises from the implementation of these non-Clifford rotations.
% Each segment is compiled into layers of local Pauli rotations of the form $\exp(-i\theta P)$, where $P$ is a tensor product of Pauli operators. At this algorithmic level, the dominant resource is the total number of such non-Clifford rotations.

% While the above algorithms deterministically prepare the time-evolved state, observable estimation in quantum systems requires repeated measurements due to the uncertainty principle, introducing a sampling overhead that must be included in any fair resource comparison. Therefore, we consider estimating the expectation value $\bra{\psi(t)}O\ket{\psi(t)}$ of a given observable $O$ to accuracy $\varepsilon$, where $\ket{\psi(t)}=U(t) \ket{\psi_0}$ is the time-evolved state.
% As an illustration, suppose we prepare $\ket{\psi(t)}$ and measure $O$ directly. Here, the number of samples required to achieve an $\varepsilon$-accurate estimation scales as $\mathcal{O}(1/\varepsilon^2)$. When the accuracy requirement is high, this will introduce a challenging sample overhead that prevents practical quantum advantage.

\begin{figure*}[tbp]
	\includegraphics[width=1\linewidth]{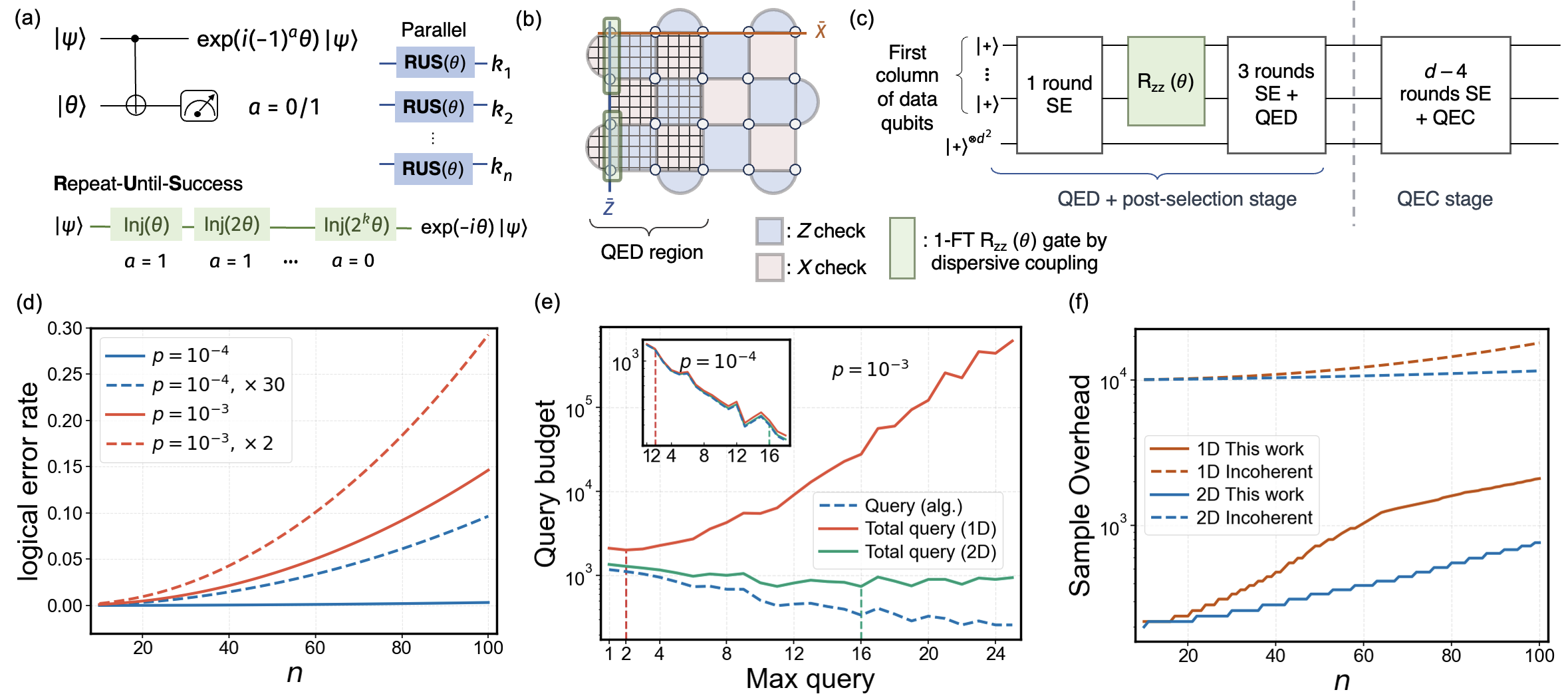}
	 \caption{\textbf{Fault-tolerant logical rotation states preparation and optimisation of the query complexity and sampling overhead.}
     (a) Protocol for implementing small-angle $Z$ rotations in parallel on $n$ qubits. \textbf{Upper left:} Basic state-injection gadget using the ancillary rotation state  $\ket{\theta}$. \textbf{Bottom:} RUS procedure: if the measurement outcome indicates failure, a rotation state with doubled angle is injected and the protocol is repeated until success at round $k + 1$.
     \textbf{Upper right:} Parallel execution of a single layer of rotation gates across $n$ qubits. The expected number of RUS rounds required for all $n$ qubits to complete their rotations scales as $\mathcal{O}(\log n)$.
     (b) Preparation of $\ket{\theta}_L$ on a $4\times 5$ rotated surface code using $1$-FT $R_{ZZ}(\theta)$ gates, where the $1$-FT $R_{ZZ}(\theta)$ gate is realised by the dispersive-coupling method of Ref.~\cite{zeng2025error}. The shaded region marks the part of the surface code on which QED is performed. (c) Full preparation circuit. The first four rounds of syndrome extraction, together with one layer of $1$-FT $R_{ZZ}(\theta)$ gates, implement QED and post-selection, while the remaining $d-4$ rounds perform QEC. 
     % (c) Preparation of the logical rotation state based on $1$-FT $R_{ZZZ}(\theta)$ rotations on a $6\times 5$ rotated surface code. Here, the $1$-FT $R_{ZZZ}(\theta)$ gate is constructed from a $1$-FT $R_{ZZ}(\theta)$ gate sandwiched between two CNOT gates.
    % Bottom panel: (a,b) Performance of logical rotation-state preparation based on the $R_{ZZZ}(\theta)$ scheme.
    % (a) Success probability of preparing the logical rotation state as a function of $\theta$ for rotated surface codes of shape $d\times(d+1)$ with distance $d=3m$, where $m$ is the number of $R_{ZZZ}(\theta)$ operations applied along the first column of data qubits. Results are shown for circuit-level noise rates $p=10^{-3}$ and $p=10^{-4}$, with $k=6$ and $k=4$, respectively. (b) The coefficient $\alpha_{\mathrm{RUS}}$ in the logical error scaling $\alpha_{\mathrm{RUS}} |\theta| p^2$ of the RUS gadget, shown as a function of $\theta$.
   % (c,d) Performance of the fault-tolerant simulation of the 1D mixed-field Ising model. (c) Total number of code cycles required for the Trotterized evolution.
   %  (d) Estimated logical error rate of the full simulation.  
    (d)  Estimated total logical error rate in 1D simulation with $n = 100$ for cases of physical error rates of $\pphys = 10^{-3}$ and $\pphys = 10^{-4}$.   The dashed line represents the amplified logical error rate incurred by querying the time-evolution circuit 2 times ($p=10^{-3}$) and 30 times ($p=10^{-4}$). These optimal maximum query numbers are determined from the minima in panel (e).
    (e) Query cost, with or without the overhead for residual error mitigation, as a function of the maximum query number $M$ for $n = 100$. The blue-dashed lines show the base algorithmic query complexities without error-mitigation overhead, which decrease monotonically as $M$ increases. 
     The solid red and green lines represent the total query costs, including the error-mitigation sampling overhead, for 1D and 2D systems, respectively. Inset: Total queries for $\pphys = 10^{-4}$.  The total number of queries strictly decreases because the accumulated logical error rate remains highly suppressed.
    (f) Comparison of the total sampling cost between the coherent estimation and direct incoherent measurement as a function of system size $n$ at $\pphys = 10^{-3}$.
    % We have incorporated the fact that larger queries are less frequent, which reduces the total cost.
    }
	\label{fig:RUS_preparation}
\end{figure*}

While Trotterisation circuit deterministically prepare of the time-evolved state $\ket{\psi(t)}=U(t) \ket{\psi_0}$, extracting physical properties requires estimating the expectation value $\bra{\psi(t)}O\ket{\psi(t)}$ to a target accuracy $\varepsilon$. As outlined previously, direct sampling incurs a demanding sampling overhead that scales as $\mathcal{O}(\varepsilon^{-2})$. When the accuracy requirement is high, this scaling imposes large measurement costs, preventing practical quantum advantage. 
% On the other hand, prior works for eigenenergy esitmation consider coherent estimation, which, although reduces the total runtime, results in longer depth in single run, which makes the error budget exploding after taking account of QEC. 
On the other hand, prior works on eigenenergy estimation with dynamics simulation as a subroutine~\cite{lee2021even,lin2021heisenberg} often rely on coherent estimation schemes, which reduce the total runtime to the so-called Heisenberg limit at the cost of much deeper circuits per run. This increased circuit depth leads to uncontrolled error accumulation in STAR architectures~\cite{akahoshi2024partially,akahoshi2025compilation}, and such errors bring exponentially increasing error mitigation cost.
To address this issue, at the algorithmic level, we apply amplitude estimation techniques, which allow us to balance coherent depth and total queries in the Hamiltonian evolution [Fig.~\ref{fig:cartoon}(c)]. At the QEC level, we leverage error-structure-tailored rotation-gate design to reduce the logical error rate~\cite{zeng2025error}.
We show how it achieves the optimal space-time resource trade-off in Fig.~\ref{fig:RUS_preparation}.

The quantum circuit, constructed at the algorithmic level, is then mapped to a fault-tolerant gate set, as shown in the right panel of Fig.~\ref{fig:cartoon}(c). Here, we consider two distinct implementations of non-Clifford operations: decomposition into Clifford and \texttt{T} gates, and the preparation and injection of small-angle rotation states. We find that the latter, implemented directly using repeat-until-success (RUS) state-injection methods, surpasses the former. 
Exploiting the identity $\exp(-i\theta P)=C\exp(-i\theta Z)C^\dagger$, where $C$ is a Clifford operator, the problem reduces to implementing logical $Z$-rotations with angle $\theta$. 
These are realised via RUS protocols, in which an ancilla prepared in the state $|\theta\rangle=\exp(-i\theta Z)|+\rangle$ is injected into the circuit. 
Each individual RUS attempt encompasses error detection, state preparation, and subsequent injection. 
While each attempt succeeds with a constant probability, failures are recursively corrected using rotations of a doubled angle. When executing a layer of parallel operations, the expected number of required RUS attempts grows logarithmically with the number of operations.
% This introduces a probabilistic overhead that scales differently from the deterministic cost of \texttt{T}-gate synthesis. 
% In the rotation-based paradigm, the cost is governed by the preparation of logical rotation states and the expected number of RUS rounds, which scales logarithmically with $n$.
The cost of this rotation-based method is governed by the preparation of logical rotation states and this logarithmic probabilistic RUS overhead, yielding a scaling behaviour that is different from the deterministic cost of \texttt{T}-gate synthesis. 
% depth. 
% Each round involves state preparation, injection, {error detection},  and conditional corrections. 
% The ability to perform many rotations in parallel, combined with the logarithmic scaling of the RUS depth, results in a distinct dependence of the total cost on system size, as shown in Fig.~\ref{fig:RUS_preparation}.

At the fault-tolerant level, time is discretised into QEC cycles. Each logical operation is measured in units of QEC cycles, whose implementation may involve multiple rounds of syndrome extraction depending on the code distance $d$, reflecting the overhead of fault-tolerant protection. The cost of a single Trotter segment, $T_{\mathrm{step}}$, is determined by the number of layers of single- and two-qubit rotations, together with the overhead of basis changes and state injection. For rotation-based implementations, $T_{\mathrm{step}}$ depends on the expected number of RUS rounds, which scales as $\cO(\log n)$ for $n$ parallel rotations, multiplied by the cost of state preparation and injection. 
The resource cost of the Clifford+\texttt{T} implementation is discussed in \suppref{sec:small-angle-rotation}.

\begin{figure*}[ht]
\centering
\includegraphics[width=\linewidth]{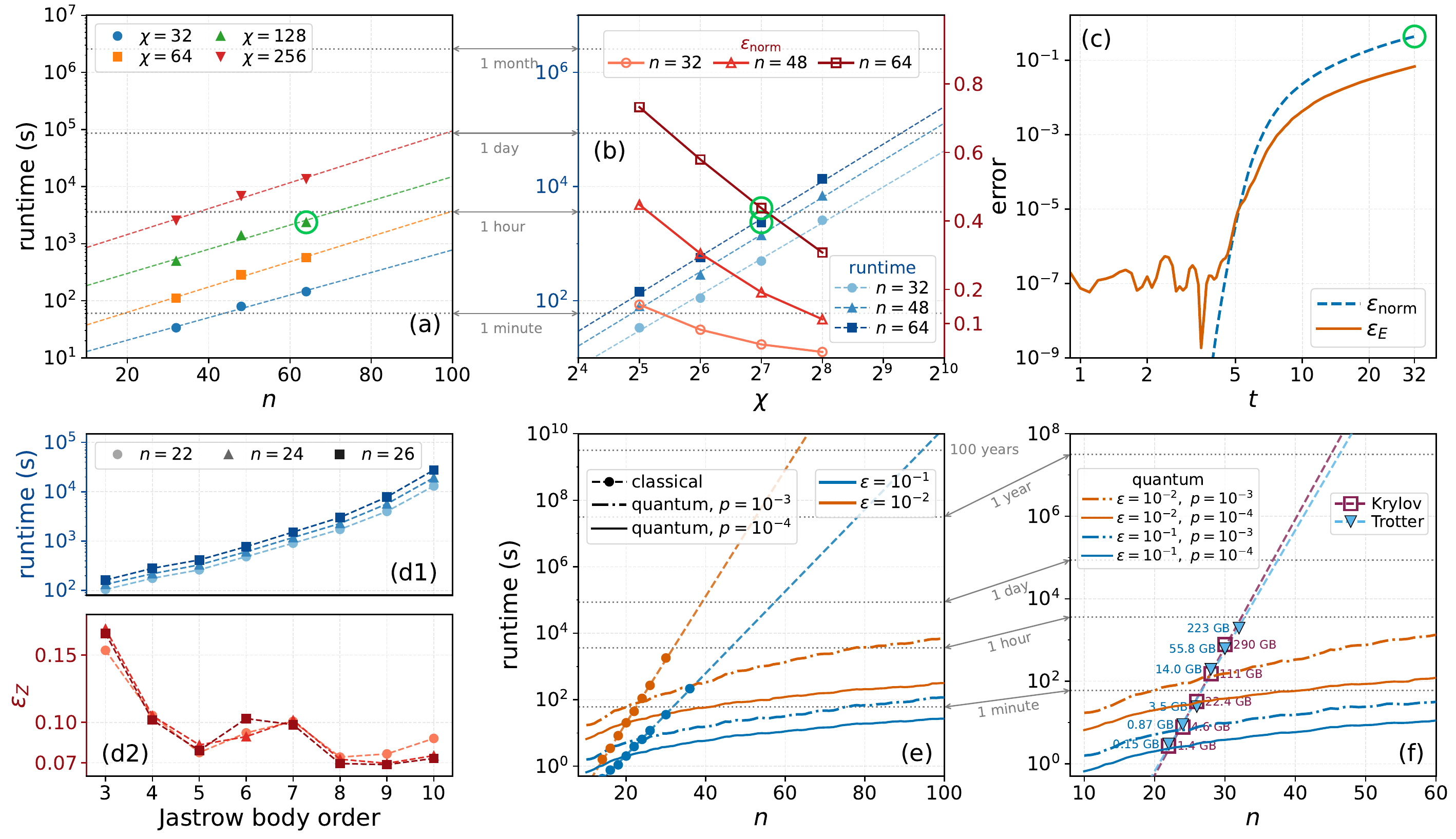}
\caption{ \textbf{Classical simulation cost and quantum-classical crossover for the 1D mixed-field Ising model.}
\textbf{Top row:} MPS benchmarks at fixed bond dimension
($n=32,48,64$ and $\chi=32,64,128,256$) evolved to $t=n/2$.
(a)~Runtime versus system size~$n$ at fixed~$\chi$.
Dashed lines are exponential fits.
(b)~Runtime (left axis) and final norm-loss error
$\varepsilon_{\mathrm{norm}}$ (right axis) versus~$\chi$.
Dashed lines are power-law fits.
(c)~Time evolution of $\varepsilon_{\mathrm{norm}}$ and the
energy-density error $\varepsilon_E$ for $n=64$ and $\chi=128$.
Green open circles in (a)--(c) mark the same run
$(n{=}64,\chi{=}128)$. This run illustrates that the truncation error grows rapidly during the
time evolution at fixed $\chi$, so bounding it requires $\chi$ to grow
exponentially with system size.
\textbf{Bottom row:} 
(d1)~Single-GPU wall time in seconds. (d2)~$\varepsilon_{Z}$ is the time-averaged error of the centre observable $C_{Z}(t)=\langle Z_{\text{center}}\rangle(t)$ (over the entire evolution) for the Jastrow ansatz with tVMC evolved up to $t=\sqrt{n}$ due to uncontrollable error. The error fluctuates and cannot be controlled below $0.07$ as Jastrow body order increases.
(e)~Runtime comparison between MPS and quantum runtime with bounded error.
The fault-tolerant quantum estimate at target accuracy
$\varepsilon=10^{-2}$ (orange) is shown for $p=10^{-3}$ (dash-dotted line) and for $p=10^{-4}$ (solid line). The looser target $\varepsilon=10^{-1}$ (blue) is shown for
the same two values of $p$ with the same line styles.
(f)~Wall times of full state-vector simulation (on up to
four GPUs with 80 GB memory) with the restarted-Krylov (squares) and
fourth-order Trotter (triangles) methods.
Each data point is labelled with its peak GPU memory usage.
Quantum runtime curves as in~(e).
Grey dotted horizontal lines mark reference timescales.}
\label{fig:1D_classical_simulation}
\end{figure*}

We consider two representative hardware regimes that capture the leading experimental platforms for fault-tolerant quantum computation, $\pphys = 10^{-4}$ and $\pphys = 10^{-3}$, corresponding to trapped-ion and superconducting qubit architectures. The unique advantage of the rotation-state-preparation method is the dependence on the physical error rate $\pphys$.
Integrating with recent small-angle-state preparation techniques~\cite{zeng2025error}, the gate error can be suppressed to $\mathcal{O}(\pphys^2)$, showing better error suppression than the conventional approach of order $\mathcal{O}(\pphys)$~\cite{akahoshi2024partially,Toshio2025PracticalQuantumAdvantage}. This feature allows us to execute more rotation gates in quantum dynamics simulation.
The logical residual error rates extending to $n = 100$ are shown in Fig.~\ref{fig:RUS_preparation}(d).

% The effect of the residual error rate is important as it induces the sampling cost.
% The distribution and the sampling overhead dependencies on the number of queries of $U(t)$ in \autoref{fig:RUS_preparation}(XXX). We have incorporated the fact that larger queries are less frequent, which reduces the total cost. 
% The total query cost over the maximum query number $M$ for $n = 100$ is shown in \autoref{fig:RUS_preparation}(XXX). This indicates an optimal maximum depth considering the sampling overhead. The total sampling cost scaling with qubit number is shown in \autoref{fig:RUS_preparation}(XXX). 

% The residual error rate indeed plays an important role, as it directly determines the sampling overhead. 
Crucially, the residual logical error rate directly dictates the error-mitigation sampling overhead. 
To explore the interplay between algorithmic and fault-tolerant considerations, we analyse the full-stack computational cost to identify the operating point that minimises the overall resource requirements.
%The distribution of queries to $U(t)$ is Gaussian, which means that deeper circuits (with large error-mitigation overhead) are less likely to appear.
The algorithm samples the number of queries to $U(t)$ from a discrete Gaussian distribution, deeper circuits, which incur larger error-mitigation overheads, are queried with exponentially suppressed probabilities making it more resource-friendly. 
% while the corresponding mitigation sampling overhead grow exponnetially with the query number are illustrated in \autoref{fig:cartoon}(b3)
The total query required for observable estimation decays as the maximum query number $M$ increases, as demonstrated for $n=100$ in Fig.~\ref{fig:RUS_preparation}(e). 
% The total number of queries required by the observable estimation algorithms decays with increasing query number. 
% The advantage of this observable estimation algorithm is that it balances the maximum depth and the sampling overhead required by the coherent estimation and incoherent, direct measurements.
The primary advantage of this observable estimation algorithm is to optimally balance the maximum coherent circuit depth against the sampling overhead, successfully interpolating between fully coherent estimation and incoherent direct sampling. 
\sun{
The total runtime is determined by the QEC cycles of a single $U(t)$ query and the overall sampling overhead, defined as the weighted sum of the query numbers and their corresponding error-mitigation sampling costs.
%which is defined relative to a single $U(t)$ query as the weighted sum of the query numbers and their corresponding error-mitigation sampling costs.}
% which  \sun{refers to the overhead compared with the single-query of $U(t)$, which is summation of query numbers mulitilied by the corresponding error-mitigation sampling cost. }
}
For $\pphys = 10^{-3}$, this hardware-algorithm co-design reveals an optimal maximum depth that minimises total sampling overhead.

\begin{figure*}[htb]
\centering
\includegraphics[width=1.0\linewidth]{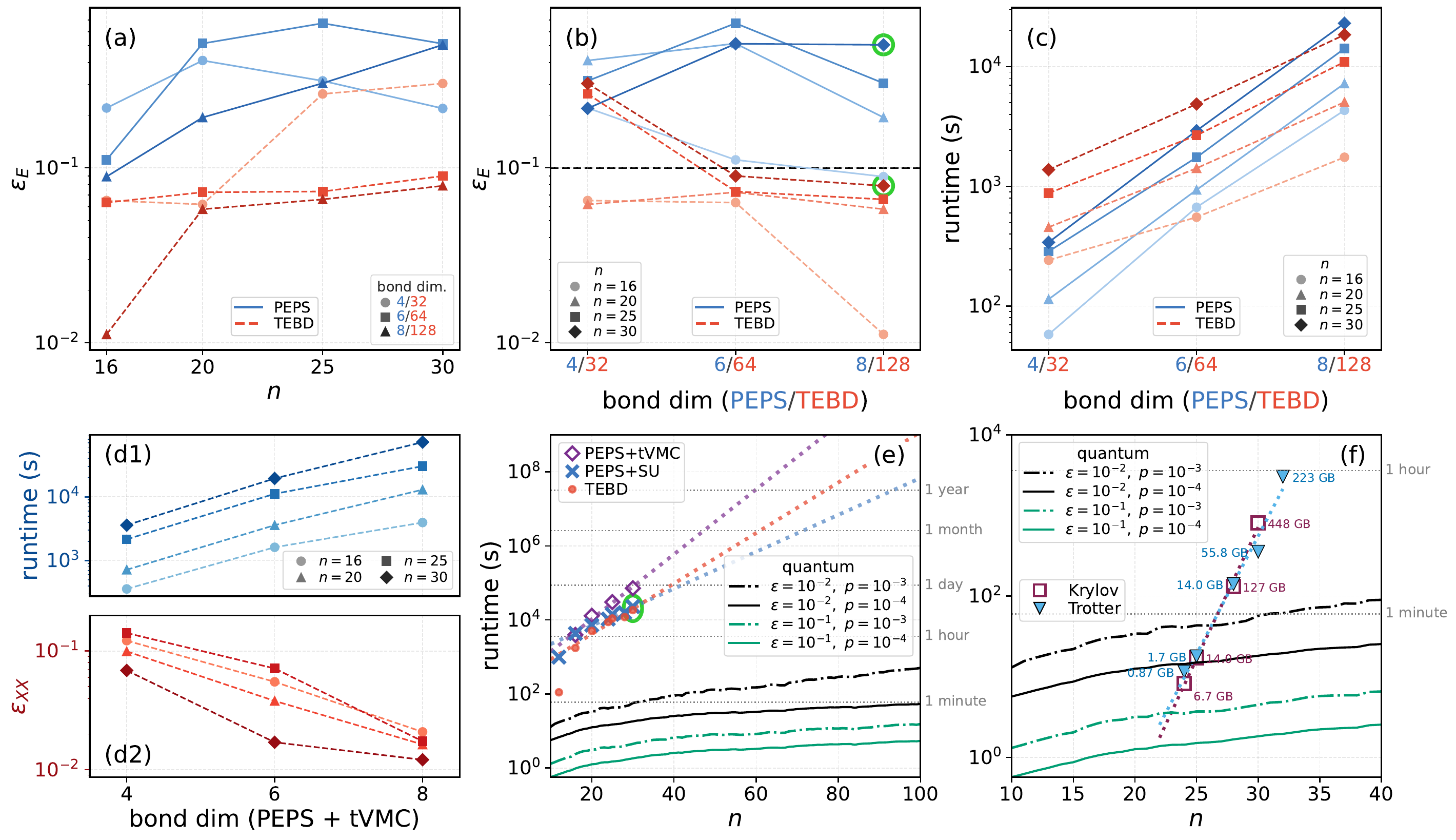}
\caption{\textbf{Classical simulation error and quantum-classical crossover for the 2D mixed-field Ising model}. The 2D system is evolved to $t=\sqrt{n}$ (open
$N_x\times N_y$ lattices with $n=N_xN_y=16,20,25,30$).
(a,\,b)~Final-time energy-density error $\varepsilon_E$ for PEPS with
simple-update truncation (blue solid lines with bond dimension
$\chi=4,6,8$) and for TEBD on a 1D snake mapping
(red dashed lines with $\chi=32,64,128$), shown versus system size~$n$
in~(a) and versus bond dimension in~(b).
(c)~Corresponding CPU wall times of the time evolution versus bond dimension.
In (b,\,c) the paired $x$-ticks list the PEPS/TEBD bond dimensions.
Green circles in~(b) mark the $n=30$ data at the largest bond dimension
(PEPS $\chi=8$, TEBD $\chi=128$), and the black dashed line marks the
target $\varepsilon_E=10^{-1}$.
(d1,\,d2)~Runtime and error $\varepsilon_{XX}$ of a third classical method, a PEPS ansatz evolved by tVMC with PEPS bond dimension $\chi=4,6,8$; marker shape and shade encode~$n$.
$\varepsilon_{XX}$ is the error of the corner-to-corner correlator
$C_{XX}(t)=\langle X_{(1,1)}X_{(N_x,N_y)}\rangle(t)$, time-averaged over the
final unit of evolution time.
% $\varepsilon_{XX}=\int_{\sqrt{n}-1}^{\sqrt{n}}\big|C_{XX}(t)-C_{XX}^{\mathrm{exact}}(t)\big|\,\mathrm{d}t$.
The runtimes are single-GPU wall times.
(e)~Runtime versus~$n$ for the PEPS ansatz with $\chi=8$ (evolved by
simple update and by tVMC) and for TEBD with $\chi=128$.
The green ellipse marks the $n=30$ PEPS and TEBD data.
Dotted lines are exponential fits for $n\ge16$.
The fault-tolerant quantum estimate at target accuracy
$\varepsilon=10^{-2}$ is shown for $p=10^{-3}$ (dash-dotted black line) and for $p=10^{-4}$ (solid black line). The looser target $\varepsilon=10^{-1}$ (green) is shown for the
same two values of $p$ with the same line styles.
(f)~Wall times of full state-vector simulation (on up to eight
GPUs with 80 GB memory) with the restarted-Krylov (squares)
and fourth-order Trotter-decomposition (triangles) methods.
The local-observable error of $\langle Z_{(2,2)}\rangle(t)$
% $|\langle Z_{(2,2)}\rangle(t)-\langle Z_{(2,2)}\rangle_{\mathrm{exact}}(t)|$
is calibrated to stay below $10^{-2}$.
Each data point is labelled with its peak GPU memory usage.
Dotted lines are exponential fits for $n\ge24$.
% Quantum curves as in~(e).
Grey dotted horizontal lines mark reference timescales.
}
\label{fig:2D_classical_quantum}
\end{figure*}

The scaling of the total sampling cost with the number of qubits for 1D and 2D for $\pphys = 10^{-3}$ is shown in Fig.~\ref{fig:RUS_preparation}(f). From there, we can see that the coherent observable estimation is strictly better than the direct incoherent measurements. The total runtime can be directly computed by multiplying the required code cycles and the sampling overhead in Fig.~\ref{fig:RUS_preparation}(f), which will be presented in Fig.~\ref{fig:1D_classical_simulation} and Fig.~\ref{fig:2D_classical_quantum}.

\section{Fault-tolerant resource estimation and quantum-classical crossover}
 
% We first present quantum resource estimates using the full-stack framework, focusing on the total runtime as the system size increases.  
% The analysis starts with the structure of the Trotterised circuit. 
% For both the 1D and 2D models, each Trotter segment consists of a constant number of layers of local Pauli rotations, and the nearest-neighbour interactions can be partitioned into constant disjoint sets, allowing parallel implementation across $\cO(n)$ sites. 
% Rotation-based methods significantly reduce the required QEC cycles compared to Clifford+\texttt{T} implementations (see \suppref{sec:small-angle-rotation} for details).
% We consider two representative physical error rates of $\pphys = 10^{-3}$ and $\pphys = 10^{-4}$.
% Each QEC cycle is assumed to take $500$~ns.
% The evolution time is set to $t = n/2$ in 1D and $t = \sqrt{n}$ in 2D, a regime in which classical computing remains capable (i.e., the classical simulation error can be controlled below $0.1$).
% Based on the results shown in Fig.~\ref{fig:RUS_preparation}, the quantum runtime can be estimated.
% The resulting estimates are shown in Fig.~\ref{fig:1D_classical_simulation} (1D) and Fig.~\ref{fig:2D_classical_quantum} (2D). 

We first present quantum resource estimates derived from our full-stack framework, focusing on how the total runtime scales with the system size $n$. 
The overall runtime is estimated by combining the execution cost of the Trotterised circuit with the query and sampling overheads dictated by the observable-estimation algorithm and error mitigation. 
At the algorithmic level, for both 1D and 2D models, nearest-neighbour interactions can be partitioned into disjoint sets, allowing each Trotter segment to be executed in parallel across all $\mathcal{O}(n)$ sites. 
At the hardware level, employing direct rotation-state injection significantly reduces the required QEC cycles compared to conventional Clifford+\texttt{T} implementations (see \suppref{sec:small-angle-rotation} for details).
We assume a QEC cycle time of $500$~ns and evaluate two representative physical error rates: $\pphys = 10^{-3}$ and $\pphys = 10^{-4}$. 
The target evolution time is set to $t = n/2$ for 1D systems and $t = \sqrt{n}$ for 2D systems, a regime in which classical computing remains capable (i.e., the classical simulation error can be bounded below $0.1$), thus enabling a rigorous quantum-classical comparison. 
By integrating the single-query circuit costs with the optimised query budget from the observable-estimation algorithm (Fig.~\ref{fig:RUS_preparation}(e,f)), we compute the total quantum runtime, shown in Fig.~\ref{fig:1D_classical_simulation} (1D) and Fig.~\ref{fig:2D_classical_quantum} (2D).

Next, we establish the classical computational cost for 1D and 2D dynamics. We evolve from the product state $\ket{\mathbf{0}}$ on a chain of $n$~sites with open boundary conditions up to $t=n/2$ in 1D, a timescale sufficient for correlations to spread across the chain, and to $t=\sqrt{n}$ in 2D, the light-cone traversal time of the linear lattice extent. 
%Note that here, for a clear demonstration of the separation, we choose $O$ to be a single Pauli-1 observable.
% \sun{
% We choose $O$ to be a single-site Pauli observable, which is relatively easy to simulate by tensor networks.}
Both dimensions use the same model parameters of Ising coupling $J=1$ and transverse and longitudinal fields $h_x=0.8$ and $h_z=0.9$, respectively.
% The classical computational cost based on TN grows exponentially with system size due to the rapid buildup of entanglement.
The time-evolved state is represented as an MPS~\cite{schollwoeckDensitymatrixRenormalization2011} in 1D and a PEPS~\cite{verstraeteRenormalizationAlgorithms2004} in 2D  with bond dimension~$\chi$.
\sun{Herein PEPS refers to PEPS with simple-update, unless used in conjunction with tVMC sampling.}
Each Trotter layer is compressed onto the MPS via a sequential singular value decomposition (SVD) sweep, which introduces truncation errors that grow with entanglement. 
% In our MPS algorithm, each Trotter gate is represented as an exact matrix product unitary, so the gate application itself preserves the state norm.
In our MPS algorithm, the only source of norm loss is the subsequent SVD truncation at each bond.
% The fourth-order decomposition suppresses the Trotter discretisation error well below the SVD truncation error, making the latter the dominant source of error. 
We therefore use the norm loss $\varepsilon_{\mathrm{norm}}=1-\langle\psi_{\mathrm{MPS}}|\psi_{\mathrm{MPS}}\rangle$ as a direct and faithful measure of the cumulative SVD truncation error throughout the evolution. We also monitor the energy density error $\varepsilon_E = {|E(t) - E(0)|}/{n}$, where $E(t)$ is the energy expectation value.
Because the exact dynamics conserve the energy, $\varepsilon_E$ directly measures the accumulated Trotter and truncation error. 

% Beyond TN methods, we benchmark two classes of classical baselines: full state-vector evolution on GPUs based on a restarted-Krylov method and time-evolving block decimation (TEBD)~\cite{vidalEfficientSimulation2004}, see \hyperref[sec:methods]{Methods}. 

Beyond standard TN methods, we benchmark with classical baselines. First, we perform exact, full state-vector evolution on GPUs based on a restarted-Krylov method and time-evolving block decimation (TEBD)~\cite{vidalEfficientSimulation2004}. Second, tVMC methods are applied: in 1D, we evolve a Jastrow wavefunction to variationally capture long-range correlation dynamics; in 2D, we apply tVMC to evolve a PEPS ansatz (PEPS-tVMC), which allows for global parameter optimisation via Monte Carlo sampling.

% In two dimensions the state is represented as a projected entangled pair state (PEPS)~\cite{orusTensorNetworks2019} on a $n=N_x\times N_y$ lattice with bond dimension~$\chi$.
% Time evolution starts from the same product state and uses the simple-update scheme~\cite{jiangAccurateDetermination2008,jahromiUniversalTensor2019} with 2nd-order Trotter splitting and time step $\Delta t=0.01$.
% The evolution runs to $t=\sqrt{n}$, the light-cone traversal time of the linear lattice extent.
% Energy expectation values in 2D are computed via single-layer boundary MPS contraction with boundary bond dimension $\chi_{\mathrm{b}}=256$.

To quantify the exponential scaling of classical resources in 1D, we first benchmark the MPS evolution at fixed bond dimension.
The top row of Fig.~\ref{fig:1D_classical_simulation} summarises the runtime and the final error for chains up to $n=64$ at bond dimensions up to $\chi=256$.
At each fixed~$\chi$ the runtime grows exponentially with~$n$ [Fig.~\ref{fig:1D_classical_simulation}(a)].
At fixed~$n$ it follows a power law $t_{\mathrm{run}}\sim\chi^{\gamma}$ with $\gamma\approx2.2$ [Fig.~\ref{fig:1D_classical_simulation}(b)].
These runs are nevertheless far from converged.
For $n=64$ and $\chi=128$ both error measures stay below $10^{-5}$ up to $t\approx5$ [Fig.~\ref{fig:1D_classical_simulation}(c)].
They then grow by many orders of magnitude once the entanglement generated by the dynamics exceeds the representational capacity of the ansatz.
The norm loss passes $10^{-2}$ around $t\approx9$ and reaches $\varepsilon_{\mathrm{norm}}\approx0.44$ at $t=32$.
Even $\chi=256$ still leaves $\varepsilon_{\mathrm{norm}}\approx0.3$ at $n=64$ [Fig.~\ref{fig:1D_classical_simulation}(b)].

The fixed-$\chi$ benchmarks thus expose rapid cost growth but leave the accuracy uncontrolled. We therefore systematically investigate the bounded-error setting, in which the objective is to keep the simulation error below a prescribed target.
For each~$n$ we determine the minimum bond dimension $\chi_{\min}$ that keeps $\varepsilon_{\mathrm{norm}}\le\varepsilon$ at $t=n/2$ for $\varepsilon=10^{-1} $ and $10^{-2}$ in which we find $\chi_{\min}\sim e^{0.12n}$, $e^{0.17n}$.
The runtime grow exponentially with~$n$ [Figs.~\ref{fig:1D_classical_simulation}(e)]: $t_{\mathrm{run}}\sim e^{0.28n}$, $e^{0.44n}$ for the target precisions, respectively.
Extrapolating these fits, a chain of $n=100$ would require about $10^{9}$~years at $\varepsilon=10^{-2}$. 

By contrast, the fault-tolerant quantum runtime at target accuracy $\varepsilon=10^{-2}$ grows far more slowly than any of the classical fits.
It remains below $2$ hours up to $n=100$ ($\approx1.8$~h at $p=10^{-3}$ and $\approx3$~min at $p=10^{-4}$).

We emphasise that 1D dynamics simulated with MPS is the setting in which classical TN methods are strongest.
It is also the setting in which the bounded-error comparison can be made cleanly, since the error target can actually be met at finite~$\chi$.
Even here, the classical cost explodes at a few tens of sites, and the quantum--classical crossover is already directly visible in 1D: the crossover lies at $n\approx18$--$22$ [Fig.~\ref{fig:1D_classical_simulation}(e)].

The conclusion is corroborated by tVMC based on the Jastrow ansatz [Fig.~\ref{fig:1D_classical_simulation}(d)] and full state-vector GPU simulations [Fig.~\ref{fig:1D_classical_simulation}(f)]. The 1D tVMC baseline using the Jastrow ansatz struggles to maintain the strict $\varepsilon \le 10^{-2}$ error threshold over long physical times as the error quickly accumulates.
The error of $\varepsilon_Z$ fluctuates and cannot be controlled below $0.07$ with increasing Jastrow body orders even for a much shorter timescale up to $t = \sqrt{n}$ [Fig.~\ref{fig:1D_classical_simulation}(d2)].
 Both the restarted-Krylov and the fourth-order Trotter-decomposition evolutions show runtimes growing as $\sim e^{0.7n}$ to control the simulation error below $\varepsilon = 10^{-2}$.
Their peak GPU memory usage grows as $2^n$ and already reaches $223$~GB at $n=32$, the largest benchmarked size of the Trotter evolution.
They cross the quantum estimates at $n\approx25$--$27$. These results imply that a practical quantum advantage can be achieved even in the 1D setting.
% does not require the 2D setting in which classical methods struggle.

% \gyt{Similarly, the 1D tVMC baseline using the Jastrow ansatz struggles to maintain the strict $\varepsilon \le 10^{-2}$ error threshold over long physical times as the error quickly accumulates.}   \gyt{While highly effective for low-energy states, the rapidly growing complexity of the dynamical phase profile makes accurate variational tracking increasingly difficult at larger system sizes. To this end, we only simulate the time evolution up to $t=\sqrt{n}$. A practical quantum advantage thereby does not require the 2D setting in which classical methods struggle.}
% A practical quantum advantage thereby does not require the 2D setting in which classical methods struggle.

In 2D, the central difficulty for classical simulation is error control rather than runtime alone.
The truncation error accumulates rapidly as the evolution time increases.
The error analysis in Fig.~\ref{fig:dim2-op-state} shows all error measures growing by orders of magnitude already within $t<\sqrt{n}$, the light-cone traversal time of the linear lattice extent.
Evolution beyond this timescale with controlled error is therefore out of reach.
This motivates restricting the 2D comparison to $t=\sqrt{n}$.
For $n\ge20$, the final-time energy-density error no longer decreases systematically with bond dimension and saturates at $\varepsilon_E\gtrsim5\times10^{-2}$ [Figs.~\ref{fig:2D_classical_quantum}(a,b)].
Only the smallest lattice reaches $\varepsilon_E\approx10^{-2}$ ($n=16$ with TEBD at $\chi=128$).
The 2D error therefore remains above $10^{-2}$ throughout the benchmarked range.
Pushing it down to $10^{-2}$ would require substantially larger bond dimensions, whose cost grows steeply [Figs.~\ref{fig:2D_classical_quantum}(c,e)].
Already at the largest affordable bond dimensions the evolution at $n=30$ takes \gyt{$20$~hours for PEPS-tVMC}, $6.5$~hours for PEPS\gyt{-SU} at $\chi=8$ and $5.1$~hours for TEBD at $\chi=128$ [Fig.~\ref{fig:2D_classical_quantum}(e)]. Although PEPS-tVMC avoids the systematic bias of local projection schemes by updating the tensors globally, the statistical sampling overhead and optimisation cost of the stochastic reconfiguration step pose challenges at higher bond dimensions.
As shown in Figs.~\ref{fig:2D_classical_quantum}(d1,e),   the bond dimension to $\chi=8$ incurs a longer runtime than PEPS.

By contrast, the fault-tolerant quantum estimate at $\varepsilon=10^{-2}$ takes $57$~seconds at $n=30$ and stays below $9$~minutes up to $n=100$ [Fig.~\ref{fig:2D_classical_quantum}(e)].
The quantum simulation at $n=30$ is thus two to three orders of magnitude faster and more accurate.
The gap widens exponentially with~$n$.
Full state-vector GPU simulation with its error controlled below $10^{-2}$ stays faster than the quantum estimate only up to $n\approx26$ [Fig.~\ref{fig:2D_classical_quantum}(f)].
Beyond this size, its runtime and GPU memory grow exponentially, and the quantum estimate takes over.

The comparison reveals a crossover regime in which quantum simulation becomes favourable. For small system sizes and short times, classical methods remain more efficient due to the large constant overhead of fault-tolerant quantum computation.
However, the classical cost scales exponentially while the quantum cost grows polynomially.
Fault-tolerant quantum simulation therefore becomes the faster approach beyond a few tens of sites.
The crossover lies at $n\approx15$--$34$ in 1D, depending on the accuracy target and physical error rate.
In 2D, it lies at $n\approx26$ compared with full-state-vector methods, while approximate TNs are already slower and less accurate at $n=16$.
For the 2D model, the quantum runtime based on rotation-state preparation is on the order of seconds to minutes at $n\sim100$.
The crossover thus occurs at modest system sizes that are well within the reach of near-term fault-tolerant devices.

\section{Discussion}

% On the quantum side, we perform a detailed resource analysis starting from the Trotterised circuit representation of the time-evolution operator and propagating it through compilation, gate synthesis, and surface-code-based error correction. We have applied state-of-the-art quantum algorithm construction and fault-tolerant compilation. 

We start with a resource analysis at the algorithmic level to determine the required non-Clifford gate count, focusing on both small-angle rotations and their \texttt{T}-gate equivalents. 
% We then consider implementing these operations onto a surface-code-based fault-tolerant architecture.  
% We directly implement fault-tolerant small-angle rotation gates without fully decomposing them into \texttt{T} gates. Here, we estimate the cost of synthesising approximate rotation gates using error-crafted or approximate non-Clifford primitives or the method by Zeng et al.~\cite{zeng2025error}. 
% This analysis captures potential savings when the rotation angles are sufficiently small and the gate number is not very large. 
We then consider implementing these operations within a surface-code-based fault-tolerant architecture through two distinct approaches: MSD for \texttt{T} gates, and direct rotation-state injection.
For the latter, rather than fully decomposing operations into \texttt{T} gates, we directly implement fault-tolerant small-angle rotations and evaluate their cost via error-crafted non-Clifford primitives~\cite{zeng2025error}.
This direct approach captures significant resource savings when rotation angles are sufficiently small, and the overall gate count is moderate.
Importantly, our approach suppresses physical errors to $\mathcal{O}(|\theta|\pphys^2)$, substantially extending the regime to which this method is applicable.   
% This work focuses on the regime where fewer than one million physical qubits are available. Although the MSD method can reduce runtime through parallelisation (e.g., using a time-optimal strategy~\cite{Fowler2013TimeOptimalComputation}), it becomes less practical as the number of qubits increases.
In contrast, while the MSD approach may significantly reduce runtime through extensive parallelisation (e.g., time-optimal strategies~\cite{Fowler2013TimeOptimalComputation}), prohibitive qubit requirements of MSD render it less practical in the early fault-tolerant regime with qubit numbers below one million. 
% Because this work focuses on near-term architectures with fewer than one million physical qubits, 

% We found a way to balance the circuit depth for coherent estimation and incoherent estimation. This way enables a co-design of QEC process and classical residual error mitigation. Our result indicates that at an intermedaiet stage where physical error is large, we need to balance the maximum amount of gates to control the logicla errrors.

We introduce a co-design of observable estimation, QEC and residual-error mitigation that balances algorithmic and fault-tolerant resource costs. In particular, for relatively high physical error rates, increasing circuit depth beyond a certain threshold becomes counterproductive: although the algorithmic error is reduced, the accumulation of logical errors leads to a substantial increase in the error-mitigation overhead. 
The optimal operating point is therefore determined by a balance between residual simulation error and fault-tolerant overhead. \sun{Furthermore, this work highlights the importance of improving gate fidelity rather than merely increasing the number of qubits. As demonstrated, achieving a physical error rate of $p=10^{-4}$ leads to significant reductions in both qubit count and runtime.}

A key finding from resource comparison is that the quantum-classical crossover for 1D systems can be identified unambiguously, as the TN bond dimension is determined at a fixed target error. This allows the classical computational cost to be quantified directly from the resources required to maintain a prescribed simulation accuracy. 
By systematically estimating the bond dimension and runtime, we establish the complexity of the classical simulation and obtain a well-defined benchmark for determining the onset of quantum advantage. 
% For 2D systems, it is hard to reduce the error below $0.01$ for classical algorithms. 
% Quantum computing is more advantageous for 2D simulations.
In stark contrast, classical simulations of 2D systems are difficult to suppress the simulation error below $0.01$. 
Consequently, 2D dynamics simulations present a much more immediate and compelling regime for demonstrating practical quantum advantage.

% Much of the existing resource-estimation literature for physics applications, however, has largely focused on ground-state energy estimation~\cite{chung2026partially,akahoshi2025compilation,lee2021even,ding2023even,yoshioka2024hunting,chung2026partially}. 
 
This work focuses on quantum simulation tasks, whereas extensive research exists on resource estimations for ground-state energy estimation~\cite{chung2026partially,akahoshi2025compilation,lee2021even,ding2023even,yoshioka2024hunting,chung2026partially,lee2021even} and cryptography tasks~\cite{zhou2025resource, webster2026pinnacle, gidney2021factor, cain2026shor}, an emerging direction of research. 
% In particular, there are considerable progress on fault-tolerant resource estimation for ground-state energy estimation~\cite{chung2026partially,akahoshi2025compilation,lee2021even,ding2023even,yoshioka2024hunting,chung2026partially,lee2021even}. 
A shared challenge in quantum eigenenergy estimation~\cite{keen2021quantum,chakraborty2024implementing,wang2023faster,ding2024quantum,wang2023quantum,lin2021heisenberg,zeng2021universal,lu2021algorithms,zhang2022computing,huo2021shallow,wang2023quantum,ding2024quantum,he2022quantum,wang2023faster,wang2024qubit,lin2020near,an2023linear}, however, is the requirement for a finite initial state overlap, which usually does not hold, as evidenced by complexity-theoretic arguments~\cite{kempe2006complexity} and empirical simulations~\cite{lee2023evaluating}.
%These algorithms often rely on empirical results to improve performance. 
To circumvent the prohibitive costs dictated by worst-case bounds, these approaches may rely on empirical estimates to justify their performance.
For example, Ref. \cite{akahoshi2025compilation}, based on the phase estimation method proposed in~\cite{ding2023even}, uses a much smaller estimate of the actual overhead, around 0.06, which is significantly lower than the scaling prediction of $\sqrt{1-p_{\operatorname{init}}^2}=0.8$ with initial state overlap $p_{\operatorname{init}}$. 
In addition, this work focuses on surface-code implementation, while implementing small-rotation gate injection in the qLDPC code~\cite{ismail2026transversal} could also be an interesting direction for future work.

\section*{Methods}
\label{sec:methods}

\subsection{Algorithm-QEC co-design and quantum resource estimation}

% The quantum resources required to simulate the 1D and 2D mixed-field Ising models are summarised in Table \ref{tab:resource}. The resource estimates reported in the table are obtained by combining the Trotterisation cost, QEC cycles, and logical error mitigation overhead (relevant to the query budget (in observable estimation)) and the QEC logical error rate. We briefly discuss the interplay between these components below.

In our analysis, the 1D system evolves to significantly longer times than the 2D system, as classical TN methods face severe limitations in 2D that restrict accurate classical simulations to much shorter timescales. 
Consequently, the increased circuit depth associated with these prolonged 1D simulations results in a higher accumulated logical error rate, directly increasing the error-mitigation sampling overhead required for accurate observable estimation.

\begin{table*}[t]
    \centering
    \small
    \caption{\textbf{Quantum resources required for simulating 1D and 2D mixed-field Ising models with different time scales and target precision $\varepsilon$.} 
    Due to the trade-off between error-mitigation overhead and query depth, we optimise the total runtime (i.e., the product of sampling overhead and the total number of queries). This table displays the resources at $n = 100$, with two columns $\cdot / \cdot$ displaying results for physical error rates of $p = 10^{-3}$ and $p = 10^{-4}$, respectively. The corresponding required physical qubits are $3.76 \times 10^5$ (for $p=10^{-3}$) and $3.12 \times 10^4$ (for $ p = 10^{-4}$). Total cycles: QEC cycles $\times$ error mitigation overhead.  }
    \label{tab:resource}
    \renewcommand{\arraystretch}{1.08}
    \begin{tabular}{@{}c|c c c c c c c@{}}
        \toprule[0.9pt]
        $\varepsilon$ & Max query & Total queries & Error-mitigation overhead & QEC cycles & Total cycles & Runtime (s)  \\
        \midrule[0.6pt]
        $\varepsilon = 0.1~ (t = n/2)$
        &  $1/1 $
        &  $20/20 $
        &  $36/20 $
        & $6.5\times10^{6}/2.7\times10^{6}$
        & $2.3\times10^{8}/5.4\times10^{7}$
        & $116.7/27.0$ 
        
        \\

        $\varepsilon = 0.01~ (t = n/2)$
        &  $2/30 $
        & $1140/200 $
        & $2094/237 $
        & $6.5\times10^{6}/2.7\times10^{6}$
        & $1.4\times10^{10}/6.4\times10^{8}$
        & $6804.0/315.6$ 
     \\

        $\varepsilon = 0.1~ (t = \sqrt{n})$
        &  $1/1 $
        & $20/20 $
        & $23/20 $
        & $1.3\times10^{6}/5.4\times10^{5}$
        & $3.0\times10^{7}/1.1\times10^{7}$
        & $15.4/5.4$ 
         \\

        $\varepsilon = 0.01~ (t = \sqrt{n})$
        & $16/30$
        & $340/200$
        & $759/205$
        & $1.3\times10^{6}/5.4\times10^{5}$
        & $9.9\times10^{8}/1.1\times10^{8}$
        & $507.4/53.8$ 
        \\

        $\varepsilon = 0.1~ (t = n)$
        & $1/1 $
        & $20/20 $
        & $67/21$
        & $1.7\times10^{7}/7.1\times10^{6}$
        & $1.2\times10^{9}/1.5\times10^{8}$
        & $591.0/73.1$ 
         \\

        $\varepsilon = 0.01~ (t = n)$
        & $1/30 $
        & $1200/200 $
        & $4045/259 $
        & $1.7\times10^{7}/7.1\times10^{6}$
        & $6.9\times10^{10}/1.8\times10^{9}$
        & $34856.6/921.5$ 
         \\
        \bottomrule[0.9pt]
    \end{tabular}
\end{table*}

% The quantum resources required to simulate the 1D and 2D mixed-field Ising models are summarised in Table~\ref{tab:resource}.

Table~\ref{tab:resource} details the maximum query number, total queries, error-mitigation overhead, QEC cycles, and the total resource cost for each simulation configuration. 
The QEC cycle count is primarily determined by the Trotterised time-evolution sequence and the number of state-injection operations required for logical rotations. 
The overall resource cost is then rigorously quantified by combining this base QEC cycle count with the optimised error-mitigation sampling overhead.

These estimates are obtained by systematically integrating the Trotterisation circuit depth, the corresponding QEC cycle count, the QEC logical error rate,   the error-mitigation sampling overhead, and the observable-estimation query budget. We briefly discuss the interplay between these components below.

% The observable-estimation algorithm seeks the optimal query budget that balances the max query and logical-error mitigation overhead. 
The observable-estimation algorithm determines an optimal query budget by balancing the maximum query depth against the overhead of logical-error mitigation. Increasing the maximum query number reduces the total number of queries required to achieve a target precision $\varepsilon$, ultimately approaching the Heisenberg limit of $\mathcal{O}(1/\varepsilon)$ samples. 
Conversely, restricting the maximum query number to one corresponds to direct measurement at the standard quantum limit, which necessitates $\mathcal{O}(1/\varepsilon^2)$ samples.
However, this algorithmic advantage incurs a severe overhead: as shown in Fig.~\ref{fig:RUS_preparation}(d), extending the maximum query number inevitably amplifies the accumulated logical error, resulting in an error-mitigation sampling overhead that grows exponentially as $e^{4p_{\mathrm{logical}}}$.

The resulting optimized total queries are depicted in Fig.~\ref{fig:RUS_preparation}. For a physical error rate of $\pphys=10^{-3}$, an optimal maximum query number emerges from the trade-off between query complexity and error-mitigation overhead. 
Even at this optimum, our co-design approach retains a significant scaling advantage over direct incoherent measurement at the standard quantum limit. 
At a lower physical error rate of $\pphys=10^{-4}$, the logical error rate is strongly suppressed, scaling as $\mathcal{O}(\pphys^2)$, which correspondingly reduces the requisite query budget.

% \subsection{Sampling overhead for logical error mitigation}

% This work takes into account that larger query numbers occur less frequently, thereby reducing the overall sampling cost. 
% The error-mitigation overhead for physical error rates of $\pphys=10^{-3}$ and $\pphys=10^{-4}$ is shown in Fig.~\ref{fig:Methods_diff_Pphys}. 
% The sampling overhead decreases dramatically as the physical error rate is reduced, as expected. For $\pphys=10^{-4}$, the additional sampling cost associated with residual logical errors becomes nearly negligible. This highlights the significant benefits of improving hardware performance: reductions in the physical error rate directly lower fault-tolerant overhead, leading to improved overall computational efficiency.

\begin{figure}[tb]
\centering
\includegraphics[width=1.0\linewidth]{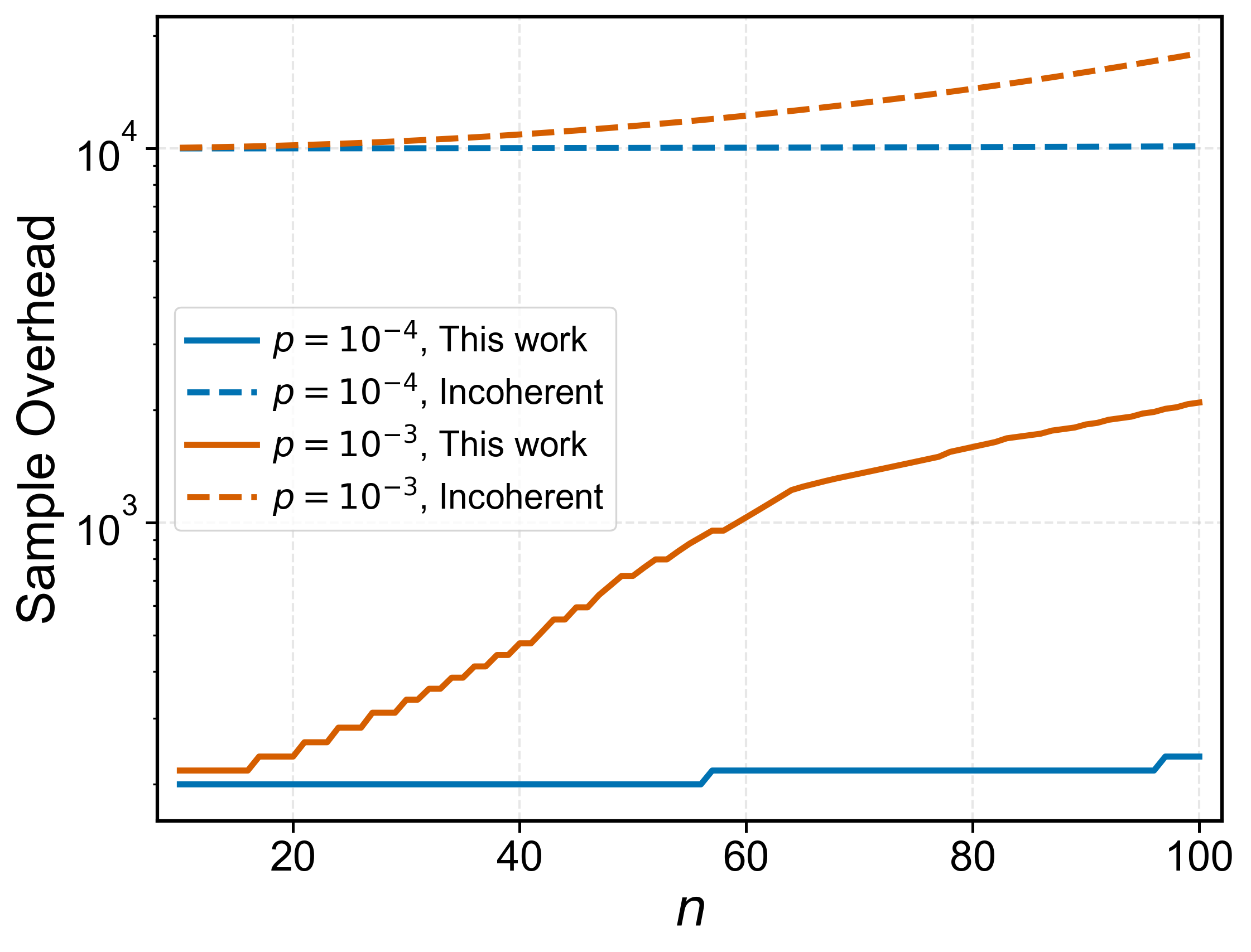}
\caption{Error mitigation sampling overhead for $\pphys = 10^{-3}$ and $\pphys = 10^{-4}$ in the 1D case.  
}
\label{fig:Methods_diff_Pphys}
\end{figure}

% We present the sampling overhead for logical error mitigation in 
% Our algorithm accounts for the fact that deeper circuits, which require larger query numbers, are sampled with exponentially suppressed probabilities, thereby reducing the overall sampling overhead. 
Fig.~\ref{fig:Methods_diff_Pphys} illustrates this error-mitigation overhead for physical error rates of $\pphys=10^{-3}$ and $\pphys=10^{-4}$. 
As expected, the sampling overhead drops dramatically as the physical error rate improves. 
Notably, at $\pphys=10^{-4}$, the additional sampling cost incurred by residual logical errors becomes practically negligible. 
This highlights the significant benefits of improving hardware performance: reducing the physical error rate not only suppresses fault-tolerant overhead but also reduces the algorithmic sampling requirements, thereby improving overall computational efficiency.

\subsection{Classical simulation methods}

In 1D, the time-evolved state is represented as a MPS with bond dimension~$\chi$.
We apply a fourth-order Suzuki-Trotter decomposition with time step $\Delta t=0.1$ which suppresses the Trotter discretisation error below the SVD truncation error.
Each Trotter layer is compressed onto the MPS via a sequential SVD sweep, which introduces truncation errors that grow with entanglement.
In our MPS algorithm, each Trotter gate is represented as an exact matrix product unitary  (MPU), so the gate application itself preserves the state norm.
The only source of norm loss is the subsequent SVD truncation at each bond.
% The fourth-order decomposition suppresses the Trotter discretisation error well below the SVD truncation error, making the latter the dominant source of error. 
% As shown in the main text, we use the norm loss $\varepsilon_{\mathrm{norm}}=1-\langle\psi_{\mathrm{MPS}}|\psi_{\mathrm{MPS}}\rangle$ as a direct and faithful measure of the cumulative SVD truncation error throughout the evolution. We also monitor the energy density error $\varepsilon_E = \frac{|E(t) - E(0)|}{n}$, where $E(t)$ is the energy expectation value.
% Because the exact dynamics conserves the energy, $\varepsilon_E$ directly measures the accumulated Trotter and truncation error.
% We evolve from the product state $\ket{\mathbf{0}}$ on a chain of $n$~sites with open boundary conditions up to $T=n/2$, a timescale sufficient for correlations to spread across the chain.
The 2D state is represented as a PEPS on a $n=N_x\times N_y$ lattice with bond dimension~$\chi$.
Time evolution starts from the same product state and uses the simple-update scheme~\cite{jiangAccurateDetermination2008} with second-order Trotter splitting and time step $\Delta t=0.01$.
The evolution runs to $t=\sqrt{n}$, the light-cone traversal time of the linear lattice extent.
Energy expectation values in 2D are computed via single-layer boundary MPS contraction with boundary bond dimension $\chi_{\mathrm{b}}=256$.

\gyt{We also benchmark tVMC using a Jastrow expansion ansatz in 1D and a PEPS ansatz in 2D. Its error, arising from the variational approximation and finite Monte Carlo sampling, is quantified by the mean absolute observable deviation from full state-vector evolution. These tVMC baselines test whether global variational optimisation can avoid the local truncation errors of TN time evolution. In practice, the same mixed-field Ising dynamics remains difficult: the rapidly growing entanglement and dynamical phase structure lead to expressivity limits, statistical noise, and increasingly unstable stochastic-reconfiguration solves. The detailed are given in the SI.}

Beyond these methods, we benchmark two further classes of classical baselines.
The first is TEBD applied to the 2D lattice through a 1D snake mapping (implemented with the ITensor package~\cite{fishmanITensorSoftware2022}).
The second is full state-vector evolution on GPUs based on a restarted-Krylov method and a fourth-order Trotter decomposition.
% ~\cite{forestFourthorderSymplectic1990,yoshidaConstructionHigherOrder1990,suzukiFractalDecomposition1990}.
Both run in single precision (complex64), which halves the memory usage and accelerates the memory-bandwidth-bound evolution.
The free parameters of both baselines are tuned to keep the simulation error below $10^{-2}$, where the error is measured against a double-precision Chebyshev reference. Implementation details and extended benchmarks are given in the SI.

% Note that here, for a clear demonstration of the separation, we choose $O$ to be a single Pauli-1 observable. This is usually considered classically easy.
% However, even in this case, the TN-based method finds it hard to bound the error.
In this work, we consider $O$ to be a few-body Pauli observable (single-site or two-point correlation) because it is usually considered classically easy. And yet, the TN or tVMC method still struggles to bound the simulation error. For estimation on a quantum circuit, however, $O$ is not restricted to be a Pauli operator. If $O$ can be expressed as a linear combination of Pauli operators, this suffices to construct a block encoding of $O$~\cite{Koizumi26Faster}, after which amplitude estimation can be directly applied. In Supplementary Sec.~\ref{sec:algorithmic_level}, we discuss how entanglement allows for an additional speed-up with respect to the Frobenius norm of the observable $\| O \|_F$~\cite{feng2025trotterizationoperatorscramblingentanglement}, which satisfies $\| O \|_F \leq \| O \|$.

% and 
% The first is  Time-Evolving Block Decimation (TEBD)~\cite{vidalEfficientClassical2003,vidalEfficientSimulation2004} applied to the 2D lattice through a 1D snake mapping.
% The second is full state-vector evolution on GPUs based on a restarted-Krylov method~\cite{parkUnitaryQuantum1986,hochbruckKrylovSubspace1997} and a fourth-order Trotter decomposition~\cite{forestFourthorderSymplectic1990,yoshidaConstructionHigherOrder1990,suzukiFractalDecomposition1990}.

\subsection{Scaling of fault-tolerant small-angle rotations}

We now estimate $T_{\mathrm{step}}$, the logical clock cost of a single Trotter segment within the rotation-based paradigm. Each logical clock spans a number of QEC cycles proportional to the code distance $d$, since it requires $d$ rounds of syndrome extraction to suppress the measurement error.

Each segment contains layers of single-qubit rotations and nearest-neighbour two-qubit rotations, all of which can be reduced to logical $Z$ rotations up to Clifford conjugations. 
Because these Clifford basis transformations incur only a constant QEC clock overhead, the dominant cost arises from the non-Clifford rotations.
Letting $T_Z(n)$ denote the QEC clock of applying a parallel layer of $n$ logical single-qubit $Z$ rotations, and $T_{ZZ}(n)$ the cost of applying a layer of nearest-neighbour two-qubit rotations, we can write
$
T_{\mathrm{step}} = \cO(1) + \alpha \, T_Z(n) + \beta \, T_{ZZ}(n),
$
where $\alpha$ and $\beta$ are constants determined by the decomposition of the Hamiltonian.
For nearest-neighbour interactions, the two-qubit rotations can be partitioned into a constant number of disjoint sets, allowing the rotations within a single set to be executed in parallel.
Because the operations within each set can be reduced to $Z$-type rotations up to a constant number of Clifford gates, we obtain the scaling relation
$
T_{ZZ}(n) = \cO(1) + \gamma\, T_Z(n/2),
$
where $\gamma$ is a constant dependent on the partitioning of the two-qubit rotations.
The dominant cost of $T_{\mathrm{step}}$ is thus governed by $T_Z(n)$.

The scaling of $T_Z(n)$ depends on the specific fault-tolerant implementation of non-Clifford operations. In the rotation-based paradigm, logical $Z$ rotations are implemented using RUS protocols. 
Each individual rotation succeeds with constant probability, and the total number of rounds required to complete a layer of $n$ parallel rotations grows logarithmically with $n$. Denoting by $\langle k_{\max} \rangle$ the expected number of RUS rounds, we have
$ 
\langle k_{\max} \rangle \sim \log n.
$
The total cost required to execute one rotation layer is therefore
$
T_Z(n) \sim \langle k_{\max} \rangle \, (t_i + 1),
$
where $t_i \leq 2$ is the cost of preparing a logical rotation state, and the $+1$ term accounts for the subsequent injection step. Combining these bounds, the overall scaling for a single Trotter segment is established as $T_{\mathrm{step}} \sim \log n$, up to constant factors.

\subsection{Magic state distillation}
In the Clifford+\texttt{T} model, logical \texttt{T} gates are realised via magic state injection. MSD protocols, typically based on concatenated codes, are implemented using lattice surgery to provide the necessary logical connectivity within the surface code. In this architecture, specialised regions of the device act as magic state factories that repeatedly generate noisy $|T\rangle$ states and distil them into high-fidelity outputs.

Since each logical \texttt{T} gate consumes one distilled $|T\rangle$ state, magic-state factories constitute a rate-limiting resource that must be scheduled with the logical circuit. The finite factory throughput therefore sets the effective timescale of non-Clifford operations, which is jointly determined by the circuit \texttt{T}-depth and the steady-state production rate of the factories. These states are then stored and routed via lattice surgery to data patches, where they are consumed through gate teleportation. We follow the lattice-surgery framework of \cite{Litinski2019gameofsurfacecodes}.

We compare two Clifford+\texttt{T} realisations in \suppref{subsec:direct_vs_spbc_scheduling}: a sequential Pauli-based-compilation (SPBC) approach and a direct \texttt{T}-gate circuit. The comparison is performed under a fixed magic-state factory budget, allowing us to isolate the effect of different non-Clifford scheduling strategies. Under the default timing model, the direct approach typically achieves lower runtime due to its \texttt{T}-depth–dominated execution and straightforward parallel magic-state injection, while SPBC becomes competitive only when reductions in magic-state consumption or optimistic PPM and correction latencies compensate for its additional overhead and increased effective non-Clifford depth.

We further use the direct implementation for the physical resource estimates in \suppref{app:time-optimal-msd-resource-model}. In this model, runtime is governed by effective \texttt{T}-depth, while factory throughput determines the number of factory lanes and thus the dominant qubit overhead. Clifford operations are subleading and neglected at leading order.

For physical error rates \(p=10^{-3}\) and \(p=10^{-4}\), we scan MSD protocols (Table~1 of Ref.~\cite{Litinski2019MSD}) and select, for each case, the smallest code distance satisfying the target logical error budget~\(\varepsilon\). The resulting configuration minimises physical-qubit cost under the required \texttt{T}-state demand. Compared to rotation-state injection, MSD offers a highly parallelised execution model that avoids significant runtime degradation, but requires substantially larger qubit overhead. For \(n=100\), the estimated cost reaches $10^8$ physical qubits at \(p=10^{-3}\), and decreases significantly at \(p=10^{-4}\) due to reduced code distance and factory overhead.

The quantum resource estimates by MSD are presented in Tables S6 and S7 in Supplementary Sec.~S3.

Magic-state cultivation could further reduce factory overhead, but is not included in the present optimisation.

\vspace{6pt}
\textbf{Author contributions.---}J.S. B.Z. J.X. Y.Y. Z.D., Z.Z. and Y.G. contributed equally to this work. J.S. B.Z. J.X., Y.Y., Z.D., Z.Z., and P.Zeng contributed to and were leading each individual component of this work.
J.S. developed the framework for fault-tolerant resource estimation and led the technical aspect of this work.
P.Zeng led the error correction aspect of this work with input from J.S., Y.Y. and Z.D.
B.Z. and P.Zhang carried out TN simulations.
J.X. carried out Trotter error analysis.
Y.Y. developed and performed resource estimates for magic state distillation. 
Z.D. developed and performed resource estimates for rotation state injection.
Z.Z. and P.Zeng designed the circuit for observable estimation and contributed to resource estimates.
Y.G. and Y.W. carried out VMC simulations.
J.H. and Y.Y. analysed the performance of  Clifford+\texttt{T} scheduling routes. 
S.Z., Z.W., X.Z., and T.L. contributed to the computation and optimisation of Trotter circuits.
T.F. contributed to the Trotter error analysis with multiple observables.
% Y.M. plotted the figures.
J.S. B.Z. J.X., Y.Y., Z.D., Z.Z., J.H., Y.G., P.Zeng, and X.Y. wrote the manuscript. 
A.Y., T.F., W.D., D.S., S.S., Y.L., Q.Z., and P.Zhang revised the manuscript.
All the authors contributed to the discussion, validation of the results, and writing up the manuscript.
P.Zeng and X.Y. supervised this project. 

\bibliographystyle{SciAdv}
\bibliography{Ref_merged}

\widetext
 
\renewcommand{\thesection}{S\arabic{section}}
\renewcommand{\theequation}{S\arabic{equation}}
\renewcommand{\thefigure}{S\arabic{figure}}
\renewcommand{\thetable}{S\arabic{table}}

\newpage

\setcounter{page}{1}
\setcounter{section}{0}

\section*{Supplementary Information}

\renewcommand{\addcontentsline}[3]{\oldaddcontentsline{#1}{#2}{#3}}

\tableofcontents

% \input{Trotter_apd}
% \input{SmallAngle}
% \input{CompilerCalibration}
% \input{TN_sim}

%----------------- Trotter Appendix -----------

\section{Quantum algorithm design and algorithmic level resource estimation}
\label{sec:algorithmic_level}

\subsection{Heisenberg-limit observable estimation based on the Gaussian-sampled  Chebyshev amplitude estimation} \label{sec:observable_est}

To fairly compare the resource costs of quantum and classical algorithms, we focus on the task of observable estimation rather than time-evolved state preparation. That is, to estimate the expectation value of $\bra{\psi(t)}O\ket{\psi(t)}$ to a given accuracy $\varepsilon$, where $\ket{\psi(t)}=U(t) \ket{\psi_0}$ is the final state after given Hamiltonian dynamics. %This tasks requries more 

In the quantum task, suppose we prepare $\ket{\psi(t)}$ and measure $O$ directly; the number of samples required to achieve an $\varepsilon$-accurate estimation scales as $O(1/\varepsilon^2)$. When the accuracy requirement is high, this introduces a challenging sample overhead that prevents practical quantum advantage. For example, to achieve an estimation accuracy of $\varepsilon = 10^{-3}$, we need about $ 10^6$ samples. When no sampling is allowed due to limited quantum resources, we will need about $2\times 10^7$ seconds to complete the estimation, with a single shot taking $20$ seconds, which is about 1.5 years. To solve this problem, we consider introducing amplitude estimation techniques~\cite{Brassard_2002,Rall2023amplitudeestimation,GiurgicaTiron2022lowdepthalgorithms,huang2026low} which allow us to reduce the query number of the Hamiltonian evolution $U(t)=e^{-iHt}$ to $O(1/\varepsilon)$. 
Inspired by a recent low-depth amplitude estimation algorithm called Gaussian least-square amplitude estimation (GLSAE)~\cite{huang2026low}, we design a new algorithm, called the Gaussian-sampled Chebyshev amplitude estimation (GCAE) algorithm~\cite{zhang2026GCAE} and show how it can be used in the task of estimating the properties of the state in a 2D mixed-field Ising Hamiltonian. 
% \sun{We would like to}

In the following section, we first briefly review the GCAE algorithms and compare it with the former GLSAE algorithm. Then, we perform numerical simulation to get the optimal query depth, as required in different scenarios by fault-tolerant quantum computation architecture. Consider an $n$-qubit target state $\ket{\psi} = U\ket{0}$ and a Pauli observable $O$. Define $P:= \frac{1}{2}(I+O)$ as the projector to the $+1$ eigenvalue space of the Pauli observable. If we can estimate the amplitude $a=\|P\ket{\psi}\|^2$ to a given accuracy of $\varepsilon/2$, we can then estimate $\bra{\psi}O\ket{\psi}$ to the accuracy of $\varepsilon$. To implement GCAE, we first randomly sample the unitary query number $m$ ($m$ should an integer) in a circuit from a truncated discrete Gaussian distribution, then implement the amplitude amplification circuit. We then generate two samples and measure the ancillary qubit in the $Z$ basis.

Fig.~\ref{fig:cartoon}(c) in the main text illustrates the overall circuit construction which will be detailed below. When the value $m$ is odd, we perform $t:=\frac{m-1}{2}$ times of the Grover circuit $Q=-(I-2\ket{\psi}\bra{\psi})(I-2P)$ and measure the observable $I-2P$. The expectation value is
\begin{equation} \label{eq:GCAE_odd}
\bra{\psi}U_t^\dagger (I-2P) U_t\ket{\psi} = T_m(1-2a).
\end{equation}
Here, $U_t := Q^t$. $T_m(x):=\cos(m \arccos(x))$ is the Chebyshev polynomial of the first kind.
Otherwise, when the value $m$ is even, we perform $t:=\frac{m}{2}$ times of the Grover circuit $Q$ and measure the observable $2\ket{\psi}\bra{\psi}-I$. The expectation value is 
\begin{equation} \label{eq:GCAE_even}
\bra{\psi}U_t^\dagger (2\ket{\psi}\bra{\psi}-I) U_t\ket{\psi} = T_m(1-2a).
\end{equation}
To estimate the amplitude $a$ using the GCAE algorithm, we randomly sample a non-negative integer $m$  from the following truncated discrete Gaussian distribution
\begin{equation} \label{eq:qTm_GCAE}
\tilde{q}_T(m) =
\begin{cases}
    \frac{2}{\sqrt{2\pi} T \tilde{\mc{Z}} } \exp\left( -\frac{m^2}{2T^2} \right), & \quad 1 \le|m|\leq \sigma T, \\
    0, & \quad |m| > \sigma T, \\
    1 - \sum_{\tau\neq 0} \tilde{q}_T(\tau), & \quad m=0.
\end{cases}
\end{equation}
Here, $\tilde{\mc{Z}}:= \sum_{m=-\rho T}^{\rho T} \frac{1}{\sqrt{2\pi}T}\exp\left( -\frac{m^2}{2T^2} \right)$ is the truncated normalisation factor of the discrete Gaussian function, and $\sigma$ is a truncation parameter determining the maximal query depth $M:=\lfloor \sigma T\rfloor$ in a single round. The parameter $\rho \geq \sigma$ is introduced to ensure that the normalisation is computed over a slightly larger symmetric interval, so that the truncation error outside $[-\rho T, \rho T]$ remains exponentially small. In our numerical experiment, we set $\sigma=4$ and $\rho=16$.

When the sampled $m$ value is $0$, we skip the quantum experiment and perform rejection sampling; when $m$ is odd, we measure the observable $I-2P$ on the state $Q^t \ket{\psi}$ (Here, $t = \frac{m-1}{2}$); otherwise when $m$ is even, we measure the observable $2\ket{\psi}\bra{\psi}-I$ on the state $(I-2P)Q^t\ket{\psi}$ (Here, $t =\frac{m-2}{2}$). In the latter two cases, we record the one-shot measurement outcome as a random variable $Z_m\in\{\pm1\}$.  After sampling $N$ non-zero rounds, we construct the observable estimator by
\begin{equation} \label{eq:GCAE_est}
    \tilde{b} = \argmin_{x\in [-1,1]} \frac{1}{N} \sum_{m\sim \tilde{q}_T} (Z_m - T_m(x))^2.
\end{equation}
And the amplitude $a$ is estimated by $\tilde{a}:= \frac{1}{2}(1-b)$.

Now, we explain why Eq. \ref{eq:GCAE_est} provides a good estimation of $b:=1-2a= -\bra{\psi}O\ket{\psi}$, which is the (negative) observable value. For a fixed sample value of $m$, from Eqs. \ref{eq:GCAE_odd} and \ref{eq:GCAE_even} we have
\begin{equation} \label{eq:LeastSquare}
\begin{aligned}
\mathbb{E}[(Z_m - T_m(x))^2] &= \mathbb{E}[Z_m^2] - 2 T_m(x)\mathbb{E}[Z_m] + T_m(x)^2 \\
&= 1 - 2T_m(x)T_m(b) + T_m(x)^2 \\
&= (1-T_m(b)^2) + (T_m(x) - T_m(b))^2.
\end{aligned}
\end{equation}
The first term $(1-T_m(b)^2) = \mathbb{E}[(Z_m - T_m(b))^2]$ indicates the variance of the estimator which is independent of the value $x$, while the second term $(T_m(x) - T_m(b))^2$ indicates the Chebyshev feature distance between the guess value $x$ and the true value $b$. Consider an arbitrary guess value $x\in [-1,1]$, we can list its Chebyshev value as a vector $\Phi_M(x) = [T_0(x), T_1(x),..., T_M(x)]$. On the other hand, the quantum circuit sample the Chebyshev value of the true value $b$, $\Phi_M(b) = [T_0(b), T_1(b),..., T_M(b)]$. From Eq. \ref{eq:LeastSquare} we can see that the GCAE estimator in Eq. \ref{eq:GCAE_est} is directly related to the following vector distance
\begin{equation}
D_M(x,b) = \sum_{m=-M}^{M} \tilde{q}_T(m) (T_m(x) - T_m(b))^2.
\end{equation}
Setting the probability distribution $\tilde{q}_T(m)$ to be the truncated discrete Gaussian distribution in Eq. \ref{eq:qTm_GCAE}, we effectively construct a Gaussian-type filter. More specifically, set $x=\cos\phi$ and $b=\cos\theta$ with $\phi, \theta \in [0,\pi]$, then we can express the vector distance $D_M(x,b)$ as
\begin{equation}
    D_{M}(x,b) = F_{M}(0) + \frac{1}{2}(F_{M}(2\phi) + F_{M}(2\theta)) - F_M(\phi - \theta) - F_M(\phi + \theta).
\end{equation}
Here, $F_{M}(\phi):= \sum_{m=-M}^M \tilde{q}_T(m) \cos(m\phi)$ is an angular filter. $D_M(x,b)$ is not a single filter peak, but a \emph{filter-induced squared distance}. Here, $F_M(0)$ is the normalization term and is constant. $\frac12F_M(2\phi)$ is the self-norm correction for the candidate $x$. It accounts for the fact that the Chebyshev feature vector $\{T_m(x)\}_m$ does not have constant norm as $x$ changes. $\frac12F_M(2\theta)$ is the self-norm term for the true value $b$. Since $\theta$ is fixed, it is constant with respect to the optimization over $x$. $-F_{M}(\phi-\theta)$ is the main matching term. It is most negative when $\phi=\theta$, because then $F_{M}(\phi-\theta)=F_{M}(0)$ is maximal. $-F_{M}(\phi+\theta)$ is the mirror-symmetry correction. This term is important near the endpoints, where cosine features have reflection symmetries.
To summarize, \(F_{M}(\phi-\theta)\) gives the main filter peak, while the other terms correct for feature norms and cosine mirror symmetry, turning the filter peak into a proper squared distance.

Unlike the GLSAE algorithm~\cite{huang2026low}, the GCAE is able to search the amplitude $b$ in the whole parameter region $b\in[-1,1]$. Below we explain the major difference between them.
In GLSAE, the fitted parameter is the angle $\lambda$ with $\cos(2\lambda)=b$, and the observable has expectation $ \mathbb{E}[Z_m]=\cos(2m\lambda) $.
The population least-squares excess loss has the form
\begin{equation}
  D_G(\vartheta,\lambda)=\sum_m p_m(\cos(2m\vartheta)-\cos(2m\lambda))^2.
\end{equation}
Using the identity $ \cos A-\cos B=-2\sin((A+B)/2)\sin((A-B)/2) $,
we obtain
\begin{equation}
    D_G(\vartheta,\lambda)
=4\sum_m p_m\sin^2(m(\vartheta+\lambda))\sin^2(m(\vartheta-\lambda)).
\end{equation}
When $\lambda$ is close to the endpoint $0$, and $\vartheta$ is also close to $0$, both factors can be small. In particular, for $\lambda=0$ and $\vartheta=\delta$, we have
$ \cos(2m\delta)-1=-2m^2\delta^2+O(m^4\delta^4) $,
and hence
\begin{equation}
    D_G(\delta,0)=4\sum_m p_m m^4\delta^4+O(\sum_m p_m m^6\delta^6).
\end{equation}
Thus the GLSAE distance is quartic in the angular error near the endpoint, rather than quadratic. Consequently, GLSAE does not have a uniform lower bound of the form
$ D_G(\vartheta,\lambda)\gtrsim M^2|\vartheta-\lambda|^2 $
over the full interval. An analogous degeneracy occurs near the other endpoint $\lambda=\pi/2$, due to the reflection and periodic symmetries of the cosine signal.

In GCAE, the fitted parameter is instead the amplitude itself
$ b=\cos(2\lambda)=1-2a\in[-1,1] $,
and the observable is written as
$ \mathbb{E}[Z_m]=T_m(b) $.
The population least-squares excess loss is
\begin{equation}
  D_M(x,b)=\sum_m p_m(T_m(x)-T_m(b))^2.  
\end{equation}
Writing $ x=\cos\phi $ and $ b=\cos\theta $, we have $ T_m(x)=\cos(m\phi) $ and $ T_m(b)=\cos(m\theta) $, so
\begin{equation}
    D_M(x,b)=4\sum_m p_m \sin^2\!\left(m\frac{\phi+\theta}{2}\right) \sin^2\!\left(m\frac{\phi-\theta}{2}\right).
\end{equation}
This expression resembles the GLSAE distance in angular variables, but the key point is that the natural estimation error in GCAE is not $|\phi-\theta|$; it is
\begin{equation}
    |x-b|=2\left|\sin\frac{\phi+\theta}{2}\sin\frac{\phi-\theta}{2}\right|.
\end{equation}
Therefore, \emph{the factor that causes degeneracy in the angular GLSAE distance is exactly part of the physical error metric in the GCAE parameter $b$.} As a result, GCAE admits a uniform metric lower bound of the form
$ D_M(x,b)\ge c_D\min\{1,M^2|x-b|^2\} $
for all $x,b\in[-1,1]$, under appropriate non-degenerate sampling weights $p_m$. This bound remains valid at the endpoints $b=\pm1$. In fact, Chebyshev polynomials are highly sensitive to changes in $b$ near the endpoints, since
$ T_m'(1)=m^2 $ and $ T_m'(-1)=(-1)^{m-1}m^2 $.
Hence the endpoint behaviour is not singular in the $b$-parametrisation.

\begin{figure}[tb!]
\centering
\includegraphics[width=1\linewidth]{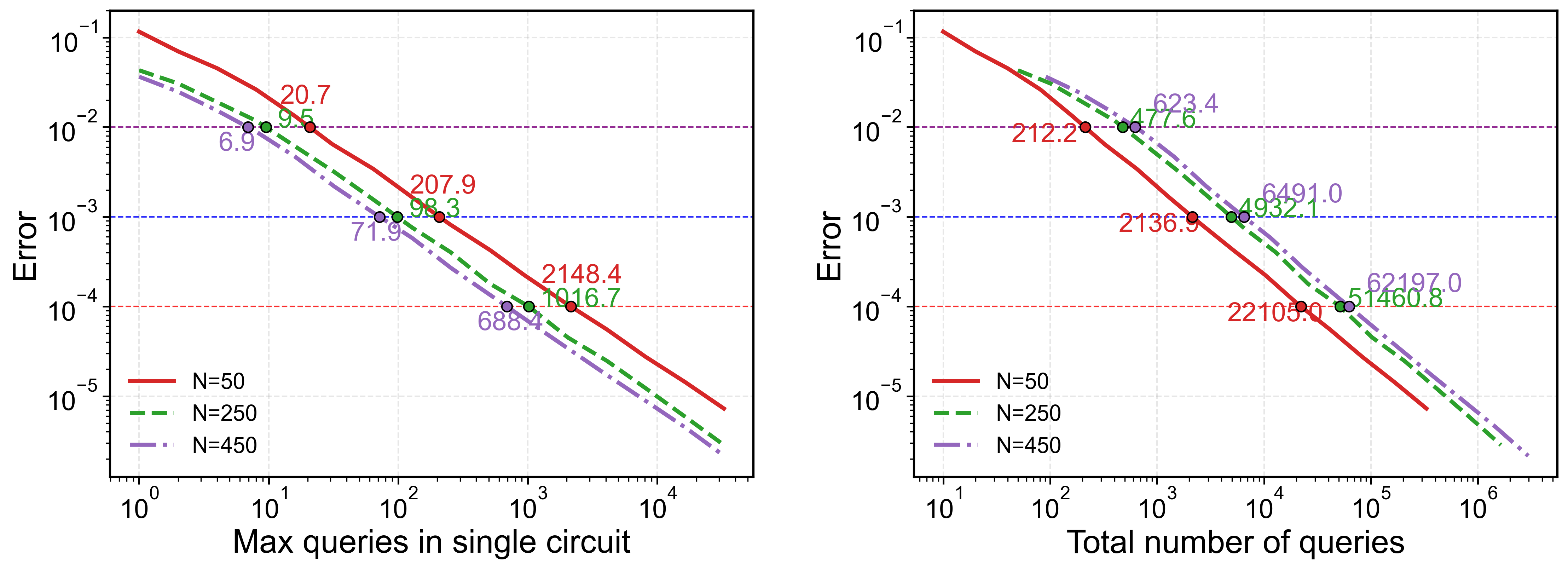}
\caption{\textbf{The dependence of the estimation error on the max queries and the total number of queries in GCAE.} The left panel shows the estimation error versus the max queries in a single circuit, while the right panel shows the error versus the total accumulated query number. Different curves correspond to different numbers of nonzero Gaussian samples, $N=50,250,450$, and the dashed lines mark representative target accuracies with the corresponding query costs. This illustrates the query-sample trade-off of GCAE, where a fixed accuracy can be achieved by balancing max queries and sampling cost.}
\label{fig:depth and queries}
\end{figure}

\begin{figure}[htb!]
\centering
\includegraphics[width=0.5\linewidth]{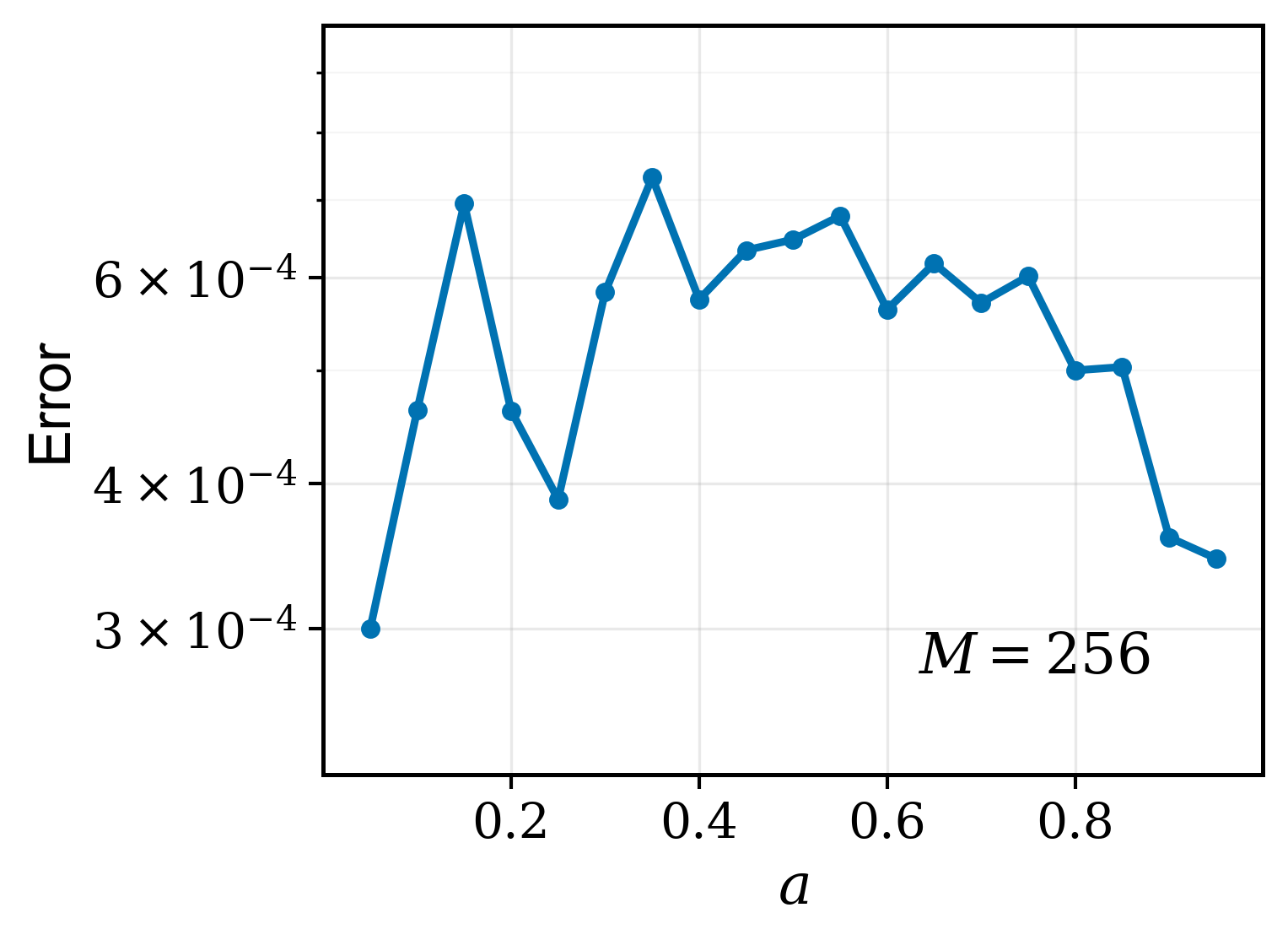}
\caption{\textbf{Error dependence on the target amplitude $a$, which is equivalent to $\frac{1+\langle O\rangle}{2}$.} Here we fix the max queries in single circuit as $M=256$ and apply the GCAE algorithm to different input amplitudes $a\in[0,1]$ under the same algorithmic parameters. The resulting estimation errors remain at a similar magnitude for different choices of $a$. Therefore, GCAE can be regarded as uniformly applicable to arbitrary target amplitudes in the full range $a\in[0,1]$.
}
\label{fig:target}
\end{figure}

Theoretically, we can prove that GCAE achieves the sample-depth trade-off
\[
M\in \widetilde{O}(\varepsilon^{-1+\beta}),
\qquad
N\in \widetilde{O}(\varepsilon^{-2\beta}),
\]
in the whole parameter region $b\in[-1,1]$~\cite{zhang2026GCAE}.
Here, the parameter $\beta\in[0,1]$ can be chosen arbitrarily to interpolate between Heisenberg and standard
shot-noise scaling.

This trade-off is illustrated in Fig.~\ref{fig:depth and queries}, where the same target accuracy can be achieved by balancing the maximum query and the number of samples.
Moreover, Fig.~\ref{fig:target} shows that the GCAE performance is nearly insensitive to the target amplitude $a$, supporting its applicability over the full amplitude range $a\in[0,1]$.

\subsection{Trotter error bounds and the Trotter cost analysis }\label{apd:sec:trotter}
This section contains the details on the improved commutator bounds of Trotter error with entangled states and the numerical results of Trotter errors in different settings.

\subsubsection{Mixed-field Ising model}
% \subsubsection{1D case}
We consider the Trotter error for the
$N$-qubit 1D QIMF Hamiltonian
\begin{equation}
    H = h_x \sum_{j=1}^N X_j + h_z \sum_{j=1}^N Z_j + J \sum_{j=1}^{N-1} X_j X_{j+1} ,
\end{equation}
where $X_j$, $Y_j$, $Z_j$ denote Pauli operators on qubit $j$, and $h_x$, $h_z$, $J$ are coupling constants. We decompose the Hamiltonian as $H = A + B$, where
\begin{align}
    A &= h_x \sum_{j=1}^N X_j + J \sum_{j=1}^{N-1} X_j X_{j+1} \tag{X-type terms} \\
    B &= h_z \sum_{j=1}^N Z_j \tag{Z-type terms}
\end{align}
Note that all terms within $A$ mutually commute (products of $X$ operators), and all terms within $B$
mutually commute. Thus, $\e^{-itA}$ and $\e^{-itB}$ can each be implemented as a product of
single- and two-qubit gates.

% \subsubsection{2D case}
% \emph{2D case}.--
% \subsubsection{2D QIMF Hamiltonian}
The 2D Hamiltonian includes nearest-neighbour interactions along both horizontal and vertical bonds
\begin{equation}
    H = h_x \sum_{i,j} X_{i,j} + h_z \sum_{i,j} Z_{i,j} + J \sum_{i,j} \left( X_{i,j} X_{i+1,j} + X_{i,j} X_{i,j+1} \right),
\end{equation}
where indices are taken modulo $L$ for periodic boundary conditions. Sites are labelled by coordinates $(i,j)$ with $i, j \in \{0, 1, \ldots, L-1\}$, giving $N = L^2$ total
qubits. Alternatively, a single linear index $k = iL + j$ can be used. The same two-summand decomposition applies in 2D
\begin{align}
    A &= h_x \sum_{i,j} X_{i,j} + J \sum_{i,j} \left( X_{i,j} X_{i+1,j} + X_{i,j} X_{i,j+1} \right) \tag{X-type terms} \\
    B &= h_z \sum_{i,j} Z_{i,j} \tag{Z-type terms}
\end{align}
All terms within $A$ mutually commute (products of $X$ operators only), and all terms within $B$
mutually commute. The decomposition structure is identical to 1D.

\subsubsection{Worst-case commutator bound at the operator level }

We first recall the Trotter formula and the worst-case commutator bound of Trotter error \cite{childsFirstQuantumSimulation2018,childs2021theory}.
The worst-case error is defined in terms of the operator norm,
$ 
\|U-\tilde{U} \|\leq \varepsilon,
$
measuring the error distance between the exact time-evolution operator and the approximate one by Trotterisation.
This guarantees an error bound for all possible input states, thus referred to as the worst-case scenario.
% The worst-case scenario means that the error is measured by the operator norm $\|\cdot \|$ looking at error of the operaotr level $\| U - \tilde{U} \| \leq \varepsilon$. In other words, it holds for arbitrary input states.

Provided the Hamiltonian $H=A+B$, the first-order Trotter formula  is 
\begin{equation}
    \mathscr{S}_1(\delta t):=
    \e^{-i\delta t A}\e^{-i\delta t B} \approx \e^{-i \delta t(A+B)}.
\end{equation}
By leveraging the error cancellation,
the second-order Trotter formula takes the symmetric form
\begin{equation}
    \mathscr{S}_2(\delta t):= \e^{-i\frac{\delta t}{2}A}\e^{-i\delta t B}\e^{-i\frac{\delta t}{2}A}.
\end{equation}
The fourth-order product formula is defined recursively
\begin{align}\label{eq:PF4_formula}
    \mathscr{S}_4(\delta t):&= \big[\mathscr{S}_2(u_2 \delta t)\big]^2 \mathscr{S}_2((1-4u_2)\delta t) \big[\mathscr{S}_2(u_2 \delta t)\big]^2 ,
\end{align}
where $u_2 = 1/(4 - 4^{1/3}) \approx 0.41449$.
While the higher-order Trotter formulas yield better scaling,
they incur substantial overhead in one Trotter step.
Taking this trade-off into account, the second-order and fourth-order Trotter formulas have the favourable overall resource requirement \cite{childsFirstQuantumSimulation2018}.

% \dzy{do we need to mention that one can shift the head $\exp(-i\frac{t}{2}A)$ to the end?}
% \jue{Thanks. I think we could say we have merged the same rotations nearby.}
% \subsubsection{Two Summands: \texorpdfstring{$H = A + B$}{H = A + B}}
From Proposition J.1 of Childs et al.~\cite{childs2021theory},
the worst-case (measured by the operator norm $\norm{\cdot}$ corresponding to the worst input state) fourth-order Trotter error has the following commutator scaling
\begin{equation}
\begin{aligned}
    \norm{\mathscr{S}_4(\delta t) - \e^{-i\delta tH}} \leq \delta t^5 \Big(
    &0.0047\norm{[A,[A,[A,[B,A]]]]} + 0.0057\norm{[A,[A,[B,[B,A]]]]} \\
    +&0.0046\norm{[A,[B,[A,[B,A]]]]} + 0.0074\norm{[A,[B,[B,[B,A]]]]} \\
    +&0.0097\norm{[B,[A,[A,[B,A]]]]} + 0.0097\norm{[B,[A,[B,[B,A]]]]} \\
    +&0.0173\norm{[B,[B,[A,[B,A]]]]} + 0.0284\norm{[B,[B,[B,[B,A]]]]} \Big).
\end{aligned}
\end{equation}
In this work, the required Trotter segment for the worst-case scenario is plotted according to the above equation.

\subsubsection{Entanglement-based Trotter error bound to saturate the average case error bound}

Recent work \cite{zhao2025entanglement} showed that if the information of the input state $\ket{\psi}$ is properly considered,
the Trotter error is actually much smaller than the worst-case, close to the average-case (random input states) \cite{zhaoHamiltonianSimulationRandom2021}.
Specifically, for a $p$th-order Trotter formula $\mathscr{U}_p$,
the Trotter error has the following entanglement bound
\begin{equation}\label{apd:eq:entanglement_Trotter_bound}
% \boxed{
    \norm{(\e^{-i\delta t H} - \mathscr{U}_p)\ket{\psi}} = \mathcal{O}\left(\delta t^{p+1}
    \sqrt{\sum_{j,j'} \norm{E_j^\dagger E_{j'}} \sqrt{\log d_{j,j'} - S(\rho_{j,j'})}}\right)
    + \mathcal{O}\left(\delta t^{p+1} \norm{E}_F\right),
% }
\end{equation}
where $E=\sum_j E_j$ is the total leading-order Trotter error with local terms $E_j$, $S(\rho_{j,j'})$ is the entanglement entropy of the subsystem on the support $\mathrm{supp}(E_jE_{j'})$, and $d_{j,j'}$ is the dimension of this subsystem. $\norm{E}_F:=2^{-n/2}\sqrt{\Tr(EE^\dagger)}$ is the normalised Frobenius norm of the Trotter error operator.
Since the normalised Frobenius norm $\norm{E}_{F}$ is also equal to 
$\qty(\sum_{P\in \mathbb{P}_n} \abs{\alpha_P}^2)^{1/2}$,
where $\alpha_P$ is the non-zero coefficients of Paulis in $E$,
this bound can be analytically evaluated regardless of the system size as long as the number of Paulis in $E$ is polynomial.

The bound Eq.~\ref{apd:eq:entanglement_Trotter_bound} means
when the entanglement of the state is large enough for all subsystems supporting the Trotter-error local operators,
the Trotter error approaches the average-case performance
\begin{equation}
\label{apd:eq:entanglement_Trotter_bound_average}
    \norm{(\e^{-i\delta t H} - \mathscr{U}_p)\ket{\psi}} = 
    \mathcal{O}\left(\delta t^{p+1} \norm{E}_F\right).
\end{equation}
For an $N$-qubit lattice Hamiltonian, 
$\norm{(\e^{-i\delta t H} - \mathscr{U}_p)\ket{\psi}} = \mathcal{O}\left(\delta t^{p+1} \sqrt{N}\right)$,
a quadratic improvement of the worst-case $\mathcal{O}\left(\delta t^{p+1} N\right)$.

% \sun{@Jue, \sout{(1) give the Trotter steps, not the Trotter error.
% (2)
% Clarify what is average case (3) why this worst-case bound saturates the average-case bound. Expand the definition and implication. All presented in a narrative way.
% } Revising}

The above bound is for the one-step Trotter error (short time $\delta t$).
For the $r$-step Trotter circuit, the total accumulated Trotter error is $r \cdot \mathcal{O}(\delta t^{p+1} \norm{E}_F)$ for the average-case (or the entangled states).
Specifically, for an $N$-qubit lattice Hamiltonian and entangled input states, the number of the fourth-order Trotter steps for total simulation precision $\epsilon$ is 
\begin{equation}
    \mathcal{O}(t^{1+1/4}(\sqrt{N}/\epsilon)^{1/4}), 
\end{equation}
which exhibits $\mathcal{O}(N^{1/8})$ improvement over the worst-case analysis.

\subsubsection{Trotter error of the expectation value of observables}

Highly entangled states have been observed to suppress errors asymptotically, achieving average-case scaling comparable to quantum simulations with random inputs \cite{zhaoHamiltonianSimulationRandom2021,chen2024average}. This achieves error suppression
compared to the worst-case analysis, and hints at a deeper connection between quantum entanglement and Trotter error \cite{zhao2025entanglement}. 
However, the generic bounds on the entangled wave function may still significantly overestimate the actual error on local observables, such as magnetisations or low-order correlation functions \cite{heyl2019quantum}. 

\sun{The key idea here is that the Trotter error can be reduced when targeting the observable expectation values.}
Heyl et al. demonstrated \cite{heyl2019quantum} that quantum localisation can constrain the Trotter error in quantum simulations, suggesting that leveraging knowledge about observables may further reduce the Trotter error. Recent investigations have explored the effects of operator growth and associated light cone phenomena \cite{childs2021theory,yuObservabledrivenSpeedupsQuantum2025}, as well as the average observable error across quantum states sampled from the Haar measure or 1-design ensembles \cite{yuObservabledrivenSpeedupsQuantum2025,zhaoHamiltonianSimulationRandom2021,LiPhysRevA.110.062614}. It is suggested that the error can be reduced from the operator norm of the observable to the normalised Frobenius norm for the average case \cite{yuObservabledrivenSpeedupsQuantum2025} or entangled case \cite{feng2025trotterizationoperatorscramblingentanglement}. 
Most recently, it has been demonstrated that the Trotter error in observable growth is fundamentally bounded by the cumulative operator scrambling, which provides a tighter upper bound of Trotter error for general observable dynamics \cite{feng2025trotterizationoperatorscramblingentanglement}.

Specifically,
      consider an ideal short-time unitary $U_0=e^{-iH  \delta t }$ and a Trotterised unitary $\mathscr{U}_p=U_0(I+\mathscr{M})$ \cite{childs2021theory}, for the Hamiltonian $H=\sum_{l=1}^L H_l$ and an input pure state $\ket{\psi}$.
      Here, one can define $\mathscr{M}$ as the multiplicative Trotter error.
      The additive Trotter error of an observable $O$ can be bound by operator scrambling as
        \begin{equation}
        \label{main:eq:scrambling_bound}
          \epsilon_O=\abs{\langle \psi| O(t)- {\mathscr{U}_p^\dagger}  O {\mathscr{U}_p}\ket{\psi}}\le \sqrt{\bra{\psi |[O( \delta t),\mathscr{M}]|^2 \ket{\psi}}}, 
        \end{equation}
        where $O(\delta t)=e^{iH \delta t}Oe^{-iH \delta t}$. Note that the Lieb-Robinson bound is the upper bound of the worst-case scenario for the square root of operator scrambling.

    Let $\mathscr{M}=M \delta t^{p+1}+\mathcal{O}(\delta t^{p+2})$, where $M=\sum_j M_j $ is the leading (multiplicative) Trotter error term. By utilising the Cauchy-Schwarz inequality, the observable error of $O$ can be further bound as \cite{feng2025trotterizationoperatorscramblingentanglement}
    \begin{equation}
    \begin{split}
         \epsilon_O 
         \le  \Bigg[\sqrt{\norm{O}^2_F+\Delta_{O}(\ket{\psi(\delta t)})} \sqrt{\norm{M}^2_F+\Delta_{M}( \ket{\psi_{O(\delta t)}})} +\sqrt{\norm{O}^2_F+\Delta_{O}(\ket{\psi_{U_0M}})} \sqrt{\norm{M}^2_F+\Delta_{M}(\ket{\psi})}\bigg] \delta t^{p+1} +\mathcal{O} (\delta t^{p+2}),
             \end{split}
    \end{equation}
    where  
    $\|A\|^2_F:= \Tr(A^{\dagger}A) /d$ is the square of the normalised Frobenius norm and
    $\Delta_A(\ket{\chi})= \sum_{j,j'} \|A_j^{\dag} A_{j'}\| ~\sqrt{2\log(d_{\text{support}(A_j^{\dag}A_{j'})})-2S(\rho_{j,j'})}.$

Consider a $k$-local operator
 $O=\sum_j O_j$. If the entanglement of state $\ket{\psi}$ is sufficiently large such that both $O(\delta t)\ket{\psi}$ and $M \ket{\psi}$ also generally exhibit substantial entanglement for system, so that all  $\Delta_{O}(\ket{\psi(\delta t)})\approx 0$, $\Delta_{M}( \ket{\psi_{O(\delta t)}})\approx 0$,  $\Delta_{O}(\ket{\psi_{U_0M}})\approx 0$, and $\Delta_{M}(\ket{\psi})\approx 0$, one obtain the following bound
  \begin{equation}
      \epsilon_O \lesssim 
      2 \norm{O}_F \norm{M}_F \delta t^{p+1}+\mathcal{O}(\delta t^{p+2}).
  \end{equation}
Compared to the worst-case result, $\norm{U^\dagger_0 O U_0-\mathscr{U}^\dagger_pO\mathscr{U}_p}\le 2\norm{O}\norm{M} \delta t^{p+1}+ \mathcal{O}(\delta t^{p+2})$, one can enjoy error reduction using the normalised Frobenius norm $\norm{O}_F \norm{M}_F $, which aligns with the behavior observed in the ($1$-design) random case \cite{LiPhysRevA.110.062614,yuObservabledrivenSpeedupsQuantum2025}. 

In this work, we aim to estimate the computational gap between quantum simulations and classical simulations of a single Pauli evolution, where $\norm{O}_F = \norm{O}=1$ for any Pauli observables. Thus, in this case, when entanglement is significant in the partitions accordingly, the Trotter error for observables can reduce to the state-dependent bound Eq. \ref{apd:eq:entanglement_Trotter_bound_average} since $\norm{M}_F=\norm{E}_F$.
For a general operator $O=\sum_{i=1}^K P_i$ ($P_i$ is Pauli operator), we may have $\norm{O}=\mathcal{O} (K)$ and $\norm{O}_F=\mathcal{O} (\sqrt{K})$. 
Consequently, different from the previous result \cite{zhao2025entanglement}, this bound suggests that entanglement allows for an additional quadratic speed-up in quantum simulation with respect to the observable. 
That is, simulating observables decomposed into a linear combination of Pauli operators in the QIMF model may save more quantum resources in Trotterisation generally. Note that this mechanism stems from the suppression of Trotter error of operator evolution by entanglement, and is independent of the estimation complexity of observables and the statistical errors of learning \cite{Huggins_2022,Koizumi26Faster}.

\subsection{Trotter steps }

\begin{figure}[htbp]        
\includegraphics[width=1\linewidth]{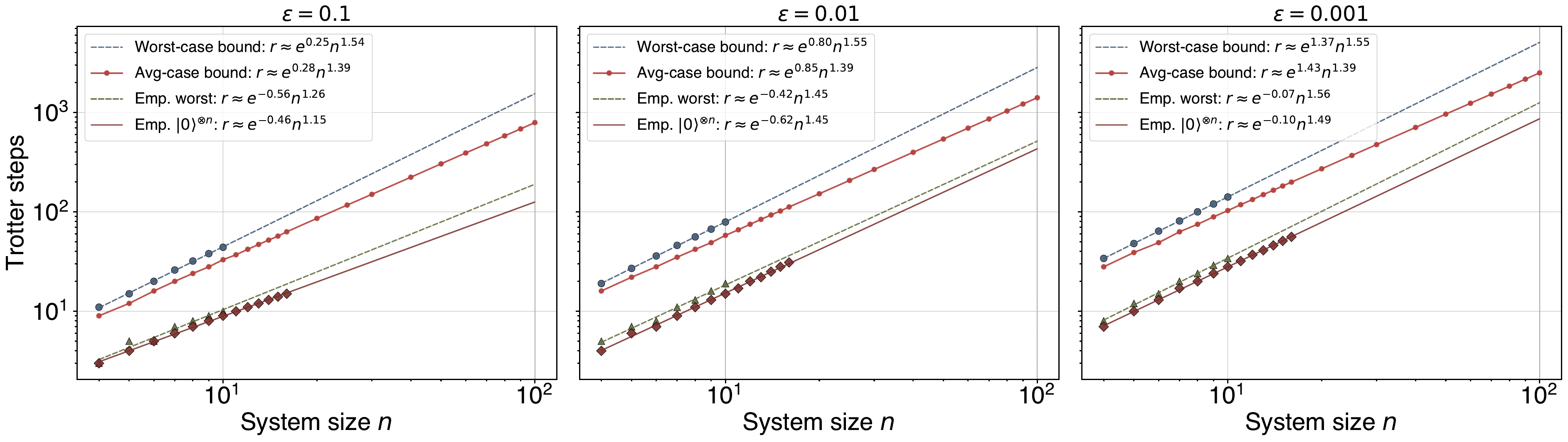}
	\caption{ 
    \textbf{The number of fourth-order Trotter steps for the 1D mixed-field Ising (MFI) Hamiltonian for simulation time $T=n/2$ and three different precisions (i.e., $\epsilon=0.1,0.01,0.001$).} 
    The worst-case (over all initial states) bound is from the commutator bound of Childs et al. \cite{childs2021theory}.
    The average-case (random input) bound is from Zhao et al. \cite{zhaoHamiltonianSimulationRandom2021} and also applicable to the entangled state \cite{zhao2025entanglement}.
    This average-case in the Frobenius norm can be evaluated analytically for large systems.
    The empirical worst line is measured by the operator norm of difference between the ideal evolution operator $e^{-i Ht}$ and the fourth-order Trotter evolution $\mathscr{S}_4$.
    The last empirical $\ket{0}^{\otimes n}$ line is the $l_2$ vector norm of difference between evolved states by the ideal evolution operator $e^{-i Ht}$ and the fourth-order Trotter evolution.
    The scalings are fitted and extrapolated by the power law $r \sim e^{c} n^{\alpha}$.
% \sun{
% @Jue Xu: Please define what we mean by 'worst', 'average', 'empirical matrix ' and state'.
% In our work, two quantities are of relavance: average case scenario and 'empirical state', as they correspond to the Trotter error bound computed by operator norm and that considering the specific input state, respectively.
% }
}
	\label{fig:Trotter_segment_1d_t2}
\end{figure} 

\begin{figure}[htbp]        
    \includegraphics[width=1\linewidth]{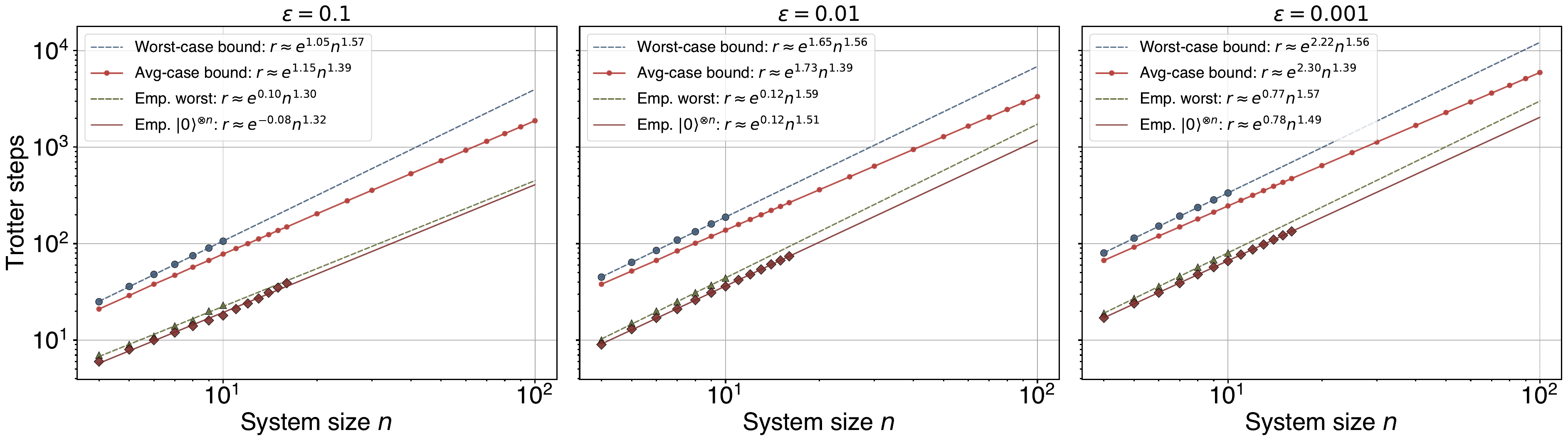}
	\caption{ 
    The same Trotter steps estimation for 1D MFI as in Fig. \ref{fig:Trotter_segment_1d_t2}, but for longer evolution time $T=n$.
}
	\label{fig:Trotter_segment_1d}
\end{figure} 

\begin{figure}[htbp]        
\includegraphics[width=1\linewidth]{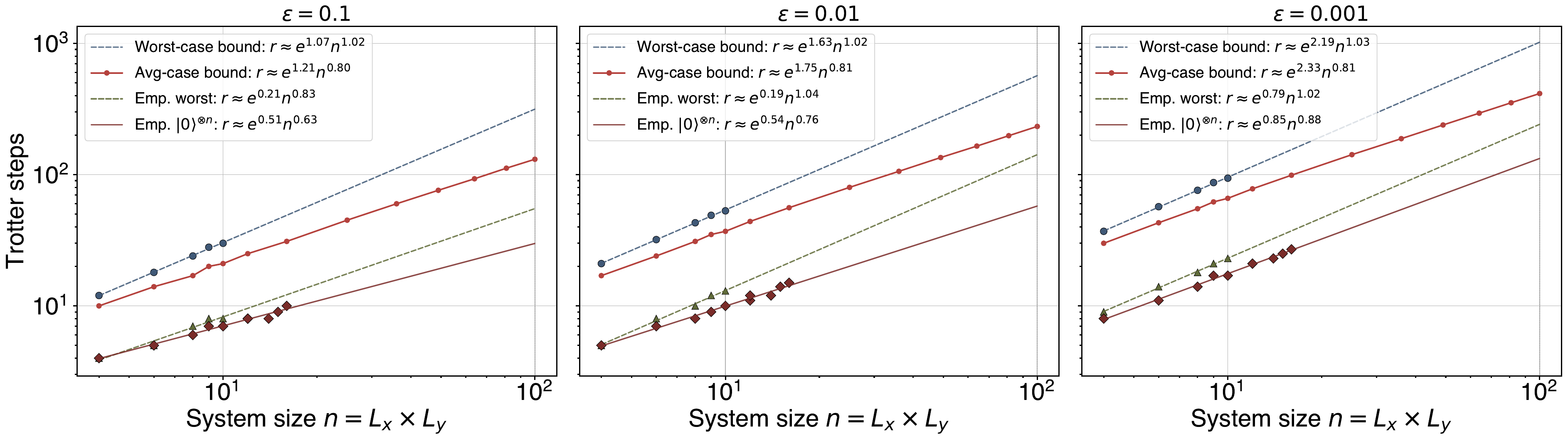}
	\caption{ 
    % Scaling of Trotter segment number with system size for the 2D mixed-field Ising model. 
    \textbf{The Trotter steps estimation for 2D MFI with $n=L_x\times L_y$ qubits for evolution time $T=\sqrt{n}$.}
    The theoretical bounds, empirical results, and the extrapolation conventions are the same as in Fig. \ref{fig:Trotter_segment_1d_t2}.
    % \sun{Empirical scaling is computed by using the state vector method, which chooses sufficiently large Trotter steps. The accuracy is verified by using the exact matrix-based method.  drop the emp matrix line.} 
% Top panels: Number of Trotter steps $r$ required to achieve fixed precision ($10^{-1}$, $10^{-2}$, and $10^{-3}$) as a function of system size $n=L\times L$, obtained using a fourth-order product formula (PF4) with different error estimates (average-case, spectral, and local bounds). The data are well described by power-law fits $r \sim e^{c} n^{\alpha}$, with exponents $\alpha \approx 1.5$--$1.7$ depending weakly on the error model and target precision. 
% Bottom panels: Comparison of Trotter step scaling under different error metrics and product formulas. Left: operator norm error $\|U-\tilde{U}\|_{\infty}$; middle: state error $\|(U-\tilde{U})|\psi\rangle\|_2$; right: observable expectation error $|\langle \psi|U^\dagger O U|\psi\rangle - \langle \psi|\tilde{U}^\dagger O \tilde{U}|\psi\rangle|$. Results are shown for first-order (PF1), second-order (PF2), and fourth-order (PF4) formulas, with both bound-based and empirical estimates. Higher-order formulas significantly reduce the required number of Trotter steps and exhibit improved scaling with system size, approaching the average-case behaviour relevant for highly entangled dynamics.
}
	\label{fig:Trotter_segment_2d}
\end{figure} 

% \dzy{add 1D, empirical, average, worst case results.}
% \jue{Done.}
For a lattice, we consider $T = \mathcal{O}(L)$, which represents the time scale for information propagation along one direction. Specifically, we consider both $T = n$ in Fig.~\ref{fig:Trotter_segment_1d} and $T = n/2$ for 1D lattices in Fig.~\ref{fig:Trotter_segment_1d_t2}.
For the 2D lattice, we consider  $T = \sqrt{n}$ in Fig. \ref{fig:Trotter_segment_2d}. With the number of Trotter steps, a lower-level resource is the number of Pauli rotation gates, which is the dominant cost of the Trotter circuit.
For the 1D MFI with $n$ qubits, one fourth-order Trotter step needs about $15n$ rotation gates,
while the 2D $n$-qubit MFI needs more rotation gates in a Trotter step, about $20n$ rotations.

% In this way, we have the empirical \texttt{T} count estimation (assume the initial state is $\ket{0}^{\otimes n}$) from the Trotter steps estimation for the 2D case in Fig. \ref{fig:Trotter_segment_2d}.

% \begin{equation}
%     T = 1200 n \times e^{0.51}  n^{0.63}~for~ \epsilon = 0.1.
% \end{equation}

% \begin{equation}
%     T = 1200 n \times e^{0.54}  n^{0.76}~for~ \epsilon = 0.01.
% \end{equation}

% \begin{equation}
%     T = 1200 n \times e^{0.85}  n^{0.88}~for~ \epsilon = 0.001.
% \end{equation}

\subsection{ Optimal query budget for observable estimation }
The overhead due to observable estimation is the following.
There are two scenarios.
First, suppose that the main objective is to minimise the total number of queries required by the observable estimation algorithm. However, this will increase the max queries in a single circuit. As a result, when error mitigation is taken into account, the larger max queries introduces a significantly increased sampling overhead, which can substantially dominate the overall computational cost. 

% Then, we only need to look at the 
% Then, extreme low depth in a single run.

% Objective 1: minimise total query:

% $$\epsilon = 0.1~~Max~query = 1,~ Total~queries =20$$

% $$\epsilon = 0.01~~Max~query = 20,~ Total~queries = 230$$

% $$\epsilon = 0.001~~Max~query =200,~ Total~queries = 2300$$

% Objective 2: minimise total running time (mitigation cost)
Second, instead of optimising purely from the perspective of query complexity, we directly incorporate error mitigation into the evaluation of the overall execution cost. In this setting, for a given target accuracy, we consider the total running time, which includes both the circuit execution cost and the additional overhead introduced by error mitigation. By scanning over different choices of max queries and comparing their corresponding total costs under the same accuracy requirement, we select the circuit configuration that minimises the overall mitigation-adjusted runtime.
% $$\epsilon = 0.1~~Max~query = 1,~ Sampling~cost = 40$$

% $$\epsilon = 0.01~~Max~query = 6,~ Sampling~cost = 2000$$

\sun{
Case I: the objective is to minimise the total query.}
The corresponding coherent-estimation parameters are

\begin{equation}
(M_{\rm coh},Q_{\rm coh})
=
\begin{cases}
(1,\;20 ), & p_{\rm alg}=10^{-1},\\
(20,\;230), & p_{\rm alg}=10^{-2},\\
(200,\;2270), & p_{\rm alg}=10^{-3}.
\end{cases}
\end{equation}
This happens when $p$ is small, for example, when $p = 10^{-4}$.

Case II: the objective is to minimise the total running time (in which logical error-mitigation cost is included).  This usually occurs at a higher physical error rate, $p = 10^{-3}$.
\begin{equation}
(M_{\rm coh},Q_{\rm coh})
=
\begin{cases}
 p_{\rm alg}=10^{-1}, & (1,\;36 )~(1D)\\
 p_{\rm alg}=10^{-1}, & (1,\;23 )~(2D)\\
p_{\rm alg}=10^{-2}, &  (2,\;2094)~(1D)\\
p_{\rm alg}=10^{-2}, &  (16,\;759)~(2D).
\end{cases}
\end{equation}

\section{Fault-tolerant implementation and resource analysis for Clifford+$\varphi$ approaches } \label{sec:small-angle-rotation}

In this section, we present the resource analysis for the small-angle rotation implementation used in the main text. 
Our goal is to estimate the number of surface-code cycles, the accumulated logical error, and the active space-time volume associated with direct rotation-state injection for a single call to the time-evolution $U(t)$. 
% In the Supplementary Information, we denote the physical error rate to be $p_{\rm phys}$.
We define $n$ as the number of logical data qubits, $p$ as the circuit-level physical error rate, and $d$ as the surface-code distance. Because one QEC clock corresponds to $d$ rounds of syndrome extraction (SE), the runtime of $T$ clocks equates to $dT$ surface-code cycles. 
The additional query overhead from the observable-estimation procedure is accounted for separately in Sec.~\ref{sec:algorithmic_level}.

\subsection{Logical rotation primitive and parallel RUS layers}
Trotter decomposition yields a circuit composed of layers of local Pauli rotations, $R_P(\theta)=\exp(-i\theta P)$, where $P$ is a local Pauli operator. 
For each such operator, we choose a constant-depth Clifford circuit $C_P$ that maps $P$ to a single logical $Z$ operator, so that
\begin{equation}
    R_P(\theta)
    =
    C_P R_Z(\theta) C_P^\dagger ,
    \qquad
    R_Z(\theta)=\exp(-i\theta Z_L).   
\end{equation}
The Clifford conjugations are implemented using constant-QEC clocks via lattice surgery. The non-Clifford part of the implementation is therefore reduced to the direct implementation of logical $Z$-rotations. A logical $Z$-rotation is implemented by injecting the logical rotation state
\begin{equation}
    \ket{\theta}_L \coloneqq R_Z(\theta)\ket{+}_L .
\end{equation}
The injection gadget applies either $R_Z(\theta)$ or $R_Z(-\theta)$ to the data qubit, depending on the measurement outcome. We call the former outcome a success. In the latter case, the data receives the inverse rotation and can be corrected by injecting a doubled-angle state $\ket{2\theta}_L$. If this correction also fails, the procedure continues with $\ket{4\theta}_L$, and so on. Thus, a single target rotation is implemented by a repeat-until-success (RUS) sequence of rotation states $\ket{\theta}_L, \ket{2\theta}_L, \ket{4\theta}_L, \cdots$. At each RUS round, the success probability is $1/2$. A schematic illustration of the RUS procedure is shown in Fig.~\ref{fig:RUS_preparation}(a).

% \begin{figure}[!htbp]
% 	\includegraphics[width=0.7\linewidth]{RUS.png}
% 	\caption{Protocol for implementing small-angle $Z$ rotations in parallel on $n$ qubits. (a) Basic state-injection gadget using the ancillary rotation state $\ket{\theta}$. (b) Repeat-until-success procedure: if the measurement outcome indicates failure, a rotation state with doubled angle is injected and the protocol is repeated until success at round $k+1$. (c) Parallel implementation of small-angle rotation on $n$ qubits, for which the maximum number of attempts scales as $k_{\max}\approx \log n$.}
% 	\label{fig:injection}
% \end{figure}
 
The relevant time cost in a Trotter segment is the total QEC clocks required to complete an entire parallel layer of $n$ independent rotations. 
The overall layer is considered complete only when all $n$ sites have successfully finished their operations.
Because each individual RUS attempt succeeds with a probability of $1/2$, successful sites halt their participation after each round, while the remaining sites advance to the next doubled-angle correction round.
This effectively halves the number of unfinished sites at each step, meaning the expected maximum number of rounds $\langle k_{\max} \rangle$ required for all $n$ sites to finish grows only logarithmically. Following the numerical analysis in Ref.~\cite{akahoshi2025compilation}, we use the fit
\begin{equation}
    \langle k_{\max}\rangle = 1.4\ln n + 1.54.
\end{equation}
Let $t_i$ denote the number of QEC clocks required to prepare a logical rotation state. Adding one QEC clock for the injection step itself [Fig.~\ref{fig:RUS_preparation}(a)], the total QEC-clock cost for a parallel logical $Z$-rotation layer of width $n$ is
\begin{equation}\label{eq:TZ-layer-cost}
    T_Z(n) = \langle k_{\max}\rangle (t_i+1).
\end{equation}
This expression serves as the basic timing primitive throughout our small-angle-rotation resource estimate.

\subsection{Rotation state preparation}\label{subsec:rotation_state_preparation}
The timing primitive in Eq.~\ref{eq:TZ-layer-cost} requires the state-supply time $t_i$, defined as the number of QEC clocks needed to make an accepted logical rotation state available for each active RUS site. 
In this subsection, we describe the rotation-state preparation protocol and explain how both $t_i$ and the logical error of the accepted state are estimated.

\begin{figure}[!htbp]
	\includegraphics[width=0.9\linewidth]{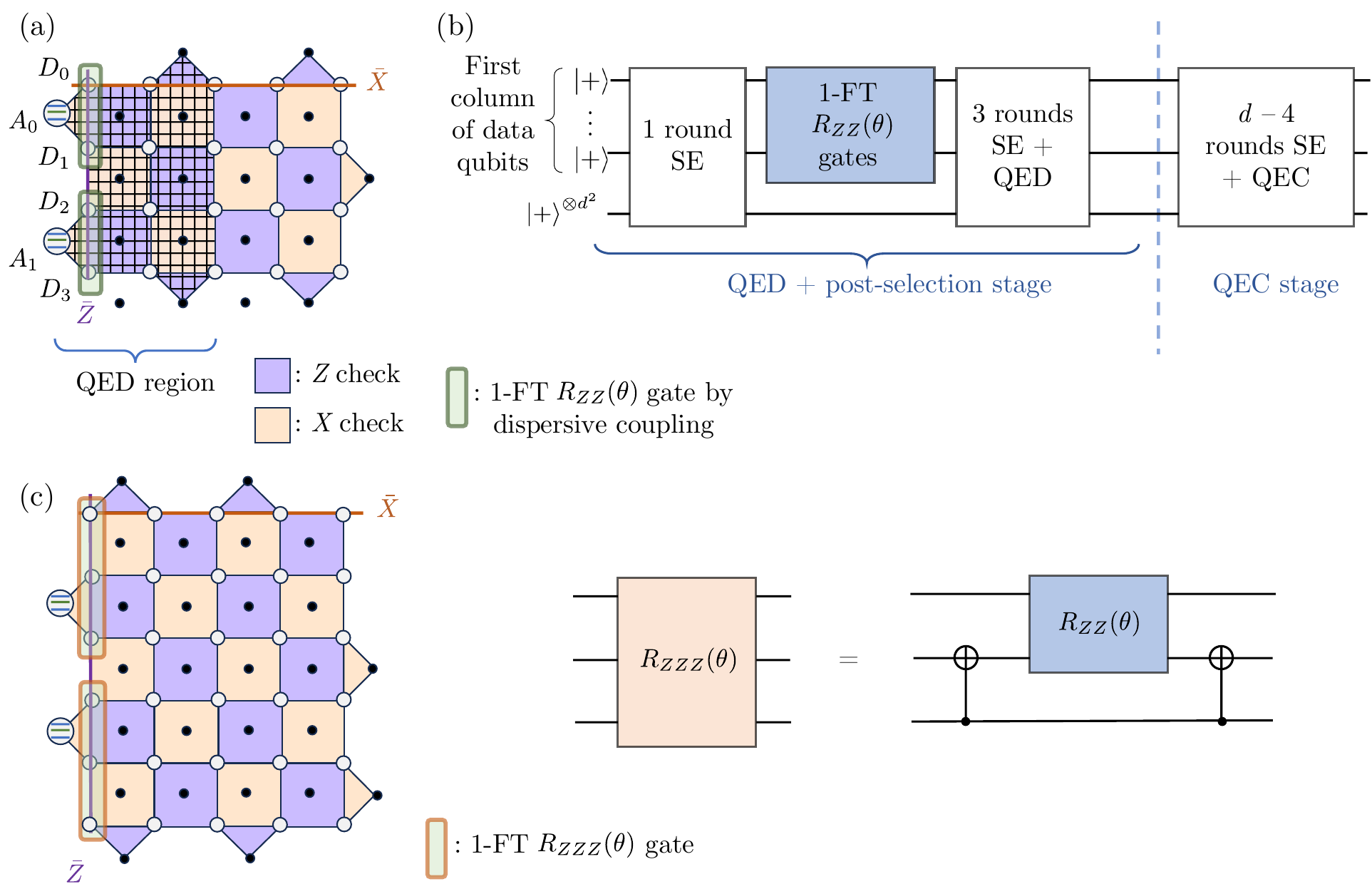}
	\caption{\textbf{Schematic illustration of logical rotation-state preparation, reproduced from Ref.~\cite{zeng2025error}.} (a) Preparation of $\ket{\theta}_L$ on a $4\times 5$ rotated surface code using $1$-FT $R_{ZZ}(\theta)$ gates, where the $1$-FT $R_{ZZ}(\theta)$ gate is realised by the dispersive-coupling method of Ref.~\cite{zeng2025error}. The shaded region marks the part of the surface code on which QED is performed. (b) Full preparation circuit. The first four rounds of syndrome extraction, together with one layer of $1$-FT $R_{ZZ}(\theta)$ gates, implement QED and post-selection, while the remaining $d-4$ rounds perform QEC. (c) Preparation of the logical rotation state based on $1$-FT $R_{ZZZ}(\theta)$ rotations on a $6\times 5$ rotated surface code. Here, the $1$-FT $R_{ZZZ}(\theta)$ gate is constructed from a $1$-FT $R_{ZZ}(\theta)$ gate sandwiched between two CNOT gates.}
	\label{fig:preparation}
\end{figure}

We prepare the logical rotation state $\ket{\theta}_L$ on a rotated surface-code patch of shape $d \times (d+1)$ with distance $d=3m$, following the partially fault-tolerant preparation scheme of Ref.~\cite{zeng2025error}. 
The basic idea is to apply tailored physical rotations along the support of the logical $\bar{Z}$ operator, as illustrated in Fig.~\ref{fig:preparation}.
We use the $R_{ZZZ}(\theta)$-based construction, where the 1-FT $R_{ZZZ}(\theta)$ primitive is built from a 1-FT $R_{ZZ}(\theta)$ rotation together with Clifford entangling gates [Fig.~\ref{fig:preparation}(c)]. 
Here, ``$1$-FT'' means that the gate suppresses first-order physical errors, such that the leading error contribution appears only at order $p^2$, where $p$ denotes the circuit-level noise strength.
The full preparation consists of $d$ rounds of SE: the first four SE rounds, together with the 1-FT gates, implement quantum error detection (QED) and post-selection. 
Conditional on successful post-selection, the accepted state undergoes the remaining $d-4$ SE rounds, which serve as QEC before the state is supplied to the injection gadget. 
We refer the reader to Ref.~\cite{zeng2025error} for further technical details regarding the physical realisation.

We set the surface-code distance to $d=12$ for a physical error rate of $p=10^{-4}$, and $d=18$ for $p=10^{-3}$. 
This choice ensures that the higher-order logical errors accumulated during the QEC rounds of rotation-state preparation remain negligible. 
Specifically, the logical failure probability per logical qubit per QEC clock is well approximated by the empirical formula $q = 0.1(100p)^{\lfloor(d+1)/2\rfloor}$~\cite{Fowler2019LowOverheadLatticeSurgery,Litinski2019gameofsurfacecodes}. 
The higher-order error accumulated during the preparation of these rotation states over the entire time evolution $U(t)$ is thus upper-bounded by $n q \widetilde{T}_{\rm clock}$, where $\widetilde{T}_{\rm clock} = T_{\rm cycle} / (d \langle k_{\max}\rangle)$ is the effective number of QEC clocks (see Fig.~\ref{fig:RotationCycle} for $T_{\rm cycle}$ estimates). 
Here, the division by $\langle k_{\max}\rangle$ reflects that only the successful preparation attempts can introduce logical errors into the main computation, while failed attempts are simply discarded. 
Under our parameter choices, this higher-order error is suppressed well below the algorithmic precision target ($\varepsilon = 0.01$). 
Consequently, we can safely neglect these higher-order errors in rotation-state preparations.

As a side remark, we note that the QEC memory error of the main computation patches does not benefit from this $1/\langle k_{\max} \rangle$ reduction. However, the surface-code distance of the main computation can be increased independently of that of the state preparation. For instance, incrementing the distance from $d=18$ to $d=19$ provides an additional error suppression factor of $0.1$, which comfortably offsets the absence of the $\langle k_{\max} \rangle$ divisor. This modification would only marginally increase the total QEC cycles by a factor of $19/18$. For simplicity, we therefore assume the same surface-code distance for both the state preparation and the main computation in all estimations below.

\begin{figure}[!htbp]
	\includegraphics[width=0.9\linewidth]{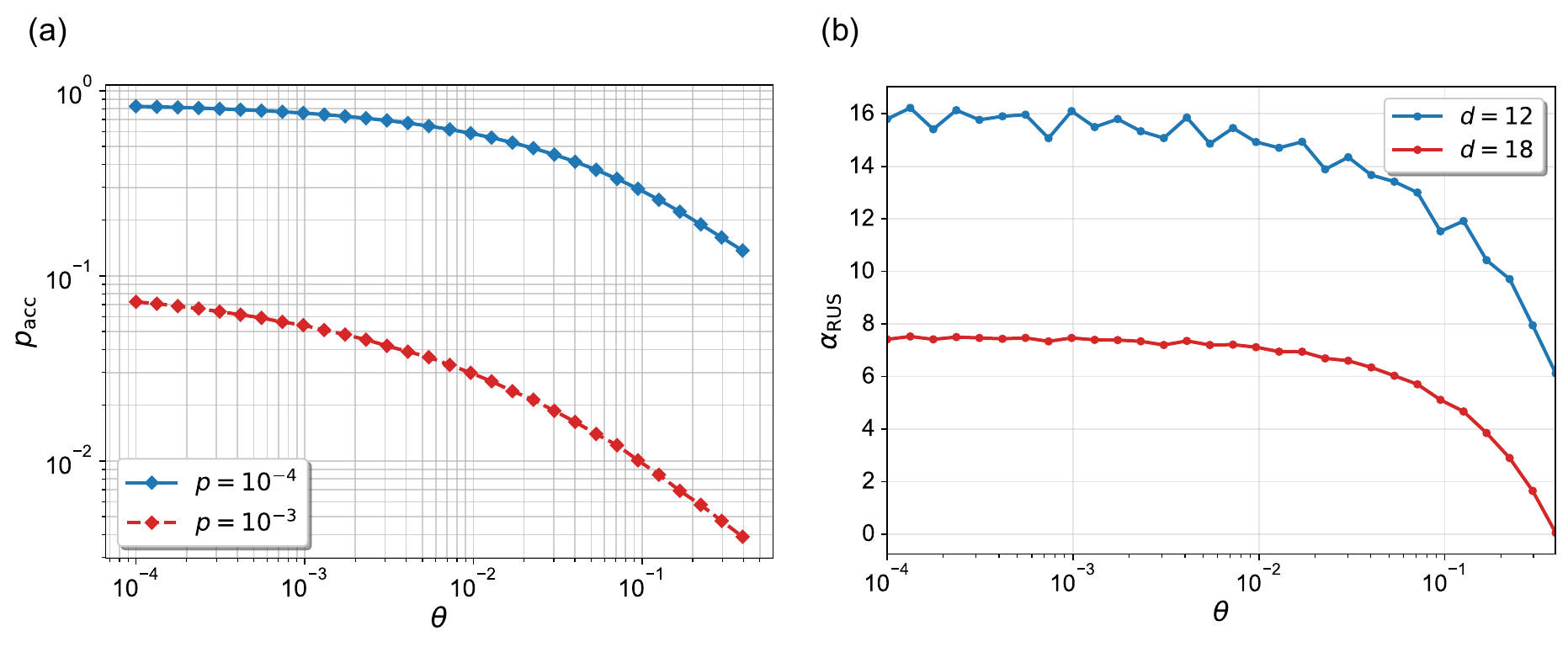}
    \caption{\textbf{Acceptance probability and logical error for the $R_{ZZZ}(\theta)$-based logical rotation-state preparation.} 
    (a) Acceptance probability $p_{\rm acc}(\theta)$ of a single preparation attempt as a function of the target rotation angle $\theta$. 
    (b) The error coefficient $\alpha_{\mathrm{RUS}}$ as a function of $\theta$. When combined with the RUS injection gadget, the logical error of the injected rotation scales quadratically as $\alpha_{\mathrm{RUS}} |\theta| p^2$, where $p$ is the physical error rate.  }
	\label{fig:rotation_performance}
\end{figure}

The acceptance probability of a single preparation attempt, $p_{\rm acc}(\theta)$, is given in Fig.~\ref{fig:rotation_performance}(a). 
Crucially, a rejected attempt neither causes the main computation to fail nor contributes to the accumulated logical error. It merely requires restarting the ancilla preparation while the data qubits wait. For $p=10^{-4}$, the acceptance probability is sufficiently high that one preparation clock is typically enough. 
This efficiency is supported by an adaptive scheduling strategy, where regions that have already completed their rotations can be dynamically reallocated to prepare rotation states for the remaining unfinished sites~\cite{akahoshi2025compilation}.
In terms of spatial overhead, each logical data qubit is paired with an ancillary code patch to facilitate rotation-state injection or lattice-surgery routing. Consequently, the architecture requires a total of $2n$ logical code patches, corresponding to $2nd(d+1)$ physical qubits. For $p=10^{-3}$, however, $p_{\rm acc}(\theta)$ is approximately one order of magnitude lower [Fig.~\ref{fig:rotation_performance}(a)]. To maintain the supply rate and prevent a time bottleneck, we trade space for time by preparing $q$ candidate rotation states in parallel for a single active RUS site. The probability that at least one accepted state is available after one preparation clock is
\begin{equation}
    p_{\rm acc}^{(q)}(\theta) =  1-\left[1-p_{\rm acc}(\theta)\right]^q.
\end{equation}
By provisioning roughly 10 times more ancilla qubits ($q \approx 10$), we restore a high acceptance probability. 
In terms of spatial overhead, assigning 10 ancillary patches to each of the $n$ logical data qubits yields a total of $11n$ logical code patches, which corresponds to $11nd(d+1)$ physical qubits. 

Once a candidate succeeds, the resulting logical rotation state must be moved to the target data qubit and used for injection. 
Because patch movement in the surface code can be implemented via lattice surgery with a clock cost largely independent of the movement distance, this spatial parallelisation has a time overhead comparable to the adaptive scheme used at lower error rates. Therefore, we set the state-supply time to
\begin{equation}
t_i =
\begin{cases}
1, & p=10^{-4},\\
2, & p=10^{-3}.
\end{cases}
\label{eq:ti-choice}
\end{equation}
We emphasise that our estimates for both $t_i$ and the associated spatial overhead remain conservative. A more detailed optimisation incorporating adaptive scheduling strategies~\cite{akahoshi2025compilation} could further suppress these requirements, suggesting that actual space overhead at $p=10^{-3}$ may not be significantly larger than at $p=10^{-4}$.

Finally, we characterise the logical error of the accepted rotation state.
Because the preparation scheme is only partially fault-tolerant, its logical error rate is not exponentially suppressed as $p^{\Omega(d)}$, but rather reduced to a lower order in $p$. 
Nonetheless, when combined with the RUS injection gadget, the leading logical error of the injected rotation scales quadratically as $\alpha_{\mathrm{RUS}} |\theta| p^2$~\cite{zeng2025error}, where $\alpha_{\mathrm{RUS}}$ is a small constant that depends only weakly on $\theta$, as shown in Fig.~\ref{fig:rotation_performance}(b).
To maintain this favourable error scaling during the RUS procedure, where the intermediate rotation angle doubles upon each failure, we must prevent the angle from growing too large.
We achieve this by mapping large angles back into a bounded interval, a procedure we refer to as wrapping. 
The specific wrapping strategy depends on the physical error rate:
\begin{itemize}
\item For $p=10^{-4}$, we apply $S$-gate wrapping to constrain all intermediate angles within the interval $[-\pi/8, \pi/8]$.
\item For $p=10^{-3}$, we apply $\texttt{T}$-gate wrapping to strictly bound the angles within $[-\pi/16, \pi/16]$.
\end{itemize}
While $S$-gate wrapping consumes at most two QEC clocks~\cite{Hirai2026SpaceVolumeSGate}, $\texttt{T}$-gate wrapping relies on magic-state preparation. Fortunately, wrapping is only triggered in later RUS rounds. By that time, the majority of sites have already successfully completed their rotations, leaving an abundant ancillary space available. Because the overhead for $\texttt{T}$-state preparation via cultivation~\cite{Gidney2024Cultivation} is comparable to the logical rotation-state preparation discussed above, we expect that dynamically reallocating this freed space will yield similar time overheads. Assuming a preparation time analogous to our state-supply time ($t_i=2$ at $p=10^{-3}$), we estimate that a complete $\texttt{T}$-gate wrapping step requires at most $4$ QEC clocks—$2$ for state cultivation and injection, and up to $2$ for potential $S$-gate corrections.

To account for the additional timing overhead of these wrapping operations, the expected layer time $T_Z(n)$ defined in Eq.~\ref{eq:TZ-layer-cost} is modified to a wrapped layer time $\widetilde{T}_Z(n)$, which we use for our total cycle estimates. 
Numerical simulations confirm that this time cost is well approximated by $\widetilde{T}_Z(n) \approx 1.5 T_Z(n)$. 
The logical error coefficient $\alpha_{\mathrm{RUS}}$ shown in Fig.~\ref{fig:rotation_performance}(b) is estimated when wrapping operations are considered.
This quadratic suppression, $\alpha_{\rm RUS}|\theta|p^2$, significantly improves upon previous schemes that exhibited first-order scaling $\cO(|\theta|p)$~\cite{akahoshi2024partially,Toshio2025PracticalQuantumAdvantage}, thereby substantially extending the accessible simulation time and system size. As we detail in the next subsection, this low-order error suppression is already sufficient to keep the accumulated logical error within the acceptable budget, precluding the need for substantial quantum error mitigation resources.

\subsection{Resource estimation for quantum simulation}
Now we proceed to analyse the resources required for simulating time evolution under the mixed-field Ising model, focusing on two key metrics: the number of QEC cycles, which directly determines the time cost, and the space-time overhead, which accounts for both time and qubit resources.

\subsubsection{QEC cycles}
To determine the total number of QEC cycles, we first evaluate the QEC-clock cost of a single Trotter segment, denoted by $T_{\rm step}$. 
For the mixed-field Ising Hamiltonian, according to Eq.~\ref{eq:PF4_formula}, one fourth-order product-formula segment consists of five layers each of $Z$-field rotations, $X$-field rotations, and nearest-neighbour $XX$ interactions.
To minimise the basis-change overhead, the $X$ and $XX$ operations are executed consecutively. 
Through a shared layer of Hadamard gates, both operations are transformed into the $Z$-basis, mapping $X$ to $Z$, and $XX$ to $ZZ$.

\begin{figure*}[!htbp]
	\includegraphics[width=0.9\linewidth]{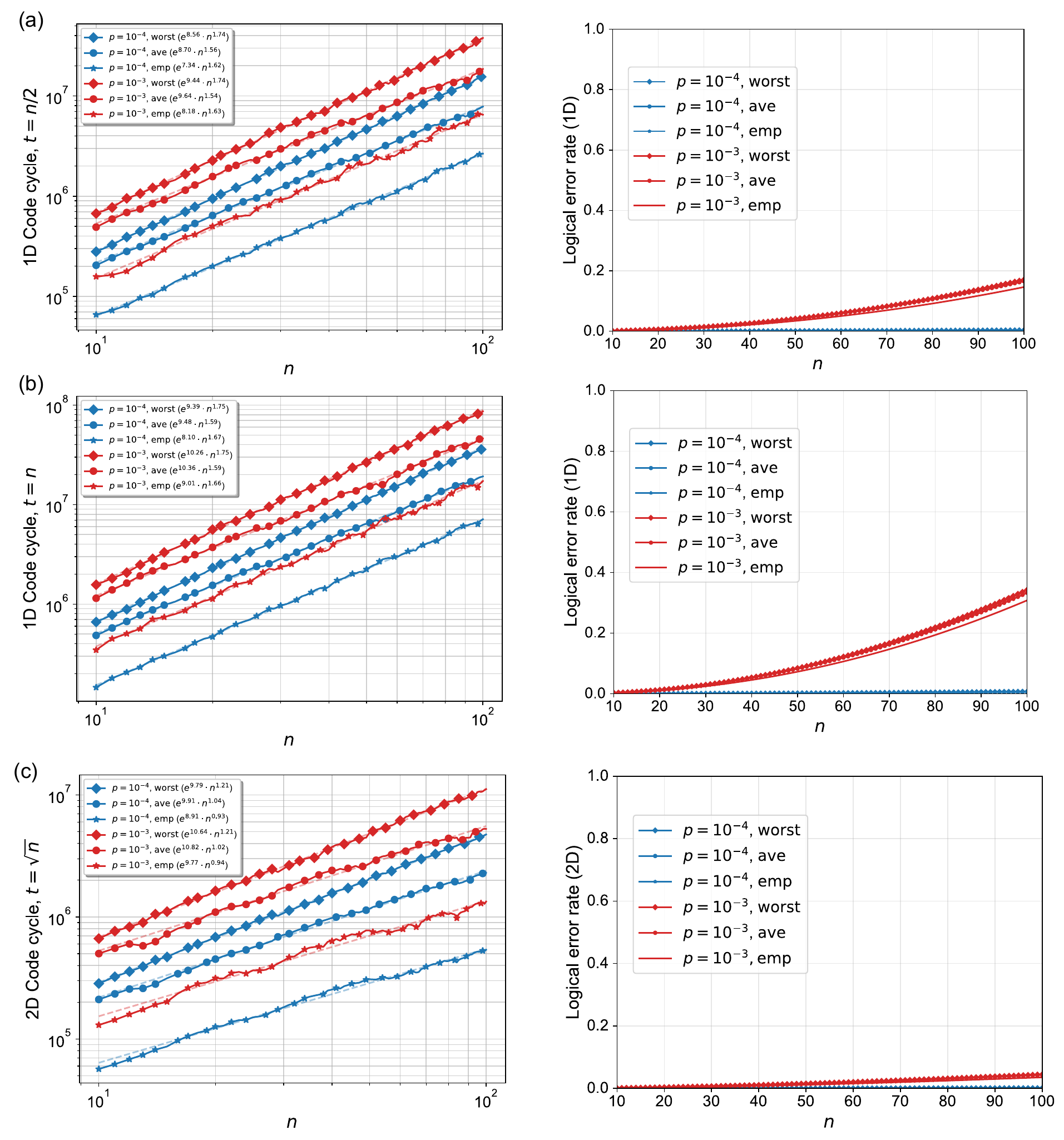}
    % \caption{\textbf{Resource estimation of the fault-tolerant simulation for the mixed-field Ising model based on small-angle rotations.} 
    % (a) Results for the 1D system, showing the total number of QEC cycles required for the evolution $\exp(-iHt)$ up to time $t=n/2$ with an algorithmic precision of $\varepsilon=0.01$ (left), and the accumulated logical error of the full simulation (right). The logical error is estimated by summing the error contribution, $\alpha_{\mathrm{RUS}} |\theta| p^2$, over all intermediate rotation states across the Trotter segments. 
    % (b) Corresponding QEC cycles (left) and accumulated logical error (right) for the 2D system on a square lattice, evaluated up to time $t = \sqrt{n}$ under the same precision target and error estimation.
    % }
    \caption{\textbf{Resource estimation of fault-tolerant simulation for the mixed-field Ising model based on small-angle rotations.} Across all cases, the left panels show the total number of QEC cycles required for the evolution $\exp(-iHt)$ to achieve an algorithmic precision of $\varepsilon=0.01$, while the right panels display the accumulated logical error of the full simulation. The logical error is estimated by summing the error contribution, $\alpha_{\mathrm{RUS}} |\theta| p^2$, over all intermediate rotation states across the Trotter segments. (a) Results for the 1D system up to $t=n/2$. (b) Results for the 1D system up to $t=n$. (c) Results for the 2D system up to time $t = \sqrt{n}$.}
	\label{fig:RotationCycle}
\end{figure*}

The single-qubit $Z$ and $X$ rotation layers each cost $\widetilde{T}_Z(n)$ QEC clocks. 
For the two-qubit $XX$ interactions, the lattice bonds must be decomposed into disjoint matchings to allow for parallel execution. 
In 1D, the chain splits into $M_{\rm 1D}=2$ matchings (even and odd bonds), with each matching containing approximately $n/2$ independent rotations. 
In 2D on a square lattice, the horizontal and vertical bonds require $M_{\rm 2D}=4$ matchings, each again encompassing roughly $n/2$ rotations. Once in the $Z$-basis, each nearest-neighbour $ZZ$ layer must be further reduced to single-qubit logical $Z$-rotations via CNOT operations. This contributes $6$ QEC clocks per layer, corresponding to two logical CNOTs~\cite{Litinski2019gameofsurfacecodes}.
The non-Clifford RUS cost is required for each matching.
Thus, the cost of a single $ZZ$-rotation layer is
\begin{equation}\label{eq:rotation_ZZ_layer_cost}
    T_{ZZ}^{(D)}(n) = 6 + M_D \widetilde T_Z(n/2),
\end{equation}
where $D \in \{\text{1D}, \text{2D}\}$ labels the spatial dimension. Summing the $10$ single-qubit rotation layers and the $5$ nearest-neighbour layers, and adding $4$ QEC clocks to account for the Hadamard basis changes~\cite{Litinski2019gameofsurfacecodes} (two clocks to enter the $X$-basis and two to return), the total QEC-clock cost per Trotter segment is
\begin{equation}
T_{\mathrm{step}}^{(D)} = 4 + 10 \widetilde T_Z(n) + 5 T_{ZZ}^{(D)}(n).
\end{equation}
Substituting Eq.~\ref{eq:rotation_ZZ_layer_cost}, we obtain
\begin{equation}
    \begin{split}
        T_{\mathrm{step}}^{\mathrm{(1D)}} &= 34 + 10 \widetilde T_Z(n) + 10 \widetilde T_Z(n/2), \\
        T_{\mathrm{step}}^{\mathrm{(2D)}} &= 34 + 10 \widetilde T_Z(n) + 20 \widetilde T_Z(n/2).         
    \end{split}
\end{equation}
To simulate the evolution with an algorithmic precision of $0.01$ up to the target times of $t=n/2$ and $t=n$ for the 1D system, and $t=\sqrt{n}$ for the 2D system, we employ the required number of Trotter segments, $\nu_D$, as derived in Sec.~\ref{apd:sec:trotter}.
Specifically, we consider three distinct estimates for $\nu_D$: (i) a worst-case bound, (ii) an average-case bound, and (iii) an empirical bound. 
Based on these segment counts, the total number of QEC cycles for a single unitary call of $U(t)$ is given by 
\begin{equation}
T_{\mathrm{cycle}}^{(D)} = d \times \nu_{D} \times T_{\mathrm{step}}^{(D)}.
\end{equation}
Furthermore, the accumulated logical error of the entire time-evolution circuit is obtained by summing the error contribution, $\alpha_{\mathrm{RUS}} |\theta| p^2$, over all intermediate small-angle rotation gates across the $\nu_{D}$ Trotter segments. The total QEC-cycle costs, evaluated under the three aforementioned bounds, and the corresponding accumulated logical errors for simulating the 1D and 2D mixed-field Ising models are presented in Fig.~\ref{fig:RotationCycle}.

\subsubsection{Space-time overhead}

\begin{figure}[!htbp]
	\includegraphics[width=0.8\linewidth]{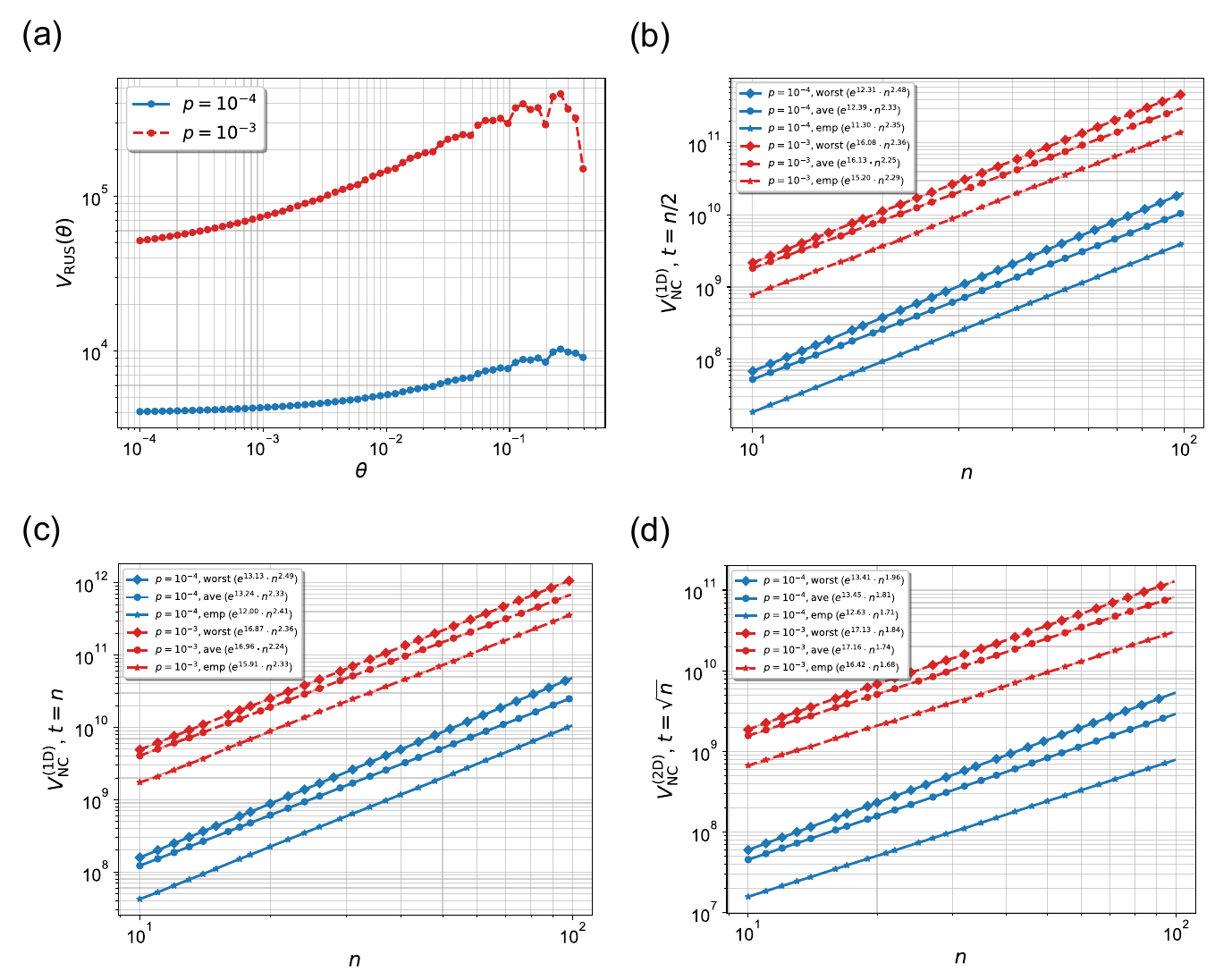}
    % \caption{\textbf{Active space-time volume of non-Clifford rotations for the fault-tolerant simulation of the mixed-field Ising models.} (a) Expected active space-time volume, $V_{\mathrm{RUS}}(\theta)$, for implementing a single small-angle rotation as a function of the target angle $\theta$. (b) Total non-Clifford active volume, $V_{\mathrm{NC}}^{\mathrm{(1D)}}$, for the full 1D simulation. This is obtained by summing the contributions from all intermediate small-angle rotations throughout the time evolution. (c) Corresponding total non-Clifford active volume, $V_{\mathrm{NC}}^{\mathrm{(2D)}}$, for the full simulation of the 2D system.}
    \caption{\textbf{Active space-time volume of non-Clifford rotations for the fault-tolerant simulation of the mixed-field Ising models.} (a) Expected active space-time volume, $V_{\mathrm{RUS}}(\theta)$, for implementing a single small-angle rotation as a function of the target angle $\theta$. Panels (b)--(d) display the total non-Clifford active volume, $V_{\mathrm{NC}}^{(D)}$, for the full simulation, which is obtained by summing the space-time contributions from all intermediate small-angle rotations throughout the time evolution. These total volumes are presented for (b) the 1D system up to time $t=n/2$, (c) the 1D system up to time $t=n$, and (d) the 2D system up to time $t=\sqrt{n}$.}
	\label{fig:RotationSpacetime}
\end{figure}

Finally, we estimate the active space-time volume associated with rotation-state injection. We focus on the active non-Clifford volume, as this isolates the dominant space-time overhead introduced by the small-angle-rotation implementation. For a single target rotation angle $\theta$, the expected RUS volume is obtained by summing over all possible RUS attempts
\begin{equation}
    V_{\rm RUS}(\theta) =  \sum_{j} 2^{-j} V(\theta_j),
\end{equation}
where $\theta_j$ is the effective angle in the $j$th attempt after applying the wrapping operations defined in Sec.~\ref{subsec:rotation_state_preparation}. 
The term $V(\theta_j)$ encompasses the active volume required to prepare the corresponding logical rotation state, inject it into the data block, and perform any necessary wrapping. The total active non-Clifford volume $V_{\mathrm{NC}}^{(D)}$ (where $D \in \{\mathrm{1D}, \mathrm{2D}\}$) for a single call to the time-evolution operator $U(t)$ is then calculated by aggregating the contributions from all rotation layers within a Trotter segment and multiplying by the total number of segments. 
The resulting space-time volume estimates for individual rotations, alongside the total active volume of the full simulation—evaluated at $t=n/2$ and $t=n$ for the 1D system, and $t=\sqrt{n}$ for the 2D system—are presented in Fig.~\ref{fig:RotationSpacetime}.

%---------------------------QEC--------------
\section{Fault-tolerant implementation and resource analysis for Clifford+\texttt{T} approaches}
\label{app:msd-resource-model}

% \subsection{From Trotter model and coherent estimation to Clifford + $\texttt{T}$ circuit implementatoin}

\subsection{Direct versus SPBC scheduling for the Clifford+\texttt{T} branch}
\label{subsec:direct_vs_spbc_scheduling}

\begin{figure}[t]
    \centering
    \includegraphics[width=\linewidth]{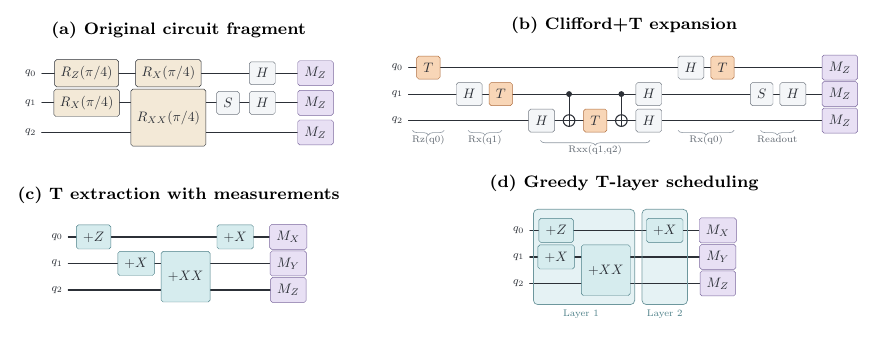}
    \caption{\textbf{Preprocessing pipeline from a native circuit fragment to the two fault-tolerant implementations compared in this appendix.} (a)~A representative circuit fragment containing Clifford gates and non-Clifford rotations. (b)~The same fragment after synthesis into a shared logical Clifford+$\texttt{T}$ circuit, which is the common input to both implementations. (c)~For the SPBC implementation, the $\texttt{T}$-gate content is extracted into signed Pauli-product phase gadgets implemented through Pauli-product measurements, following Ref.~\cite{Litinski2019gameofsurfacecodes}. (d)~The extracted gadgets are grouped by \textit{greedy $\texttt{T}$-layer scheduling} to form the logical SPBC standard form.
    The direct implementation keeps the shared Clifford+$\texttt{T}$ dependency structure.}
    \label{fig:spbc_pipeline}
\end{figure}

The logical \texttt{T}-count alone does not determine the cost of the Clifford+\texttt{T} branch. In this appendix, the logical \texttt{T}-count denotes the number of explicit \texttt{T} gates produced by our Clifford+\texttt{T} synthesis pipeline; each such gate is treated as consuming one injected magic state in the direct implementation. Once the native-rotation circuit has been synthesised into a Clifford+\texttt{T} circuit, the resulting logical circuit must still be mapped to a fault-tolerant surface-code implementation.  This mapping involves choices such as lattice-surgery scheduling, measurement grouping, routing, and pipelining, all of which can change the final runtime and footprint.

We use the PF4 XYZ benchmark family as a fixed logical workload to expose this dependence.  Here PF4 denotes the fourth-order product-formula circuit family used in our simulations, and XYZ denotes the Hamiltonian instance.  Holding this workload fixed allows us to ask how much of the final resource estimate comes from the logical \texttt{T}-count, and how much comes from the fault-tolerant implementation of the same Clifford+\texttt{T} circuit.

The comparison starts from the same synthesised logical Clifford+\texttt{T} circuit on both sides, as shown in Fig.~\ref{fig:spbc_pipeline}. In the direct Clifford+\texttt{T} scheduling route, we keep the circuit representation and construct a directed acyclic dependency graph from the gate sequence. In the sequential Pauli-based computation (SPBC) route, we rewrite the \texttt{T}-gate content into signed Pauli-product phase gadgets, following the Pauli-based computation framework of Ref.~\cite{Litinski2019gameofsurfacecodes}. These gadgets are first grouped into logical \texttt{T}-layers and then further refined into support-disjoint primitive Pauli-product-measurement (PPM) layers.

Thus, the two routes implement the same synthesised logical Clifford+\texttt{T} circuit, but expose different scheduled objects to the surface-code timing model: the direct route schedules circuit-level gate dependencies, whereas the SPBC route schedules layers of primitive Pauli-product measurements.

The runtime estimates use a common magic-state-factory, inventory, and waiting-time model, together with implementation-specific scheduled blocks. All times are measured in logical time steps (LTS), and the primitive block durations are listed in Table~\ref{tab:primitive_timing_model}. We use duration \(\tau_C=1\) for one layer of explicit Clifford gates, \(\tau_D=13\) for one magic-state factory cycle as a default benchmark value motivated by recent logical-level 15-to-1 factory cadences~\cite{beverland2022assessing,Silva2024MultiLevelFactories}, \(\tau_{T,\mathrm{app}}=1\) for one batch of single-qubit $\texttt{T}/\texttt{T}
^\dagger$ applications via magic-state injection, and \(\tau_M=1\) for one final single-qubit measurement layer. Single-qubit Clifford updates tracked classically are assigned zero duration.
For the default SPBC timing model, one primitive Pauli-product-measurement (PPM) layer has duration \(\tau_{\mathrm{PPM}}=8\). A correction layer denotes the measurement-dependent correction step following a primitive PPM layer and has raw duration \(\tau_{\mathrm{corr}}=8\). In the expected-time model used below, each correction layer contributes \(p_{\mathrm{corr}}\tau_{\mathrm{corr}}=4\), where \(p_{\mathrm{corr}}=1/2\) is the assumed probability that a nontrivial correction layer is required.
Within one layer, ready operations acting on disjoint qubit sets are executed in parallel, whereas different layers remain serial.
The number of parallel magic-state factories is denoted by \(F\). These
factories run continuously during the logical computation, and their distilled
outputs may accumulate in a magic-state inventory. No free initial inventory is
assumed. If the inventory is insufficient for the next magic-state-consuming
block, that block stalls until enough states have been produced. This stall is
represented by \(\sigma_\ell(F)\), so factory parallelism is included through
the waiting term: larger \(F\) increases the supply rate and reduces the
accumulated waiting time.

This is a fixed-factory-budget runtime comparison. We do not optimise over
\(F\), nor do we account for the extra physical footprint needed to increase
the factory count and eliminate waiting. Accordingly, the results compare
runtimes at fixed \(F\), rather than fully space-time-optimised layouts.

\begin{table}[t]
    \centering
    \small
    \caption{\textbf{Block-level timing model used in the implementation comparison.} All durations are measured in logical time steps (LTS).}
    \label{tab:primitive_timing_model}
    \begin{tabular}{@{}lllp{0.44\linewidth}@{}}
        \toprule[0.9pt]
        operation block & symbol & time (LTS) & execution rule \\
        \midrule[0.6pt]
        explicit Clifford-gate layer
        & $\tau_C$
        & $1$
        & operations in the same layer run in parallel when they act on disjoint qubit sets \\

        classically tracked single-qubit Clifford update
        & $\tau_{C,\mathrm{frame}}$
        & $0$
        & tracked in software and not scheduled as a physical operation \\

        magic-state factory cycle
        & $\tau_D$
        & $13$
        & up to $F$ factories produce in parallel; production continues while other blocks run \\

        single-qubit $\texttt{T}$ application
        & $\tau_{T,\mathrm{app}}$
        & $1$
        & ready $\texttt{T}/\texttt{T}^\dagger$ gates in one batch run in parallel if enough stored magic states are available \\

        single-qubit measurement layer
        & $\tau_M$
        & $1$
        & compatible final measurements run in parallel \\

        Pauli-product-measurement layer
        & $\tau_{\mathrm{PPM}}$
        & $8$
        & gadgets in the same layer run in parallel when they act on disjoint qubit sets \\

        correction layer
        & $\tau_{\mathrm{corr}}$
        & $8$
        & follows the corresponding PPM layer; expected contribution is $p_{\mathrm{corr}}\tau_{\mathrm{corr}}=4$ \\
        \bottomrule[0.9pt]
    \end{tabular}
\end{table}
Given the scheduled block sequence of length \(M\) produced by either implementation, let
\(d_\ell\) be the nominal duration of block \(\ell\) before magic-state
waiting, and let \(\sigma_\ell(F)\) be the additional stall time inserted
before that block when \(F\) factories are available. We include a single
initial factory warm-up time \(\tau_D\). The total logical runtime is
\begin{equation}
    T_{\mathrm{LTS}}(F)
    =
    \tau_D
    +
    \sum_{\ell=1}^{M}
    \left[
        d_\ell+\sigma_\ell(F)
    \right].
    \label{eq:lts_runtime}
\end{equation}
Here \(\sigma_\ell(F)\) accounts only for magic-state-supply stalls; it is
zero for blocks that do not require unavailable magic states. For correction
blocks, \(d_\ell\) denotes the expected-time contribution
\(p_{\mathrm{corr}}\tau_{\mathrm{corr}}\).

For the decomposition tables below, we regroup the same runtime into warm-up,
stalling, non-Clifford execution, Clifford or correction execution, and final
measurement contributions
\begin{equation}
    T_{\mathrm{LTS}}(F)
    =
    T_{\mathrm{warm}}
    +
    T_{\mathrm{stall}}(F)
    +
    T_{T/\mathrm{PPM}}
    +
    T_{\mathrm{Cliff}/\mathrm{corr}}
    +
    T_{\mathrm{meas}} .
    \label{eq:lts_decomposition}
\end{equation}
The five terms are defined by
\begin{align}
    T_{\mathrm{warm}}
    &=
    \tau_D, \\
    T_{\mathrm{stall}}(F)
    &=
    \sum_{\ell=1}^{M}\sigma_\ell(F), \\
    T_{T/\mathrm{PPM}}
    &=
    \sum_{\ell\in\mathcal{B}_{T/\mathrm{PPM}}} d_\ell, \\
    T_{\mathrm{Cliff}/\mathrm{corr}}
    &=
    \sum_{\ell\in\mathcal{B}_{\mathrm{Cliff}/\mathrm{corr}}} d_\ell, \\
    T_{\mathrm{meas}}
    &=
    \sum_{\ell\in\mathcal{B}_{\mathrm{meas}}} d_\ell .
\end{align}
For compactness in the representative decomposition table, we combine the warm-up and magic-state stalling terms as \(T_{\mathrm{wait}}(F)\equiv T_{\mathrm{warm}}+T_{\mathrm{stall}}(F)\).
The block sets form a partition of the nominal scheduled blocks. For the direct Clifford+\texttt{T} implementation, \(\mathcal{B}_{T/\mathrm{PPM}}\) contains explicit $\texttt{T}$ and $\texttt{T}^\dagger$ application blocks, while \(\mathcal{B}_{\mathrm{Cliff}/\mathrm{corr}}\) contains explicit Clifford layers. Classically tracked Clifford-frame updates are assigned zero duration and are not counted as scheduled physical blocks. For the SPBC implementation, \(\mathcal{B}_{T/\mathrm{PPM}}\) contains primitive PPM layers, and \(\mathcal{B}_{\mathrm{Cliff}/\mathrm{corr}}\) contains the corresponding correction layers. Thus, Eq.~\ref{eq:lts_decomposition} is only a regrouping
of Eq.~\ref{eq:lts_runtime}; both equations describe the same logical
runtime for a fixed number \(F\) of factories.

We first compare the direct implementation with the optimised SPBC implementation.
The optimised SPBC implementation starts from the standard-form SPBC rewrite of the same Clifford+\texttt{T} circuit. We then simplify the resulting Pauli-product gadget sequence by removing redundant frame-trackable operations and correction blocks where possible, and we optimise the layer structure by grouping support-disjoint primitive PPMs into parallel layers subject to the required measurement dependencies. This produces a refined SPBC schedule without changing the underlying synthesised logical circuit.

\begin{table*}[t]
    \centering
    \small
    \caption{\textbf{Matched-factory-budget runtimes for the direct and optimised SPBC implementations under the default timing model, including the common terminal observable $M_z=n^{-1}\sum_j Z_j$.}
    Each entry is reported as \textbf{direct/SPBC} in LTS, and boldface marks the faster implementation in each matched-budget comparison.}
    \label{tab:matched_factory_runtime_matrix}
    \renewcommand{\arraystretch}{1.08}
    \begin{tabular}{@{}c|c c c c c c@{}}
        \toprule[0.9pt]
        $n$ & $F=1$ & $F=2$ & $F=4$ & $F=8$ & $F=12$ & $F=16$ \\
        \midrule[0.6pt]
        $n=4$
        & $35000/\mathbf{34832}$
        & $\mathbf{17502}/25531$
        & $\mathbf{8753}/24939$
        & $\mathbf{4385}/24897$
        & $\mathbf{2929}/24897$
        & $\mathbf{2201}/24897$ \\

        $n=8$
        & $\mathbf{74156}/74228$
        & $\mathbf{37080}/46439$
        & $\mathbf{18542}/43133$
        & $\mathbf{9273}/42483$
        & $\mathbf{6192}/42438$
        & $\mathbf{4645}/42437$ \\

        $n=12$
        & $115600/\mathbf{113792}$
        & $57802/\mathbf{57360}$
        & $\mathbf{28903}/43319$
        & $\mathbf{14460}/40670$
        & $\mathbf{9637}/40332$
        & $\mathbf{7232}/40283$ \\

        $n=16$
        & $159540/\mathbf{159372}$
        & $79772/\mathbf{79735}$
        & $\mathbf{39888}/50403$
        & $\mathbf{19946}/45856$
        & $\mathbf{13303}/45386$
        & $\mathbf{9975}/45166$ \\

        $n=20$
        & $202283/\mathbf{201836}$
        & $101143/\mathbf{100986}$
        & $\mathbf{50573}/56921$
        & $\mathbf{25288}/50241$
        & $\mathbf{16864}/49407$
        & $\mathbf{12652}/49117$ \\
        \bottomrule[0.9pt]
    \end{tabular}
\end{table*}

Table~\ref{tab:matched_factory_runtime_matrix} reports the matched-factory-budget runtime matrix under the default timing model.
Counting the bold entries shows that the direct implementation is faster in \(23\) of the \(30\) tested \((n,F)\) instances, while the optimised SPBC implementation is faster in the remaining \(7\) cases.
Thus, for this benchmark set and default timing model, direct Clifford+\texttt{T} execution is favoured in most fixed-factory-budget comparisons. 

The comparison in Table~\ref{tab:matched_factory_runtime_matrix} shows that the optimised SPBC implementation can be
faster in a small number of low-factory-budget cases. However, this does not
by itself identify why SPBC wins in those cases. After the standard-form SPBC
rewrite is simplified, the resulting SPBC schedule may consume fewer magic
states than the direct Clifford+\texttt{T} implementation. Therefore, an SPBC
advantage can come either from its PPM-based scheduling structure, from a
reduction in magic-state waiting due to a smaller magic-state count, or from
both effects. To separate these effects, we perform a control comparison in which the SPBC
schedule is kept unchanged, but its total magic-state consumption is matched
to the direct implementation. In this control, the SPBC PPM layers, PPM timing,
and correction timing are unchanged, while the number of consumed magic states
is set equal to the direct implementation's \(\texttt{T}/\texttt{T}^{\dagger}\) count. This removes
the possible advantage coming only from simplification-induced magic-state
count reduction. With this matched-count control and the same default timing model, the direct
implementation is faster in all \(30\) tested instances. This indicates that
the SPBC-favourable cases observed for the optimised SPBC implementation rely
at least partly on the reduced magic-state consumption produced by SPBC
simplification, rather than on the PPM-based execution pattern alone.

The time decomposition in representative cases clarifies the source of the
crossover. Table~\ref{tab:representative_decomposition} reports the partial
runtime contributions defined in Eq.~\ref{eq:lts_decomposition}. In the
direct Clifford+\texttt{T} implementation, the explicit Clifford contribution is
modest, and the dominant contribution at small factory budget comes from
magic-state waiting. This term decreases rapidly as the number of factories
\(F\) increases.
By contrast, the SPBC implementation contains a substantial PPM-plus-correction
contribution. This contribution is largely independent of \(F\), because it is
set by the scheduled PPM and correction layers rather than by the magic-state
production rate. Consequently, SPBC can be competitive in very low-budget cases,
where reduced magic-state consumption can lower the waiting time, but it retains
a large fixed execution overhead after magic-state waiting has been suppressed.
The direct implementation therefore becomes clearly faster once the factory
capacity is large enough to remove magic-state supply as the dominant
bottleneck.

\begin{table*}[t]
    \centering
    \small
    \caption{\textbf{Representative time decomposition under the default timing model for the direct and optimised SPBC implementations.}
    All entries are reported in LTS and follow the decomposition in Eq.~\ref{eq:lts_decomposition}.
    The column $T_{\mathrm{wait}}$ includes the initial factory warm-up time and all subsequent magic-state waiting time.
    For the direct implementation, $T_{T/\mathrm{PPM}}$ is the time spent on explicit $\texttt{T}/\texttt{T}^\dagger$ applications and $T_{\mathrm{Cliff}/\mathrm{corr}}$ is the time spent on explicit Clifford-gate layers.
    For the SPBC implementation, these columns correspond to Pauli-product-measurement layers and correction layers, respectively.}
    \label{tab:representative_decomposition}
    \begin{tabular}{@{}c c|r r r r|r@{}}
        \toprule[0.9pt]
        case & impl. & $T_{\mathrm{wait}}$ & $T_{T/\mathrm{PPM}}$ & $T_{\mathrm{Cliff}/\mathrm{corr}}$ & $T_{\mathrm{meas}}$ & $T_{\mathrm{LTS}}$ \\
        \midrule[0.6pt]
        $n=8,\;F=1$  & direct & 72175  & 957   & 1023  & 1  & 74156 \\
                     & SPBC   & 31804  & 28240 & 14120 & 64 & 74228 \\
        \addlinespace

        $n=8,\;F=4$  & direct & 16561  & 957   & 1023  & 1  & 18542 \\
                     & SPBC   & 709    & 28240 & 14120 & 64 & 43133 \\
        \addlinespace

        $n=8,\;F=16$ & direct & 2664   & 957   & 1023  & 1  & 4645 \\
                     & SPBC   & 13     & 28240 & 14120 & 64 & 42437 \\
        \addlinespace

        $n=12,\;F=1$ & direct & 113557 & 992   & 1050  & 1  & 115600 \\
                     & SPBC   & 73524  & 26808 & 13404 & 56 & 113792 \\
        \addlinespace

        $n=12,\;F=4$ & direct & 26860  & 992   & 1050  & 1  & 28903 \\
                     & SPBC   & 3051   & 26808 & 13404 & 56 & 43319 \\
        \addlinespace

        $n=12,\;F=16$ & direct & 5189  & 992   & 1050  & 1  & 7232 \\
                      & SPBC   & 15    & 26808 & 13404 & 56 & 40283 \\
        \addlinespace

        $n=20,\;F=1$ & direct & 200165 & 1025  & 1092  & 1  & 202283 \\
                     & SPBC   & 152968 & 32536 & 16268 & 64 & 201836 \\
        \addlinespace

        $n=20,\;F=4$ & direct & 48455  & 1025  & 1092  & 1  & 50573 \\
                     & SPBC   & 8053   & 32536 & 16268 & 64 & 56921 \\
        \addlinespace

        $n=20,\;F=16$ & direct & 10534 & 1025  & 1092  & 1  & 12652 \\
                      & SPBC   & 249   & 32536 & 16268 & 64 & 49117 \\
        \bottomrule[0.9pt]
    \end{tabular}
\end{table*}

We further test whether the comparison depends on the assumed PPM and correction
latencies. Before varying these primitive costs, we emphasise that the factory
budget is already treated as a parallel resource. The parameter \(F\) denotes
the number of simultaneously running magic-state factories, and the waiting
term \(\sigma_\ell(F)\) is computed from the magic-state inventory supplied by
these \(F\) factories. Thus, increasing \(F\) increases the magic-state supply
rate and directly reduces the waiting contribution. The sensitivity study below
keeps this fixed-\(F\) comparison framework and varies only the assumed PPM and
correction latencies. Since the main fixed overhead of SPBC comes from its PPM and
correction layers, we vary these two primitive costs and repeat the same
direct-versus-SPBC comparison.  First, we scale them together as
\[
    (\tau_{\mathrm{PPM}},\tau_{\mathrm{corr}})
    \in
    \{(1,1),(4,4),(8,8),(12,12)\}.
\]
At the most optimistic point,
\((\tau_{\mathrm{PPM}},\tau_{\mathrm{corr}})=(1,1)\), the optimised SPBC
implementation is faster in \(23\) cases, the direct implementation is faster
in \(6\) cases, and one case is tied.  Thus, if both PPM and correction layers
are made sufficiently cheap, the optimised SPBC schedule can become broadly
favourable. However, the matched-magic-count control shows that this advantage is not due
to the PPM schedule alone.  At the same optimistic timing point, the direct
implementation is faster in \(21\) cases, the SPBC control is faster in \(5\)
cases, and \(4\) cases are tied.  Therefore, much of the broad SPBC advantage
at \((1,1)\) relies on retaining the simplification-induced reduction in
magic-state consumption. Finally, we isolate the role of correction latency by reducing only the PPM
latency to \(\tau_{\mathrm{PPM}}=1\), while keeping the correction latency at
its default value \(\tau_{\mathrm{corr}}=8\).  In this case, the optimised SPBC
implementation is faster in only \(11\) cases, and the matched-magic-count
control shows no SPBC-favourable cases.  Hence, a broad SPBC speedup over direct
execution requires both PPM and correction layers to be cheap, and the effect is
strongest when the magic-count reduction from SPBC simplification is retained.

Overall, these results show that the total fault-tolerant time within the Clifford+\texttt{T} branch depends on the implementation choice even when the underlying synthesised logical circuit is fixed.
Under the default timing model, direct Clifford+\texttt{T} execution is faster in most PF4 cases because its explicit Clifford-gate cost is modest compared with the PPM-plus-correction overhead of SPBC.
The limited SPBC advantage at small factory budget is partly due to reduced magic-state consumption from standard-form simplification.
Only under strongly optimistic assumptions for both PPM and correction latency does the optimised SPBC implementation become broadly favourable, and that advantage shrinks once the magic-state count is matched to the direct $\texttt{T}/\texttt{T}^\dagger$ count.

\subsection{Magic-state supply in the surface-code layout}
\label{app:time-optimal-msd-resource-model}

We estimate the fault-tolerant resource cost of the Clifford+\texttt{T} circuit by
combining a fast surface-code layout for the data block with magic-state
distillation (MSD) factories.  The layout fixes the active data-tile footprint,
the parallel \texttt{T}-layer width, and the code-cycle cost of consuming one
effective \texttt{T}-layer.  The MSD factory protocols are taken from Table~1 of
Ref.~\cite{Litinski2019MSD}, which provides the output error probability
\(p_{\rm out}\), physical-qubit footprint \(Q_{\rm fac}\), latency
\(\tau_{\rm fac}\), and output multiplicity \(m_{\rm out}\).  The tabulated
latency is treated as the effective latency for an accepted output batch, so
factory rejection is not applied a second time.

\begin{figure}[t]
    \centering
    \includegraphics[width=0.8\linewidth]{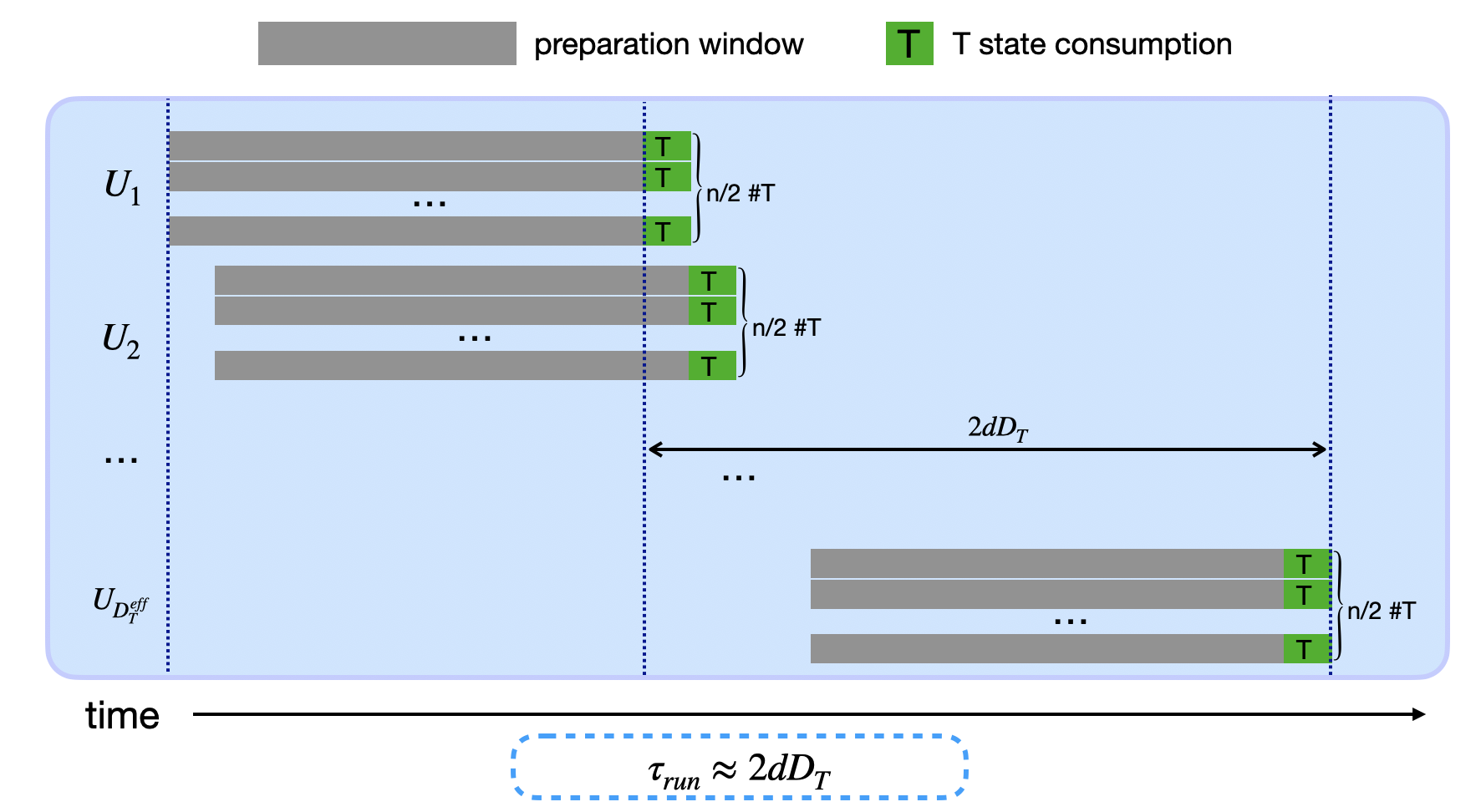}
    \caption{\textbf{Time structure of the spatial pipeline.}  Preparation
    windows for different physical units overlap in time.  After the initial
    fill, the computation consumes one effective \texttt{T}-layer every \(2d\)
    code cycles.}
    \label{fig:time-space-pipeline}
\end{figure}

\paragraph{Workload model.}
For each system size \(n\), error target \(\varepsilon\), and
workload label, the fitted \texttt{T}-count for one circuit block is
\begin{equation}
    T_{\rm block}
    =
    A\,n\,e^b n^\gamma ,
    \label{eq:t-count-fit}
\end{equation}
where \(A\), \(b\), and \(\gamma\) are taken from the corresponding empirical
or average fit.  The labels \(1{\rm D}, T=n/2\), \(1{\rm D}, T=n\), and
\(2{\rm D}, T=\sqrt n\) specify different \texttt{T}-count fits only; they do not
set the number of \texttt{T} gates that can be consumed simultaneously.  In the
resource model, the hardware-limited parallel \texttt{T}-width is fixed to
\begin{equation}
    w=\left\lceil\frac{n}{2}\right\rceil .
    \label{eq:parallel-t-width}
\end{equation}
Let \(q(\varepsilon)\) denote the query count associated with the target error
budget \(\varepsilon\).  The total \texttt{T}-state demand used to compute the runtime
is
\begin{equation}
    N_T^{\rm run}
    =
    T_{\rm block}q(\varepsilon)
    =
    A\,n\,e^b n^\gamma q(\varepsilon).
    \label{eq:t-demand-runtime}
\end{equation}
The factory-selection and magic-state-error budget use the corresponding
depth-weighted block count
\begin{equation}
    N_T^{\rm err}
    =
    \frac{N_T^{\rm run}}{q(\varepsilon)}\Delta
    =
    T_{\rm block}\Delta ,
    \label{eq:t-demand-error-budget}
\end{equation}
where \(\Delta\) is the independent circuit-depth factor.  The effective
\texttt{T}-depth for the runtime is therefore
\begin{equation}
    D_T
    =
    \left\lceil \frac{N_T^{\rm run}}{w}\right\rceil .
    \label{eq:effective-t-depth}
\end{equation}
In the numerical scan below, we use
\[
    (\Delta,q)=(1,20),(20,230),(200,2270)
\]
for \(\varepsilon=10^{-1},10^{-2},10^{-3}\), respectively.
\paragraph{Layout model and tile count.}
The layout is treated as a spatial pipeline.  This should not be interpreted
as dividing the \(n\) logical data qubits among different units. Each
pipeline unit represents a physical layout region capable of carrying one
time slice of the same \(n\)-qubit logical computation, while successive
effective \texttt{T}-layers are staggered in time. Fig.~\ref{fig:time-space-pipeline} shows the layout abstraction. The
active full-distance non-factory tile count contains one fast-block data
layout and one buffer for a parallel \texttt{T}-layer
\begin{equation}
    S_{\rm layout}
    =
    2n+\left\lceil\sqrt{8n}\right\rceil+1,
    \qquad
    S_{\rm buff}=w,
    \qquad
    S_{\rm act}=S_{\rm layout}+S_{\rm buff}.
    \label{eq:time-optimal-layout-tiles}
\end{equation}
The first expression counts the fast-block data layout, while \(S_{\rm buff}\)
is the buffer for the \(w\) magic states consumed in one effective \texttt{T}-layer.
This active tile count is not multiplied by the timing-pipeline unit count.
The timing unit count is used only to represent the overlap between
preparation and consumption; the factory lanes are accounted for separately
through \(F_{\rm tot}Q_{\rm fac}\).
\paragraph{Runtime pipeline.}
We model one effective \texttt{T}-layer as preparing and consuming \(w\) magic
states in parallel.  It is a simplified parallel
\texttt{T}-injection model.  Let \(\tau_{\rm inj}\) denote the clock used for
injecting one effective parallel \texttt{T}-layer, and let \(\tau_{\rm def}\)
denote the additional deformation or joint-measurement clock used to complete
the injection.  Both are measured in surface-code time steps.  Since one
surface-code time step is counted as \(d\) code cycles, the numerical model
uses \(\tau_{\rm inj}=\tau_{\rm def}=1\), so each effective \texttt{T}-layer costs
\(2d\) code cycles
\begin{equation}
    \tau_{\rm run}
    =
    D_Td(\tau_{\rm inj}+\tau_{\rm def})
    =
    2dD_T .
    \label{eq:time-optimal-runtime}
\end{equation}
Local movement needed to align a prepared magic state with the injection
region is absorbed into this \(2d\) injection/deformation window.  A separate
movement term would only be needed for an explicit long-range routing model.
Fig.~\ref{fig:time-space-pipeline} illustrates this time structure.

\paragraph{Global factory throughput and physical-qubit footprint.}
We use a global factory-throughput approximation rather than assigning a
separate local factory block to every pipeline unit.  Factory outputs enter a
shared magic-state pool.  One factory lane produces \(m_{\rm out}\) magic
states every \(\tau_{\rm fac}\) code cycles, so the number of lanes required
to supply one effective \texttt{T}-layer is
\begin{equation}
    F_{\rm tot}
    =
    \left\lceil
        \frac{w\,\tau_{\rm fac}}
             {m_{\rm out}}
    \right\rceil .
    \label{eq:global-throughput-factory-count}
\end{equation}
The division by \(m_{\rm out}\) is a throughput factor for multi-output
protocols.  It is not a second division of the table's per-output
space-time cost: the factory footprint is still added as
\(F_{\rm tot}Q_{\rm fac}\).  This is the conservative global-supply model used
in the numerical results; it does not amortize factory production over the
\(d\)-cycle consumption window.

The total physical-qubit footprint is
\begin{equation}
    Q_{\rm tot}
    =
    2d^2S_{\rm act}
    +
    F_{\rm tot}Q_{\rm fac}.
    \label{eq:time-optimal-total-qubits}
\end{equation}
The factor \(d^2\) converts one full-distance tile into a distance-\(d\)
surface-code patch.  The additional factor of two accounts for the
measurement/deformation ancilla area needed to carry out joint measurements
and lattice deformations during \texttt{T}-state injection.  Thus the non-factory
layout cost is counted as data/support patches plus the ancilla space used by
the deformation operations, while \(F_{\rm tot}Q_{\rm fac}\) adds the
protocol-specific MSD factory footprint.  The corresponding space-time volume
is
\begin{equation}
    V_{\rm ST}
    =
    Q_{\rm tot}\tau_{\rm run}.
    \label{eq:time-optimal-spacetime-volume}
\end{equation}

\paragraph{Error budget and protocol selection.}
For each candidate factory protocol and odd code distance \(d\), the estimate
is re-solved under the total failure constraint.  We approximate the logical
storage error rate per full-distance layout tile per code cycle by
\begin{equation}
    p_L(p,d)
    =
    0.1\left(100p\right)^{(d+1)/2}.
    \label{eq:surface-code-logical-error-rate}
\end{equation}
The total failure probability is estimated as
\begin{equation}
    p_{\rm fail}
    =
    S_{\rm act}\tau_{\rm run}p_L(p,d)
    +
    N_T^{\rm err}p_{\rm out}.
    \label{eq:time-optimal-total-failure}
\end{equation}
The first term estimates accumulated storage error over the active layout, and
the second term estimates the accumulated error from consuming all distilled
magic states.  A candidate code distance and factory protocol are accepted
only if
\begin{equation}
    p_{\rm fail}<\varepsilon.
    \label{eq:time-optimal-error-acceptance}
\end{equation}
For each factory protocol, we choose the smallest odd \(d\) satisfying this
condition.  We then select the feasible protocol with minimum
\(Q_{\rm tot}\), using \(Q_{\rm tot}\tau_{\rm run}\) as a tie-breaker.

\paragraph{Factory protocol key.}
Table~\ref{tab:selected-msd-protocol-key} lists the factory protocols that are
selected by at least one point in the numerical scan.  The \(\tau_{\rm fac}\)
column is the latency of one factory batch, and \(m_{\rm out}\) is the number
of output \texttt{T} states in that batch.  Table~1 of
Ref.~\cite{Litinski2019MSD} also reports a qubit-cycle cost per output state;
we do not use that column directly in Eq.~\ref{eq:global-throughput-factory-count}. The Table~\ref{tab:selected-msd-protocol-key} should be read as a protocol dictionary: the numerical tables below
refer to the selected factories by the IDs \(P1,\ldots,P11\), while the actual
optimisation uses the full \(p_{\rm out}\), \(Q_{\rm fac}\), \(\tau_{\rm fac}\),
and \(m_{\rm out}\) values listed here.

\begin{table*}[t]
    \centering
    % \scriptsize
    \setlength{\tabcolsep}{4pt}
    \caption{\textbf{Factory protocol key for the plotted points.}
    Protocol data are adapted from Table~1 of Ref.~\cite{Litinski2019MSD}.}
    \label{tab:selected-msd-protocol-key}
    \begin{tabular}{c c l c c c c}
        \hline
        ID & \(p\) & MSD protocol
        & \(p_{\rm out}\) & \(Q_{\rm fac}\)
        & \(\tau_{\rm fac}\) & \(m_{\rm out}\) \\
        \hline
        P1 & \(10^{-3}\) &
        \((15\textrm{-to-}1)_{17,7,7}\)
        & \(4.5\times10^{-8}\) & \(4.62\times10^{3}\)
        & 42.6 & 1 \\
        P2 & \(10^{-3}\) &
        \begin{tabular}[t]{@{}l@{}}
        \((15\textrm{-to-}1)^6_{13,5,5}\times\)\\
        \((20\textrm{-to-}4)_{23,11,13}\)
        \end{tabular}
        & \(1.4\times10^{-10}\) & \(4.33\times10^{4}\)
        & 130 & 4 \\
        P3 & \(10^{-3}\) &
        \begin{tabular}[t]{@{}l@{}}
        \((15\textrm{-to-}1)^6_{13,5,5}\times\)\\
        \((20\textrm{-to-}4)_{27,13,15}\)
        \end{tabular}
        & \(2.6\times10^{-11}\) & \(4.68\times10^{4}\)
        & 157 & 4 \\
        P4 & \(10^{-3}\) &
        \begin{tabular}[t]{@{}l@{}}
        \((15\textrm{-to-}1)^6_{13,5,5}\times\)\\
        \((15\textrm{-to-}1)_{25,11,11}\)
        \end{tabular}
        & \(2.7\times10^{-12}\) & \(3.07\times10^{4}\)
        & 82.5 & 1 \\
        P5 & \(10^{-3}\) &
        \begin{tabular}[t]{@{}l@{}}
        \((15\textrm{-to-}1)^6_{13,5,5}\times\)\\
        \((15\textrm{-to-}1)_{29,11,13}\)
        \end{tabular}
        & \(3.3\times10^{-14}\) & \(3.91\times10^{4}\)
        & 97.5 & 1 \\
        P6 & \(10^{-3}\) &
        \begin{tabular}[t]{@{}l@{}}
        \((15\textrm{-to-}1)^6_{17,7,7}\times\)\\
        \((15\textrm{-to-}1)_{41,17,17}\)
        \end{tabular}
        & \(4.5\times10^{-20}\) & \(7.34\times10^{4}\)
        & 128 & 1 \\
        P7 & \(10^{-4}\) &
        \((15\textrm{-to-}1)_{7,3,3}\)
        & \(4.4\times10^{-8}\) & \(8.10\times10^{2}\)
        & 18.1 & 1 \\
        P8 & \(10^{-4}\) &
        \((15\textrm{-to-}1)_{9,3,3}\)
        & \(9.3\times10^{-10}\) & \(1.15\times10^{3}\)
        & 18.1 & 1 \\
        P9 & \(10^{-4}\) &
        \((15\textrm{-to-}1)_{11,5,5}\)
        & \(1.9\times10^{-11}\) & \(2.07\times10^{3}\)
        & 30 & 1 \\
        P10 & \(10^{-4}\) &
        \begin{tabular}[t]{@{}l@{}}
        \((15\textrm{-to-}1)^4_{9,3,3}\times\)\\
        \((20\textrm{-to-}4)_{15,7,9}\)
        \end{tabular}
        & \(2.4\times10^{-15}\) & \(1.64\times10^{4}\)
        & 90.3 & 4 \\
        P11 & \(10^{-4}\) &
        \begin{tabular}[t]{@{}l@{}}
        \((15\textrm{-to-}1)^4_{9,3,3}\times\)\\
        \((15\textrm{-to-}1)_{25,9,9}\)
        \end{tabular}
        & \(6.3\times10^{-25}\) & \(1.86\times10^{4}\)
        & 67.8 & 1 \\
        \hline
    \end{tabular}
\end{table*}

\subsection{Space-time resource estimates}
\begin{figure}[t]
    \centering
    \includegraphics[width=\linewidth]{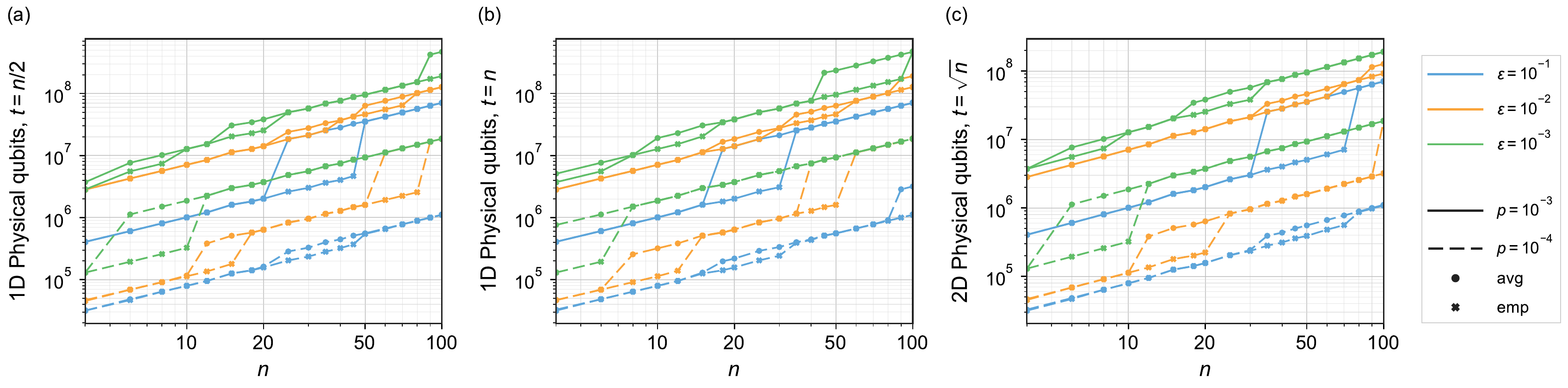}
    \caption{\textbf{Physical-qubit resource estimates for the surface-code
    layout with global MSD factory throughput.}  For each point, the estimate
    scans the MSD factory table, chooses the smallest odd code distance
    satisfying the error budget, and reports the feasible protocol with
    minimum \(Q_{\rm tot}\).
    }
    \label{fig:time-optimal-global-qubits}
\end{figure}

The qubit-resource curves in Fig.~\ref{fig:time-optimal-global-qubits} should
be read as the sum of a layout term and a factory term.  The layout term,
\(2d^2S_{\rm act}\), changes when the selected code distance changes.  The
factory term, \(F_{\rm tot}Q_{\rm fac}\), changes when either the required
throughput or the selected MSD protocol changes.  In the present global-supply
model
\[
    F_{\rm tot}=\left\lceil \frac{w\tau_{\rm fac}}{m_{\rm out}}\right\rceil ,
\]
so the factory footprint is not divided by the code distance.  Consequently,
several points can have similar physical-qubit counts if they select the same
factory protocol, even when their \texttt{T}-counts and runtimes are different.
The sharp changes in the qubit curves correspond to discrete protocol or
distance changes.  In particular, tightening \(\varepsilon\) or increasing
\(N_T\) makes the magic-state term \(N_Tp_{\rm out}\)
harder to satisfy, which can force the scan to choose a lower-error and more
expensive factory protocol.

\begin{figure}[ht]
    \centering
    \includegraphics[width=\linewidth]{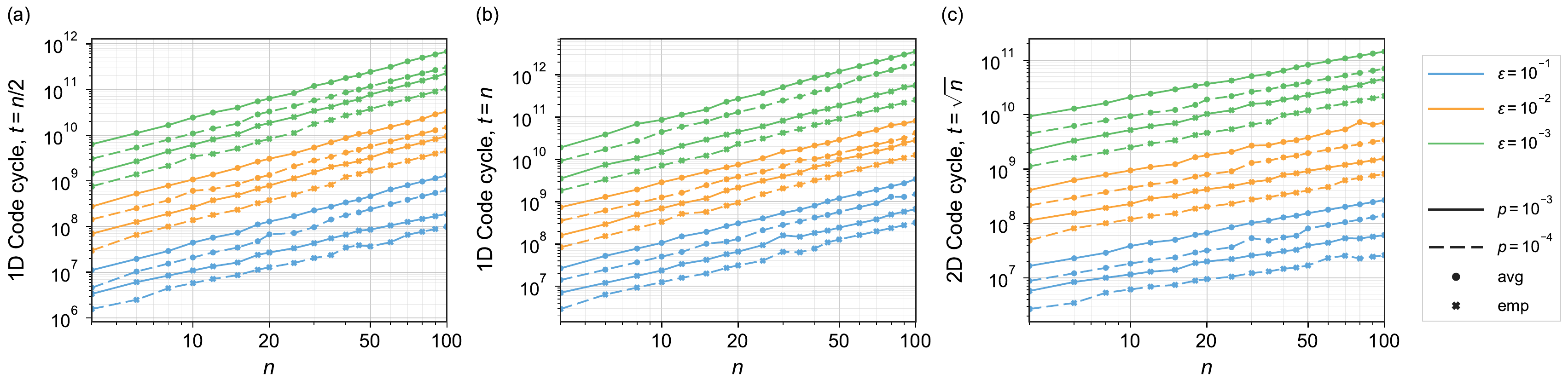}
    \caption{\textbf{Runtime estimates in code cycles.}  The runtime is
    \(2dD_T\), so it depends on the fitted $\texttt{T}$-count, the parallel width
    \(w=\lceil n/2\rceil\), and the selected code distance \(d\).}
    \label{fig:time-optimal-global-runtime}
\end{figure}

The runtime curves in Fig.~\ref{fig:time-optimal-global-runtime} should be
read differently.  Once enough factories are allocated to meet the global
throughput constraint, the factory count does not add a separate waiting time.
The runtime is instead controlled by the effective \texttt{T}-depth and the
distance-scaled layer-consumption time
\[
    \tau_{\rm run}
    =
    2dD_T .
\]
Thus, the average and empirical \texttt{T}-count fits mainly separate the runtime
curves through \(D_T\), while \(p\) enters through the selected distance
\(d\).  The labels \(T=n/2\), \(T=n\), and \(T=\sqrt n\) refer to different fitted \texttt{T}-count models; the hardware parallel width used in the resource estimate is fixed as \(w=\lceil n/2\rceil\).

Table~\ref{tab:resource-n100} gives a representative numerical slice at
\(n=100\), using the empirical \(T\)-count fit and the three algorithmic error
budgets \(\varepsilon=10^{-1},10^{-2},10^{-3}\).  Each entry compares the same
workload and error target at two physical error rates, \(p=10^{-3}\) and
\(p=10^{-4}\).  The code-cycle column is the native runtime from
Eq.~\ref{eq:time-optimal-runtime}, while the final column converts the same
value to seconds using a 500 ns code cycle.
Table~\ref{tab:resource-n100-metadata} records the discrete choices underlying
Table~\ref{tab:resource-n100}.  For each entry, it reports the selected code
distance, the factory protocol ID from Table~\ref{tab:selected-msd-protocol-key},
the number of global factory lanes \(F_{\rm tot}\), and the resulting total
failure estimate \(p_{\rm fail}\).  This table is useful for identifying when
changes in the resource table come from a distance change versus a factory
protocol change.

The main conclusion is that the estimate is space-limited by global
magic-state supply and time-limited by effective \texttt{T}-depth. Reducing
\(p\) can lower the layout contribution by reducing \(d\), but it
does not remove the factory-throughput cost.  Conversely, reducing the
\texttt{T}-count strongly improves runtime and space-time volume, but it may not
substantially reduce the qubit footprint unless it also changes the selected
factory protocol or code distance.

\begin{table*}[htb]
    \centering
    \small
    \caption{\textbf{MSD resource requirements at \(n=100\).}
    The table uses the empirical \(T\)-count fit.  Each entry is
    \(p=10^{-3}/p=10^{-4}\), and bold indicates the smaller value.  Runtime is
    converted from code cycles using 500 ns per cycle.}
    \label{tab:resource-n100}
    \renewcommand{\arraystretch}{1.08}
    \begin{tabular}{@{}c|c c c c@{}}
        \toprule[0.9pt]
        \(\epsilon\) & Code cycles & Physical qubits & Space-time cost
        & Running time \\
        \midrule[0.6pt]
        \(\epsilon=10^{-1}~(T=n/2)\)
        & \(1.904\times10^{8}/\mathbf{9.976\times10^{7}}\)
        & \(7.061\times10^{7}/\mathbf{1.110\times10^{6}}\)
        & \(1.345\times10^{16}/\mathbf{1.107\times10^{14}}\)
        & \(95.22/\mathbf{49.88}\) \\
        \(\epsilon=10^{-2}~(T=n/2)\)
        & \(9.553\times10^{9}/\mathbf{4.600\times10^{9}}\)
        & \(1.270\times10^{8}/\mathbf{1.861\times10^{7}}\)
        & \(1.214\times10^{18}/\mathbf{8.560\times10^{16}}\)
        & \(4776/\mathbf{2300}\) \\
        \(\epsilon=10^{-3}~(T=n/2)\)
        & \(2.330\times10^{11}/\mathbf{1.059\times10^{11}}\)
        & \(1.912\times10^{8}/\mathbf{1.864\times10^{7}}\)
        & \(4.456\times10^{19}/\mathbf{1.975\times10^{18}}\)
        & \(1.165\times10^{5}/\mathbf{5.296\times10^{4}}\) \\
        \(\epsilon=10^{-1}~(T=\sqrt n)\)
        & \(6.109\times10^{7}/\mathbf{2.618\times10^{7}}\)
        & \(7.061\times10^{7}/\mathbf{1.087\times10^{6}}\)
        & \(4.314\times10^{15}/\mathbf{2.847\times10^{13}}\)
        & \(30.55/\mathbf{13.09}\) \\
        \(\epsilon=10^{-2}~(T=\sqrt n)\)
        & \(1.568\times10^{9}/\mathbf{8.155\times10^{8}}\)
        & \(9.222\times10^{7}/\mathbf{3.200\times10^{6}}\)
        & \(1.446\times10^{17}/\mathbf{2.609\times10^{15}}\)
        & \(784.1/\mathbf{407.8}\) \\
        \(\epsilon=10^{-3}~(T=\sqrt n)\)
        & \(4.548\times10^{10}/\mathbf{2.200\times10^{10}}\)
        & \(1.912\times10^{8}/\mathbf{1.864\times10^{7}}\)
        & \(8.693\times10^{18}/\mathbf{4.102\times10^{17}}\)
        & \(2.274\times10^{4}/\mathbf{1.100\times10^{4}}\) \\
        \(\epsilon=10^{-1}~(T=n)\)
        & \(6.673\times10^{8}/\mathbf{3.191\times10^{8}}\)
        & \(7.066\times10^{7}/\mathbf{1.110\times10^{6}}\)
        & \(4.715\times10^{16}/\mathbf{3.541\times10^{14}}\)
        & \(333.6/\mathbf{159.6}\) \\
        \(\epsilon=10^{-2}~(T=n)\)
        & \(2.835\times10^{10}/\mathbf{1.271\times10^{10}}\)
        & \(1.271\times10^{8}/\mathbf{1.861\times10^{7}}\)
        & \(3.603\times10^{18}/\mathbf{2.365\times10^{17}}\)
        & \(1.417\times10^{4}/\mathbf{6354}\) \\
        \(\epsilon=10^{-3}~(T=n)\)
        & \(5.618\times10^{11}/\mathbf{2.554\times10^{11}}\)
        & \(4.704\times10^{8}/\mathbf{1.864\times10^{7}}\)
        & \(2.643\times10^{20}/\mathbf{4.760\times10^{18}}\)
        & \(2.809\times10^{5}/\mathbf{1.277\times10^{5}}\) \\
        \bottomrule[0.9pt]
    \end{tabular}
\end{table*}

\begin{table*}[t]
    \centering
    % \scriptsize
    \caption{\textbf{Selected distances and factories for the \(n=100\),
    empirical-fit entries in Table~\ref{tab:resource-n100}.}
    Each entry is \(d/\mathrm{protocol}/F_{\rm tot}/p_{\rm fail}\), reported
    separately for \(p=10^{-3}\) and \(p=10^{-4}\).}
    \label{tab:resource-n100-metadata}
    \renewcommand{\arraystretch}{1.05}
    \begin{tabular}{@{}c c c c@{}}
        \toprule[0.9pt]
        Workload & \(\varepsilon\) & \(p=10^{-3}\) & \(p=10^{-4}\) \\
        \midrule[0.6pt]
        \(1\mathrm{D},~T=n/2\) & \(0.1\)
        & \(21/\mathrm{P2}/1625/5.49\times10^{-2}\)
        & \(11/\mathrm{P8}/906/1.33\times10^{-2}\) \\
        \(1\mathrm{D},~T=n/2\) & \(0.01\)
        & \(27/\mathrm{P4}/4125/4.75\times10^{-3}\)
        & \(13/\mathrm{P10}/1129/1.29\times10^{-3}\) \\
        \(2\mathrm{D},~T=\sqrt n\) & \(0.1\)
        & \(21/\mathrm{P2}/1625/1.76\times10^{-2}\)
        & \(9/\mathrm{P8}/906/7.67\times10^{-2}\) \\
        \(2\mathrm{D},~T=\sqrt n\) & \(0.01\)
        & \(25/\mathrm{P3}/1963/7.94\times10^{-3}\)
        & \(13/\mathrm{P9}/1500/2.82\times10^{-3}\) \\
        \(1\mathrm{D},~T=n\) & \(0.1\)
        & \(23/\mathrm{P2}/1625/2.38\times10^{-2}\)
        & \(11/\mathrm{P8}/906/4.27\times10^{-2}\) \\
        \(1\mathrm{D},~T=n\) & \(0.01\)
        & \(29/\mathrm{P4}/4125/6.53\times10^{-3}\)
        & \(13/\mathrm{P10}/1129/3.56\times10^{-3}\) \\
        \bottomrule[0.9pt]
    \end{tabular}
\end{table*}

% \sun{This follows standard T state injection.}
% They use   asymmetric surface-code patches with different distances for the logical $X$ and $Z$ operators, as well as a distinct temporal distance that governs the number of code cycles used for lattice-surgery measurements. This enables fine-grained control over logical error rates, ensuring that each error mechanism is suppressed only to the level required to meet the target output fidelity. Faulty logical $T$ operations are realised using either state injection or faulty $T$-measurement protocols. 
% Resource costs are quantified primarily in terms of {qubit cycles}, defined as the product of the total number of physical qubits used (including ancillas) and the number of surface-code cycles required. This metric provides a meaningful measure of space--time overhead that is independent of the code distance required for storing data qubits in a full computation. For reference, costs are also sometimes reported in units of the ``full-distance'' volume $d^{3}$, where $d$ is the code distance required to store logical data qubits for the duration of an algorithm with a given number of $T$ gates. In this work, qubit cycles are taken as the primary figure of merit, as they directly capture the cost of magic state production itself.

%-------------Classical TN --------------
% \clearpage
% \newpage
% \clearpage

\section{ Resource analysis for classical simulation  }

\subsection{TN and tVMC methods}
% We perform classical simulations of the mixed-field Ising model.  We
% track the truncation errors generated by TN compression\gyt{, and mean observable errors in tVMC}. 
\gyt{We perform classical simulations of the mixed-field Ising model using TN (MPS, PEPS-SU, and TEBD), tVMC based on a Jastrow ansatz in 1D and a PEPS ansatz in 2D, and full state-vector propagation. For the TN methods, we track the errors generated by bond compression. The performance of tVMC is assessed from the deviation of the observable dynamics from full state-vector simulations.} The Hamiltonian is
\begin{equation}
H=J\sum_{\langle n,m\rangle}X_{n}X_{m}
+h_{x}\sum_{n}X_{n}+h_{z}\sum_{n}Z_{n}\,,
\label{eq:H-Ising}
\end{equation}
where $\langle n,m\rangle$ runs over nearest-neighbour pairs.  We use
the same couplings in all simulations.  We set $J=1$, $h_{x}=0.8$, and
$h_{z}=0.9$. Because the transverse and longitudinal fields compete,
the model is non-integrable; its entanglement therefore grows rapidly
under time evolution.

\begin{figure}[ht]
  \centering
  \includegraphics[width=1\linewidth]{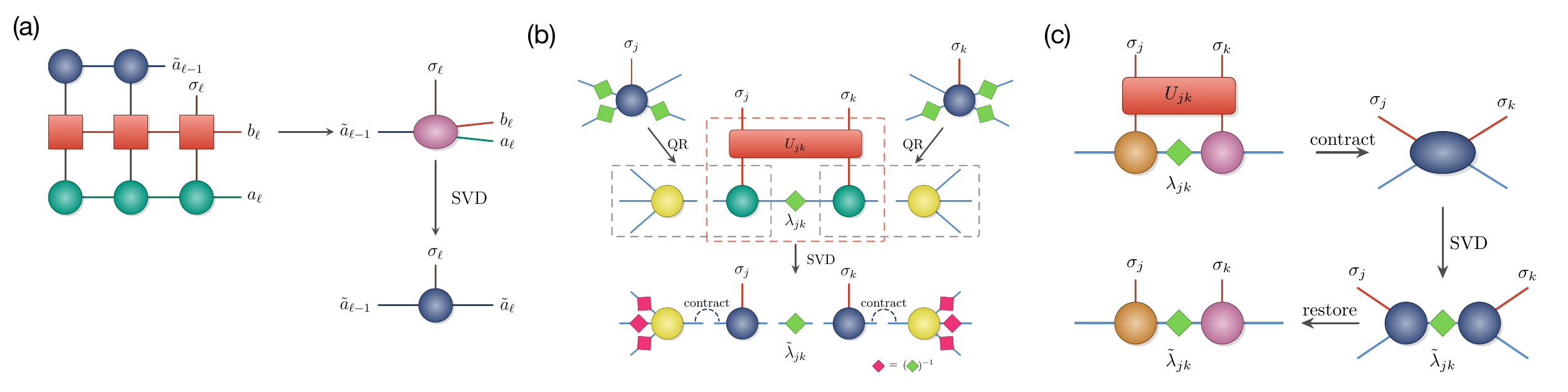}
  \caption{
  \textbf{Local TN update procedures.} (a) Local MPU update of the $\ell$th MPS tensor.
  (b) PEPS simple update on the bond $(j,k)$.
  (c) Two-site TEBD update on the bond $(j,k)$. For a 2D lattice represented by a 1D MPS ordering, the same TEBD update is applied after mapping the 2D sites onto the MPS chain. Vertical bonds become non-nearest-neighbour two-site gates.
  }
  \label{fig:tn-updates}
\end{figure}

Fig.~\ref{fig:tn-updates} summarises the local update structures used
in the TN calculations.  Fig.~\ref{fig:tn-updates}(a) shows the MPU update.
The local MPU tensor is contracted with the current MPS tensor and the
previously updated neighbouring leg.  The two outgoing legs are fused
and compressed by SVD.  This produces the updated effective MPS leg
while preserving the physical spin leg.  Fig.~\ref{fig:tn-updates}(b) shows the PEPS
simple update.  The two PEPS tensors connected by the bond $(j,k)$ are
first dressed by the surrounding bond weights~$\lambda$.  These weights
are distinct on different virtual bonds.  After QR factorisations the
central two-site object is evolved with the gate~$U_{jk}$ and truncated
by SVD.  The updated bond weight~$\tilde\lambda_{jk}$ and the updated
local PEPS tensors are reconstructed by contracting back the
corresponding $Q$ factors.  Inverse external weights are restored on
the non-updated legs.  Fig.~\ref{fig:tn-updates}(c) shows the two-site TEBD
update
used for comparison.  The two-site gate~$U_{jk}$ is applied to neighbouring
MPS tensors together with the bond weight~$\lambda_{jk}$.  The
resulting two-site tensor is split by a truncated SVD to obtain the
updated tensors and $\tilde\lambda_{jk}$.  To apply TEBD to a 2D lattice, the lattice is embedded into a 1D MPS ordering. Horizontal bonds are typically mapped to nearest-neighbour two-site gates.  Vertical bonds become non-nearest-neighbour gates along the MPS chain.  In our implementation,
this 2D TEBD calculation is performed using the ITensor package. These three updates collect the common local compression steps used for the TN simulations.  Their concrete instantiation
(geometry, Trotter order, time horizon and diagnostics) is given in
the 1D and 2D benchmark subsections below.

\gyt{In tVMC, the time-evolved state is approximated by a variational
  wavefunction $|\psi_{\boldsymbol{\theta}}(t)\rangle$ with time-dependent
  parameters $\boldsymbol{\theta}(t)$. The equations of motion are
  obtained by projecting the Schrödinger equation onto the tangent space
  of the variational manifold. Using Monte Carlo samples drawn from
  $|\psi_{\boldsymbol{\theta}}(\boldsymbol{\sigma})|^2$, the parameter
  updates satisfy
  \begin{equation}
  \sum_j S_{ij}\dot{\theta}_j=-iF_i,
  \label{eq:tvmc-eom}
  \end{equation}
  where
  \begin{align}
  S_{ij}
  &=
  \langle O_i^*O_j\rangle
  -\langle O_i^*\rangle\langle O_j\rangle,\\
  F_i
  &=
  \langle O_i^*E_{\mathrm{loc}}\rangle
  -\langle O_i^*\rangle\langle E_{\mathrm{loc}}\rangle.
  \end{align}
  Here $O_i(\boldsymbol{\sigma})
  =\partial_{\theta_i}\log
  \psi_{\boldsymbol{\theta}}(\boldsymbol{\sigma})$ and
  $E_{\mathrm{loc}}(\boldsymbol{\sigma})$ is the local energy.}

\gyt{Before applying tVMC, we rotate every spin by a Hadamard gate, so that
  \(X\) and \(Z\) are exchanged. The Hamiltonian used by the Monte Carlo
  estimator is therefore
  \(J\sum_{\langle n,m\rangle}Z_nZ_m+h_x\sum_n Z_n+h_z\sum_n X_n\).
  This basis choice avoids a support-mismatch problem. In the original
  computational basis, the product initial state is supported on a single
  bit string, whereas the real-time evolution immediately generates
  amplitude on configurations outside this initial support. A Markov chain
  sampling only the initial support would therefore give poor estimates of
  the tangent-space metric and force at early times. After the Hadamard
  rotation, the same physical initial state is an equal-amplitude state in
  the sampling basis, so the Monte Carlo distribution has full support from
  the start.} \gyt{For the tVMC calculations, the operator estimators \(O\) and the local
  energy \(E_{\mathrm{loc}}\) are evaluated from \(81920\) Monte Carlo
  samples. The SR linear system is solved differently for the two ansatz
  families. For the 1D Jastrow-Feenberg ansatz, the SR matrix is treated by
  a direct eigenvalue pseudoinverse with a cutoff on small singular values,
  without adding a diagonal shift. For the 2D PEPS ansatz, we instead solve
  a damped positive-definite system \((S+\lambda I)\dot{\theta}=-iF\) using Cholesky solver which is much more GPU friendly for large parameter count.}

\gyt{We integrate the resulting ordinary differential equation for the
  variational parameters with a fourth-order Runge-Kutta scheme and an
  adaptive physical step size
  \begin{equation}
  \Delta t =
  \sqrt{\frac{I_c}{\mathrm{Var}(H)}} ,
  \label{eq:tvmc-adaptive-step}
  \end{equation}where
  \(\mathrm{Var}(H)=
  \langle |E_{\mathrm{loc}}|^2\rangle
  -|\langle E_{\mathrm{loc}}\rangle|^2\).} \gyt{For the 1D calculations, we use a local multi-body
  Jastrow-Feenberg ansatz. In the rotated spin basis used for Monte Carlo
  sampling, a configuration is denoted by
  \(\boldsymbol{\sigma}=(\sigma_1,\ldots,\sigma_L)\), with
  \(\sigma_i=\pm 1\). The variational wavefunction is
  \begin{equation}
  \psi_{\boldsymbol{\theta}}(\boldsymbol{\sigma})
  =
  \exp\left[
  \sum_{k=1}^{R}
  \sum_{\alpha \in \mathcal{W}_k^{(R)}}
  J^{(k)}_{\alpha}
  \prod_{i\in \alpha}\sigma_i
  \right],
  \label{eq:jf-ansatz-1d}
  \end{equation}
  where \(R\) is the maximum body order. The set
  \(\mathcal{W}_k^{(R)}\) contains all \(k\)-site subsets that lie within
  a contiguous window of \(R\) sites along the chain. Thus the ansatz
  includes local one-body, two-body, and higher-body correlators up to
  order \(R\), while keeping the number of variational parameters
  controlled by locality. For open boundary conditions, only windows
  contained inside the chain are included. Spatial symmetries are
  imposed at the parameter level: terms related by the reflection operations share the same coefficient
  \(J^{(k)}_{\alpha}\).}

\gyt{For the 2D tVMC baseline, unlike PEPS-SU, PEPS-tVMC does not apply and truncate each local gate separately. Instead, all PEPS tensor parameters are evolved by the global
  TDVP equations using Monte Carlo estimates of the metric and force. This
  removes the local-projection bias of simple update, but it introduces a
  different bottleneck: each stochastic-reconfiguration step requires PEPS
  amplitude ratios, logarithmic derivatives with respect to all tensor
  entries, and approximate boundary contractions for many Monte Carlo
  samples. The resulting cost grows rapidly with \(\chi\), the boundary
  dimension and the number of samples.}

\gyt{For the 1D tVMC runs, the ansatz is the multi-body Jastrow-Feenberg
  form in Eq.~\ref{eq:jf-ansatz-1d}, motivated by its demonstrated
  effectiveness in simulations of spin-glass quantum
  annealing~\cite{mauron2025challenging}. We use open boundary conditions
  and impose reflection symmetry, so Jastrow terms related by reversing the
  chain share the same variational coefficient. The body order \(R\) is
  swept in the benchmark. For a fixed \(R\), the ansatz includes every
  \(k\)-body product, \(1\le k\le R\), whose sites fit inside a contiguous
  \(R\)-site window. Increasing \(R\) therefore expands the variational
  manifold by adding progressively larger local correlation clusters. The
  centre-site observable is compared with exact diagonalisation, and the
  reported error is the time-averaged absolute deviation over the
  trajectory
  \begin{equation}
  \varepsilon_Z =
  \frac{1}{T}\int_0^T
  \left|
  \langle Z_{\mathrm{center}}(t)\rangle_{\mathrm{tVMC}}
  -
  \langle Z_{\mathrm{center}}(t)\rangle_{\mathrm{exact}}
  \right|\,\mathrm{d}t .
  \label{eq:eps-z-tvmc}
  \end{equation}
  This metric is deliberately based on a simple local observable, which is
  usually favourable to classical simulation. The fact that the error
  remains difficult to reduce therefore gives a conservative classical
  baseline.}

\gyt{For the 2D tVMC runs, PEPS-tVMC is shown to be accurate on 2D lattice~\cite{wu2025real}. During
  real-time evolution, off-diagonal local-energy terms are evaluated from
  PEPS amplitude ratios, and logarithmic derivatives are computed
  analytically from PEPS environment contractions. To make the calculation
  feasible, we use compressed single-layer boundary contractions with a
  finite boundary dimension. The stochastic-reconfiguration solve updates
  all PEPS tensors globally rather than applying a sequence of local
  truncations.}

\gyt{The 2D PEPS-tVMC benchmark is compared against full state-vector data
  using the endpoint correlation. In the notation of the rotated basis used
  for the reference data, the measured error is the mean absolute deviation
  of the endpoint correlator over the final window of physical time
  \begin{equation}
  \varepsilon_{XX} =
  \frac{1}{\Delta T}
  \int_{T-\Delta T}^{T}
  \left|
  C_{XX}^{\mathrm{tVMC}}(t)
  -
  C_{XX}^{\mathrm{exact}}(t)
  \right|\,\mathrm{d}t ,
  \qquad
  C_{XX}(t)=
  \langle X_{(1,1)}X_{(N_x,N_y)}\rangle(t),
  \label{eq:eps-xx-tvmc}
  \end{equation}
  with \(\Delta T=1\) in our plots. This endpoint correlator probes
  correlations across the largest distance on the finite 2D lattice.}

\subsection{Classical baseline by exact simulation methods}
To set the classical baseline reported alongside TN and tVMC in Figs.~\ref{fig:dim1}
and~\ref{fig:dim2}, we additionally use four exact methods of
{full state-vector simulation} that keeps the entire wavefunction
$|\psi\rangle\in\mathbb{C}^{2^{n}}$ in memory and apply $e^{-iHt}$ to it: (i) a restarted Krylov subspace
propagator, (ii) a
fourth-order Trotter product formula, (iii) a Chebyshev polynomial
expansion, and (iv) a truncated Taylor
series scheme.  The first three are run on multiple GPUs in parallel to accelerate the computation. The Taylor scheme runs on a single CPU as an independent reference. The Krylov and Trotter methods serve as the
single-precision (Complex64) baselines for the GPU
runtime curves. The Chebyshev expansion is run in double precision
(Complex128) to give the reference $\psi_{\mathrm{Cheb}}$ used in the
operator-error metric below.  All four methods apply to both the
1D and 2D lattices. The algorithm details of the four propagators are as follows.

\paragraph{Restarted Krylov subspace propagator.}
The interval $[0,T]$ is split into $N$ outer sub-steps of size
$\tau=T/N$.  Within each sub-step we build an $m$-dimensional Krylov
subspace
\begin{equation}
\mathcal{K}_{m}(H,\psi)=\mathrm{span}\{\psi,H\psi,\ldots,H^{m-1}\psi\}
\end{equation}
via the Lanczos three-term recurrence with full re-orthogonalisation.
The orthonormal basis $V_{m}\in\mathbb{C}^{N\times m}$ and the real
tridiagonal matrix $T_{m}=V_{m}^{\dagger}HV_{m}\in\mathbb{R}^{m\times m}$
give
\begin{equation}
\psi(\tau)\;\approx\;V_{m}\,e^{-iT_{m}\tau}\,e_{1},
\label{eq:krylov-step}
\end{equation}
where the small matrix exponential $e^{-iT_{m}\tau}$ is evaluated in
double precision via the eigendecomposition of $T_{m}$.  The Krylov
subspace is discarded at the end of each sub-step and rebuilt around
the updated state, so memory grows with $m$ but not with $N$.  Each
sub-step costs $m$ applications of $H$.  The single-step truncation
error obeys the Hochbruck-Lubich bound
\begin{equation}
\bigl\|\psi_{m}(\tau)-e^{-iH\tau}\psi\bigr\|
\;\le\;12\,\exp\!\left(-\frac{(m-\tau\|H\|)^{2}}{16\,\tau\|H\|}\right),
\label{eq:krylov-bound}
\end{equation}
valid for $m\ge\tau\|H\|$.  The free parameters are $(\tau,m)$.  We
adjust them per system size so that the final-time operator error
matches a chosen band.

\paragraph{Fourth-order Trotter product formula.}
An alternative that does not require a Krylov subspace and that
matches the gate-based structure of the TN update
exploits a splitting of the Hamiltonian.  The Hamiltonian splits
into a diagonal block
$H_{z}=h_{z}\sum_{n}Z_{n}$ and an off-diagonal block
$H_{xc}=h_{x}\sum_{n}X_{n}+J\sum_{\langle n,m\rangle}X_{n}X_{m}$.
The two blocks do not commute, but every term inside $H_{xc}$ commutes
with every other term inside $H_{xc}$. The second-order Trotter splitting is
\begin{equation}
S_{2}(dt)=e^{-iH_{z}dt/2}\,e^{-iH_{xc}dt}\,e^{-iH_{z}dt/2},
\end{equation}
and the Forest-Ruth fourth-order
composition is
\begin{equation}
\mathrm{PF4}(dt)=S_{2}(\gamma_{1}dt)\,S_{2}(\gamma_{0}dt)\,S_{2}(\gamma_{1}dt),
\qquad
\gamma_{1}=\frac{1}{2-2^{1/3}},\;\;
\gamma_{0}=-2^{1/3}\gamma_{1}.
\label{eq:pf4}
\end{equation}
Each $S_{2}$ factor is exact up to floating-point round-off.  $H_{z}$
acts as a diagonal phase.  $H_{xc}$ is a product of mutually
commuting one-site $X_{n}$ rotations and two-site $X_{n}X_{n+1}$
rotations applied sequentially.  After $N_{\mathrm{steps}}=T/dt$
steps the accumulated error scales as $\mathcal{O}(T\,n\,dt^{4})$,
dominated by the nested commutator
$\bigl[H_{z},[H_{z},H_{xc}]\bigr]$.  The free parameter is $dt$.
We choose it so that the final-time operator error lies in the same
band as the Krylov method.

\paragraph{Chebyshev polynomial expansion.}
For the reference run we use a third propagator that expresses
$e^{-iHt}$ as a polynomial in $H$, so that the truncation error is
controlled by a single explicit upper bound rather than by an
empirical step size.  We introduce the rescaled Hamiltonian
\begin{equation}
H_{s}=(H-bI)/a,
\end{equation}
where $a=R$ and the centre offset $b$ is chosen so that the spectrum
of $H_{s}$ lies in $[-1,1]$.  The propagator then admits the
Jacobi-Anger expansion
\begin{equation}
e^{-iH\,dt}=e^{-ib\,dt}\sum_{k=0}^{\infty}c_{k}(a\,dt)\,T_{k}(H_{s}),
\qquad
c_{k}(z)=\begin{cases}J_{0}(z),& k=0,\\ 2(-i)^{k}J_{k}(z),& k\ge 1,\end{cases}
\label{eq:cheb-expansion}
\end{equation}
where $J_{k}$ is the Bessel function of the first kind and $T_{k}$
the Chebyshev polynomial of the first kind.  We use the
operator-norm bound
\begin{equation}
R=|h_{z}|n+|h_{x}|n+|J|(n-1),
\end{equation}
so that no eigenvalue estimation is required.  Truncating
Eq.~\ref{eq:cheb-expansion} at order $K$ leaves a remainder bounded
by
\begin{equation}
\bigl\|\psi_{K}-\psi_{\mathrm{exact}}\bigr\|
\;\le\;2\sum_{k>K}|J_{k}(a\,dt)|,
\label{eq:cheb-bound}
\end{equation}
since $\|T_{k}(H_{s})\|\le 1$ on the rescaled spectrum.  For
$k>a\,dt$ the Bessel tail decays faster than a geometric series, so
the order $K$ is fixed analytically by requiring the right-hand side
of Eq.~\ref{eq:cheb-bound} below a chosen tolerance ($10^{-12}$ in
our reference runs).  The polynomial states $T_{k}(H_{s})\psi$ are
generated by the three-term recurrence
\begin{equation}
T_{k+1}(H_{s})\psi=2H_{s}\,T_{k}(H_{s})\psi-T_{k-1}(H_{s})\psi,
\end{equation}
so only three intermediate vectors are stored, independent of $K$.
Each outer step costs $K$ applications of $H$.  Running this
propagator in double precision yields a reference whose energy and
norm drift remain at the FP64 round-off floor across all system
sizes we report.

\paragraph{Truncated Taylor series.}
The fourth propagator evaluates the action $e^{-iHt}|\psi\rangle$
through a truncated Taylor series without building the matrix
exponential.  The evolution time is split into $s$ equal sub-steps and
the action of $e^{-iHt/s}$ on each sub-step is approximated by an
order-$m$ Taylor polynomial in $H$.  The two integers $(s,m)$ are
chosen automatically from an estimate of the $1$-norm of $H$ so that
the truncation error stays at the double-precision round-off floor.
The scheme needs only the matrix-vector products $H|\psi\rangle$ and
we evaluate it on a single CPU with the \texttt{expm\_multiply}
routine of SciPy's sparse linear-algebra module.  It serves only in
this appendix as an independent cross-check of the other full
state-vector methods and the TN simulations.

The four propagators above specify how each state is advanced in
time.  We next define how their accuracy is quantified.  The full
state-vector baselines and the TN simulations are limited by different error sources.  We therefore use a separate set of diagnostics for each, defined in the two paragraphs below.

\setcounter{paragraph}{0}
\paragraph{Error metrics for the full state-vector simulations.}
Throughout the paper we report two diagnostics for the
single-precision baselines $\tilde\psi$ (the restarted Krylov
propagator and the fourth-order Trotter product formula) against the
double-precision Chebyshev reference $\psi_{\mathrm{Cheb}}$
\begin{equation}
  \varepsilon_{\mathrm{norm}}\equiv\bigl|\,\|\tilde\psi\|^{2}-1\bigr|,
  \qquad
  \varepsilon_{O}\equiv\bigl|\langle Z_{s}\rangle_{\tilde\psi}
                              -\langle Z_{s}\rangle_{\psi_{\mathrm{Cheb}}}\bigr|.
  \label{eq:eps-norm-op-GPU}
\end{equation}
The Chebyshev expansion is truncated so that the Bessel-tail bound in
Eq.~\ref{eq:cheb-bound} stays below $10^{-12}$.  The probed
observable is the single-site operator $Z_{s}$ with $s=5$ in 1D and $s=(2,2)$ in 2D.  For each system size we tune the free parameters of the two baselines (the Trotter step
size $dt$ and the Krylov sub-step $\tau$, inner dimension $m$) so that $\varepsilon_{O}$ at the final time lies in the band $10^{-4}$--$10^{-3}$, matching the $10^{-3}$ target reported in Figs.~\ref{fig:dim1}(b) and \ref{fig:dim2}(a).

\paragraph{Error metrics for the TN simulations.}
For the MPS and PEPS runs we benchmark against a full state-vector
reference $|\psi_{\mathrm{exact}}(t)\rangle = e^{-iHt}|\psi_{0}\rangle$
and report four diagnostics. The state distance is
\begin{equation}
  \varepsilon_{\mathrm{state}}
  \equiv \bigl\| |\psi_{\mathrm{TN}}\rangle
                -|\psi_{\mathrm{exact}}\rangle\bigr\|,
  \label{eq:eps-state}
\end{equation}
where $|\psi_{\mathrm{TN}}\rangle$ stands for $|\psi_{\mathrm{MPS}}\rangle$
in 1D or $|\psi_{\mathrm{PEPS}}\rangle$ in 2D.  To separate the splitting
error of the Trotter scheme from the bond-compression error at the
current $\chi$ (or $D$) we additionally run a full state-vector Trotter
reference $|\psi_{\mathrm{Trot}}(t)\rangle$ on the same Trotterised
Hamiltonian but without any SVD truncation.  Comparing this reference
with both the exact and the TN state defines
\begin{equation}
  \varepsilon_{\mathrm{Trot}}
  \equiv \bigl\| |\psi_{\mathrm{Trot}}\rangle
                -|\psi_{\mathrm{exact}}\rangle\bigr\|,
  \qquad
  \varepsilon_{\mathrm{SVD}}
  \equiv \bigl\| |\psi_{\mathrm{TN}}\rangle
                -|\psi_{\mathrm{Trot}}\rangle\bigr\|,
  \label{eq:eps-Trot-trunc}
\end{equation}
which isolate the two error sources separately. A complementary
diagnostic is the per-gate fidelity loss
\begin{equation}
  \varepsilon_{F}
  = 1-\prod_{i=1}^{N_{\mathrm{total}}}(1-\varepsilon_{i}),
  \qquad
  \varepsilon_{i}
  =1-\frac{\sum_{k=1}^{\chi}s_{k}^{2}}{\sum_{k=1}^{r}s_{k}^{2}},
  \label{eq:eps-F}
\end{equation}
accumulated combinatorially from the per-bond truncation weights
$\varepsilon_{i}$.  In 1D the MPU gates are exactly unitary and the
singular values are kept unnormalised, so each truncation reduces
$\|\psi_{\mathrm{MPS}}\|^{2}$ by the factor $1-\varepsilon_{i}$ and
$\varepsilon_{F}$ collapses to the directly measurable norm deficit
\begin{equation}
  \varepsilon_{F}
  = 1-\langle\psi_{\mathrm{MPS}}|\psi_{\mathrm{MPS}}\rangle.
  \label{eq:eps-F-1D-norm}
\end{equation}
In 2D the singular values are renormalised at every update so the
wavefunction stays normalised, and $\varepsilon_{F}$ must instead be
accumulated explicitly from the per-bond $\varepsilon_{i}$ recorded
during the sweep.  The complementary energy diagnostic is the energy-density error
\begin{equation}
  \varepsilon_{E}
  = \bigl|\langle H\rangle_{\mathrm{TN}}(t) - E_{0}\bigr| / n,
  \qquad E_{0}=\langle\psi_{0}|H|\psi_{0}\rangle,
  \label{eq:eps-E-TN}
\end{equation}
with $n$ the number of sites ($n=N_{x}N_{y}$ in 2D) and
$E_{0}$ the initial energy conserved under exact evolution.
Here $\langle H\rangle_{\mathrm{TN}}
=\langle\psi_{\mathrm{TN}}|H|\psi_{\mathrm{TN}}\rangle
/\langle\psi_{\mathrm{TN}}|\psi_{\mathrm{TN}}\rangle$ is evaluated on the normalised state.

\subsection{One-dimensional dynamics benchmarks}

We now benchmark the TN method against the full state-vector propagators, tracking how their runtime and accuracy scale with the
system size~$n$. We evolve from the all-spin-up initial state
$|\psi_{0}\rangle=\ket{\mathbf{0}}$ on a chain of $n$~sites with open
boundary conditions.  The evolution runs to $T=n$, longer than the $T=n/2$ used in the main text.  We represent the chain
as an MPS of bond dimension $\chi$ and evolve it with the MPU
update in Fig.~\ref{fig:tn-updates}(a).
Time evolution uses a fourth-order Suzuki-Trotter decomposition with time step $dt=0.1$.  The
Hamiltonian is split into odd and even bond layers.  Each resulting
MPU layer is applied by a sequential left-to-right or right-to-left
sweep. SVD truncation is performed at every bond.
We validate the MPS evolution against the Taylor reference
introduced in the \hyperref[sec:methods]{Methods} subsection and report the four diagnostics defined
in Eqs.~\ref{eq:eps-state}-\ref{eq:eps-E-TN}.  Because the MPU
gates are exactly unitary and the MPS norm is left unnormalised,
$\varepsilon_{F}$ is read off directly from the residual norm deficit
of $|\psi_{\mathrm{MPS}}\rangle$ via Eq.~\ref{eq:eps-F-1D-norm}.  It
quantifies the weight lost to SVD truncation and is a complementary
diagnostic to the state-distance pair
$(\varepsilon_{\mathrm{Trot}},\varepsilon_{\mathrm{SVD}})$ of
Eq.~\ref{eq:eps-Trot-trunc}.

\begin{figure}[t!]
  \centering
  \includegraphics[width=0.7\linewidth]{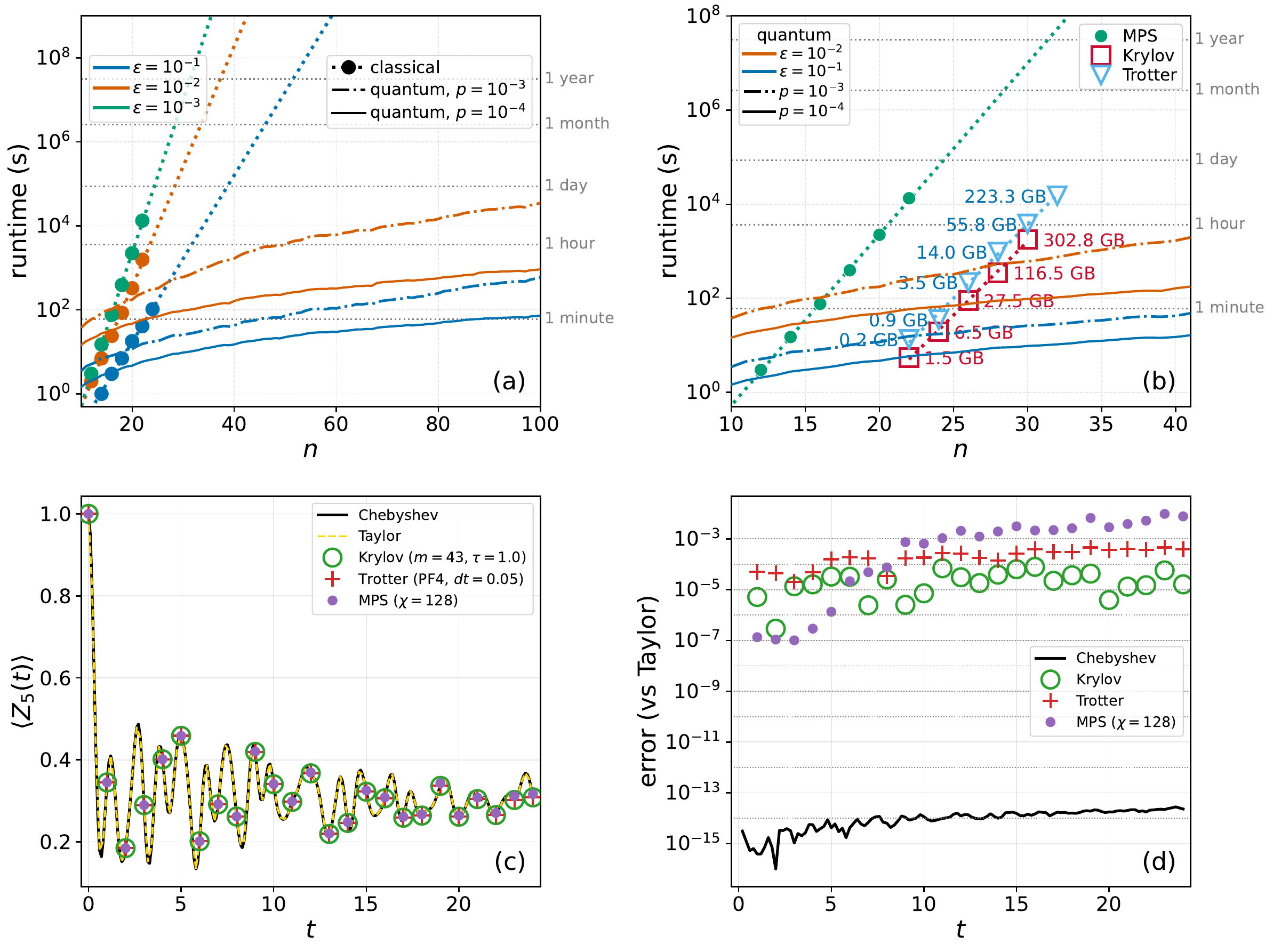}
  \caption{
    \textbf{Classical simulation benchmarks for the 1D mixed-field Ising model
    evolved to $T=n$.}
    \textbf{Top row:} Runtime versus system size.
    (a)~MPS runtime with the bond dimension chosen at each~$n$ to keep
    the truncation error below $\varepsilon=10^{-1}$, $10^{-2}$ and
    $10^{-3}$.
    Dotted lines are exponential fits over all available~$n$.
    (b)~Full state-vector evolution (on up to four GPUs with 80 GB memory)
    with the restarted-Krylov and fourth-order Trotter methods
    compared with MPS at
    $\varepsilon=10^{-3}$.
    Both GPU methods are calibrated to a final operator error of
    $10^{-3}$ against the Chebyshev reference.
    Each GPU data point is labelled with its peak GPU memory usage.
    \textbf{Bottom row:} Single-site magnetisation
    $\langle Z_{5}(t)\rangle$ for $n=24$.
    (c)~Trajectories obtained with the MPS, the two GPU full-state
    methods and the exact references.
    All curves coincide on the scale of the plot.
    (d)~Errors against the Taylor reference.}
  \label{fig:dim1}
\end{figure}

\begin{figure}[tb]
  \centering
  \includegraphics[width=\linewidth]{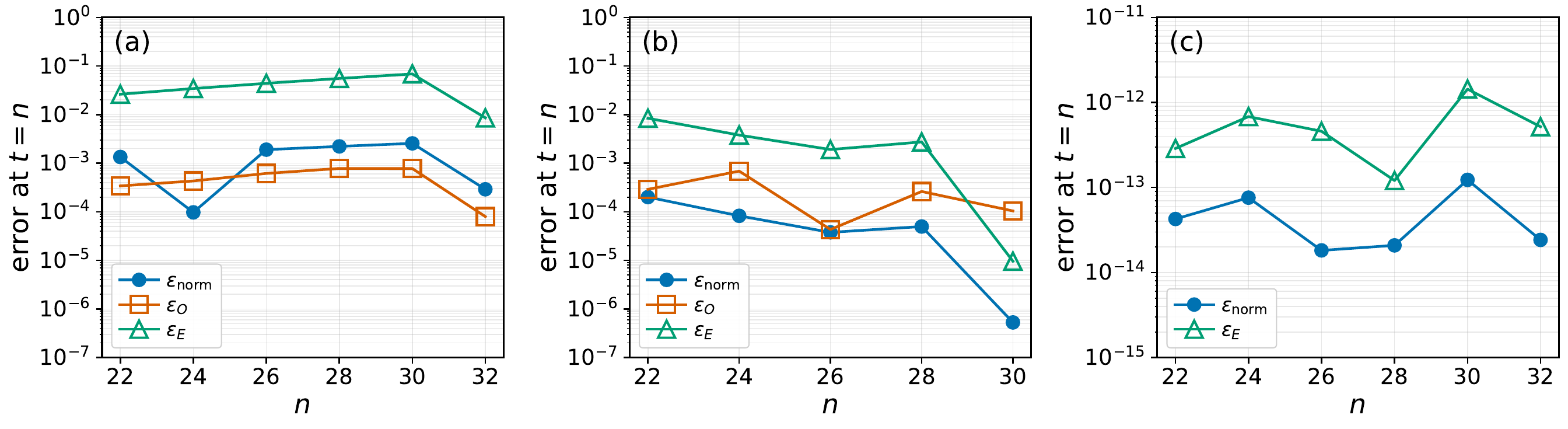}
  \caption{\textbf{Final-time errors on the 1D mixed-field Ising chain.}
    (a) Fourth-order Trotter product formula (Complex64).
    (b) Restarted Krylov subspace propagator (Complex64).
    (c) Chebyshev polynomial expansion (Complex128, used as the reference $\psi_{\mathrm{Cheb}}$) showing only
    $\varepsilon_{\mathrm{norm}}$ and $\varepsilon_{E}$.
    The errors in (a) and (b) are measured against this Chebyshev reference.}
  \label{fig:err-vs-n-1D}
\end{figure}

Fig.~\ref{fig:dim1}(a) presents the runtime when the bond dimension is
raised at each~$n$ until the truncation error falls below the
target~$\varepsilon$.  The cost grows exponentially with~$n$ and the
growth rate steepens as the target tightens.  Fig.~\ref{fig:dim1}(b)
compares the MPS method with the restarted-Krylov and fourth-order Trotter methods at $\varepsilon=10^{-3}$.
Fig.~\ref{fig:err-vs-n-1D} shows the corresponding final-time errors.
Figs.~\ref{fig:dim1}(c,d) benchmark the single-site magnetisation
$\langle Z_{5}(t)\rangle$ at $n=24$ against the Taylor reference.
The Chebyshev reference stays at the round-off floor throughout.

\begin{figure}[t]
  \centering
  \includegraphics[width=0.7\linewidth]{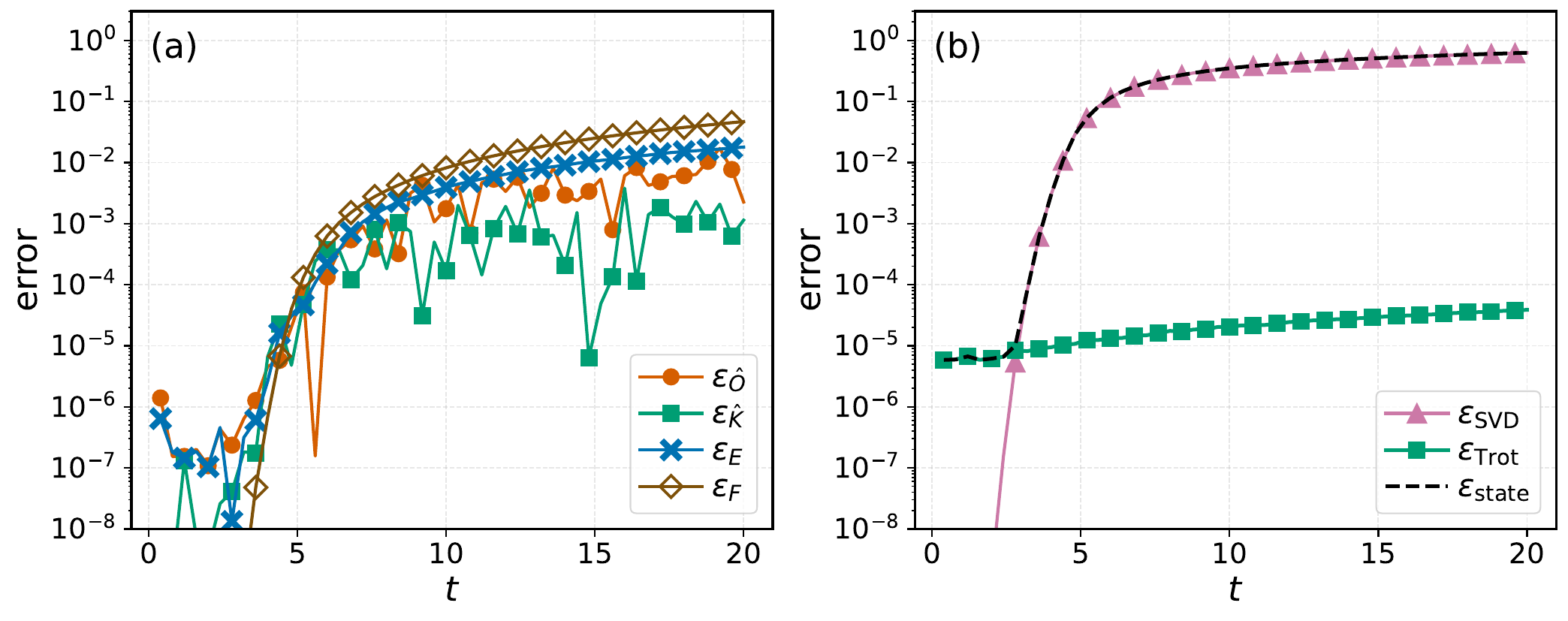}
  \caption{\textbf{Error analysis for a 1D MPS algorithm on a mixed-field
    Ising chain with $n=20$ spins.} The parameters are $\chi=64$,
    fourth-order Trotter $dt=0.1$, and $T=n$.
    (a) Operator-level errors $\varepsilon_{\hat O}$
    (local Pauli at $\hat O = Z_5$) and $\varepsilon_{\hat K}$
    (Pauli string $\hat K = \prod_j Z_j$). The panel also shows
    $\varepsilon_E$ and the per-gate fidelity loss $\varepsilon_F$.
    (b) Decomposition of the state distance $\varepsilon_{\rm state}$
    into $\varepsilon_{\rm Trot}$ and $\varepsilon_{\rm SVD}$.
    All errors are measured against the Taylor reference.}
  \label{fig:dim1-op-state}
\end{figure}

Fig.~\ref{fig:dim1-op-state} resolves the error sources for a single
$n=20$ run at $\chi=64$.  Fig.~\ref{fig:dim1-op-state}(a) tracks the local and global
operator errors $\varepsilon_{\hat O}$ and $\varepsilon_{\hat K}$
together with the energy error $\varepsilon_E$ and the fidelity loss
$\varepsilon_F$, all of which stay below $10^{-1}$ and rise once the
entanglement exceeds the $\chi=64$ capacity around $t\approx4$.
Fig.~\ref{fig:dim1-op-state}(b) splits the state distance $\varepsilon_{\mathrm{state}}$ into the
Trotter error $\varepsilon_{\mathrm{Trot}}$ and the SVD truncation
error $\varepsilon_{\mathrm{SVD}}$.  The state distance is dominated by
the SVD truncation, which saturates near $0.5$.  The Trotter error
stays four orders of magnitude smaller.  The accuracy at fixed $\chi$
is therefore limited by bond compression rather than by the Trotter time step.

\subsection{Two-dimensional dynamics benchmarks}

In 2D the comparison extends to two TN methods,
PEPS and TEBD, both benchmarked against the full state-vector propagators. The PEPS runs use the simple-update
procedure
of Fig.~\ref{fig:tn-updates}(b).  Each two-body gate is absorbed into
the neighbouring site tensors, and the resulting two-site
tensor~$\Theta$ is decomposed by SVD and truncated.  The retained
singular values are renormalised as
$\lambda_{k}\leftarrow s_{k}/\sum_{j}s_{j}$, so
$|\psi_{\mathrm{PEPS}}\rangle$ stays normalised and the per-gate
fidelity loss is accumulated combinatorially from the per-bond weights
$\varepsilon_{i}$, giving $\varepsilon_{F}$ in Eq.~\ref{eq:eps-F}.  The
second method is the two-site TEBD update of Fig.~\ref{fig:tn-updates}(c),
which represents the lattice as a snaked 1D MPS and is run with ITensor. 

\begin{figure}[b]
  \centering
  \includegraphics[width=\linewidth]{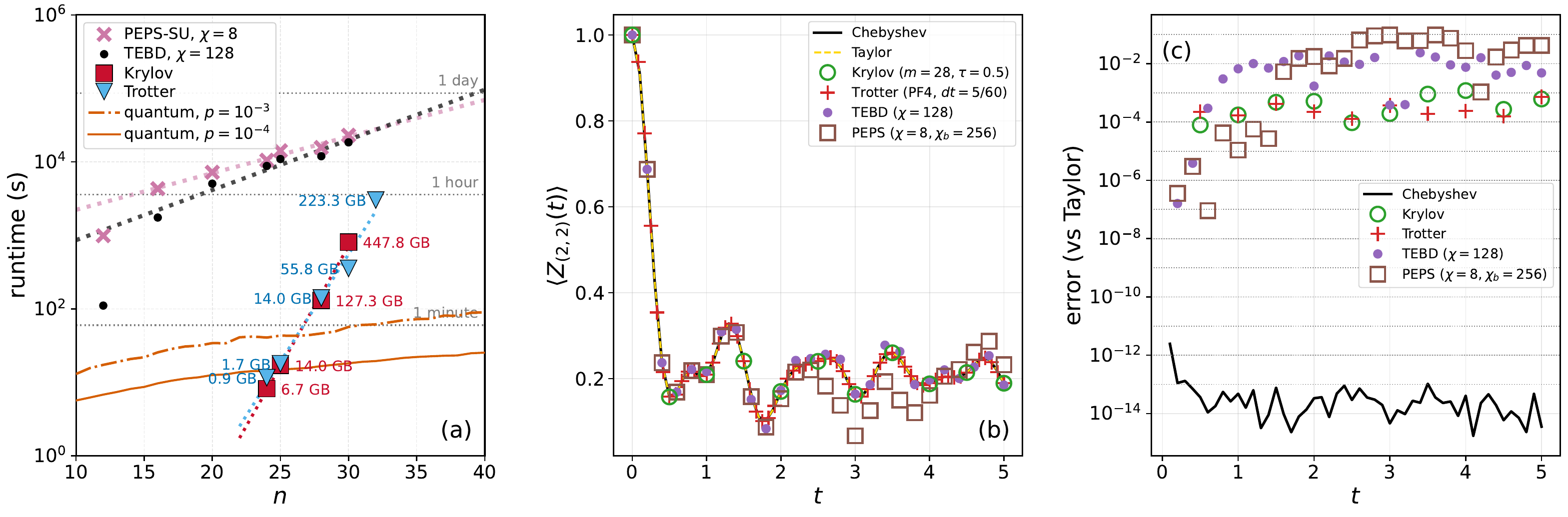}
  \caption{
    \textbf{Classical simulation benchmarks for the 2D mixed-field Ising model
    on open $N_{x}\times N_{y}$ lattices evolved to $T=\sqrt{n}$.}
    (a)~Runtime versus system size for the PEPS simple update at $\chi=8$,
    TEBD at $\chi=128$ and full state-vector evolution (on up to eight
    GPUs with 80 GB memory) with the restarted-Krylov and fourth-order Trotter
    methods. Dotted lines are exponential fits ($n\ge16$ for the TN methods and $n\ge24$ for the GPU methods). The GPU methods are calibrated to a final operator error of
    $10^{-3}$ against the Chebyshev reference.
    Each GPU data point is labelled with its peak GPU memory usage.
    (b)~Single-site magnetisation $\langle Z_{(2,2)}(t)\rangle$ for
    $n=5\times5$.
    The trajectories from the PEPS, the TEBD, the two GPU full-state
    methods and the exact references all coincide on the scale of the
    plot.
    (c)~Errors against the Taylor reference.}
  \label{fig:dim2}
\end{figure}

\begin{figure}[htb]
  \centering
  \includegraphics[width=\linewidth]{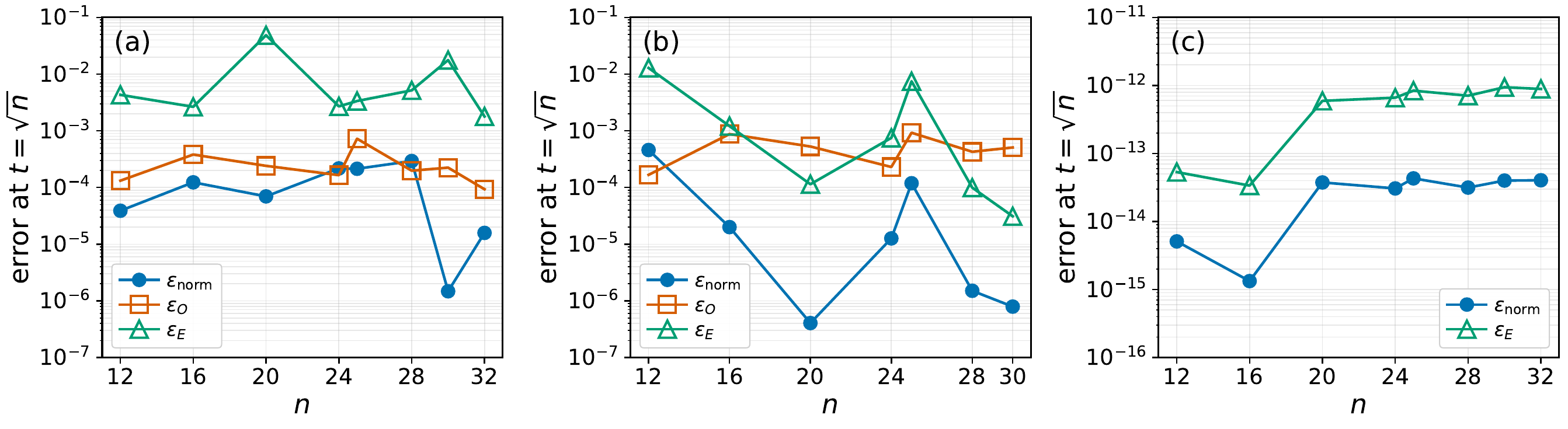}
  \caption{\textbf{Final-time errors on the 2D mixed-field Ising
    $N_x\times N_y$ lattice with $n=N_xN_y$.}
    (a) Fourth-order Trotter product formula (Complex64).
    (b) Restarted Krylov subspace propagator (Complex64).
    (c) Chebyshev polynomial expansion (Complex128, used as the reference $\psi_{\mathrm{Cheb}}$) showing only
    $\varepsilon_{\mathrm{norm}}$ and $\varepsilon_{E}$.
    The errors in (a) and (b) are measured against this Chebyshev reference.}
  \label{fig:err-vs-n-2D}
\end{figure}

\begin{figure}[H]
  \centering
  \includegraphics[width=0.7\linewidth]{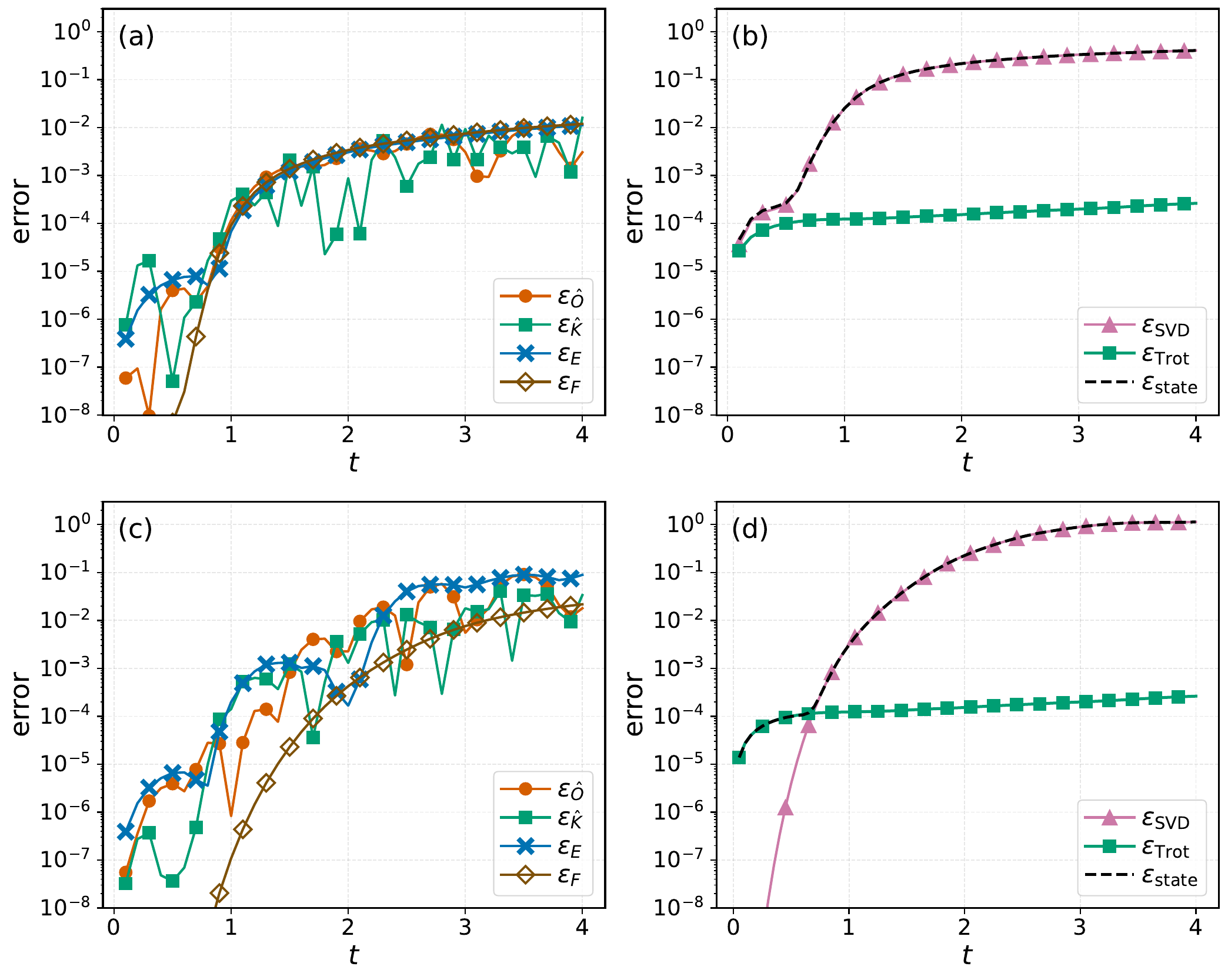}
  \caption{\textbf{Error analysis of two classical TN algorithms on a 2D mixed-field Ising model with $n=4\times4$ spins.} The
    parameters are second-order Trotter $dt=0.01$ and $T=\sqrt{n}$. (a), (b) TEBD at $\chi=128$. (c), (d) PEPS at $\chi=8$ and $\chi_b=256$. The algorithm state is denoted as $\psi_{\rm TN}$ in both cases.
    (a), (c) Operator-level errors
    $\varepsilon_{\hat O}$ (local Pauli at $\hat O = Z_{(2,2)}$) and
    $\varepsilon_{\hat K}$ (Pauli string $\hat K = \prod_j Z_j$). The
    panel also shows $\varepsilon_E$ and the per-gate fidelity loss
    $\varepsilon_F$.
    (b), (d) Decomposition of the state distance
    $\varepsilon_{\rm state}$ into $\varepsilon_{\rm Trot}$ and
    $\varepsilon_{\rm SVD}$.
    All errors are measured against the Taylor reference.}
  \label{fig:dim2-op-state}
\end{figure}

We evolve the all-spin-up product state
$|\psi_{0}\rangle=\ket{\mathbf{0}}$ on $N_{x}\times N_{y}$ square
lattices with open boundary conditions, taking a second-order Trotter step
$dt=0.01$ up to $T=\sqrt{n}$.  For the PEPS runs the energy error
$\varepsilon_{E}$ follows from a single-layer boundary MPS contraction
with $\chi_{\mathrm{b}}=256$.  The resulting runtimes for PEPS at
$\chi=8$ and TEBD at $\chi=128$ are
shown in Fig.~\ref{fig:dim2}(a) together with exponential fits for
$n\ge16$.  The same figure also includes the results of the restarted-Krylov and
fourth-order Trotter full state-vector methods probed through
$Z_{(2,2)}$.  Their Trotter step $dt$ and Krylov parameters
$(\tau,m)$ are set so that the operator error $\varepsilon_{O}$
[Eq.~\ref{eq:eps-norm-op-GPU}] stays below $10^{-3}$.  The corresponding
final-time errors appear in Fig.~\ref{fig:err-vs-n-2D}, and
Figs.~\ref{fig:dim2}(b,c) benchmark the magnetisation
$\langle Z_{(2,2)}(t)\rangle$ on the $5\times5$ lattice against the
Taylor reference. Figs.~\ref{fig:dim2-op-state}(a,b) and~(c,d) resolve the error sources
for $n=4\times4$, with TEBD at $\chi=128$ and PEPS at $\chi=8$, respectively. Figs.~\ref{fig:dim2-op-state}(a,c) track the operator errors
$\varepsilon_{\hat O}$ and $\varepsilon_{\hat K}$ together with
$\varepsilon_E$ and the fidelity loss $\varepsilon_F$, all of which
grow with time and stay below the state distance.  Figs.~\ref{fig:dim2-op-state}(b,d)
split $\varepsilon_{\mathrm{state}}$ into $\varepsilon_{\mathrm{Trot}}$
and $\varepsilon_{\mathrm{SVD}}$.  For both methods the SVD truncation
dominates while the Trotter error stays near $10^{-4}$, and this
bond-compression error grows large by $T=\sqrt{n}$.

\end{document}